%% file: PRL_v2_wnt-YOpolymer.tex
\documentclass[reprint,aps,prl,groupedaddress,10pt]{revtex4-2}

\usepackage{packages}
\usepackage{notations}


\begin{document}

\setlength{\abovedisplayskip}{4pt}
\setlength{\belowdisplayskip}{4pt}

\title{The weak noise theory of the O'Connell-Yor polymer as an integrable discretisation of the nonlinear Schrodinger equation
}

\author{Alexandre Krajenbrink}
\email{alexandre.krajenbrink@quantinuum.com}
\affiliation{Quantinuum, Terrington House, 13–15 Hills Road, Cambridge CB2 1NL, United Kingdom}
\affiliation{Le Lab Quantique, 58 rue d'Hauteville, 75010, Paris, France}

\author{Pierre Le Doussal}
\email{ledou@lpt.ens.fr}
\affiliation{Laboratoire de Physique de l'\'Ecole Normale Sup\'erieure, CNRS, ENS $\&$ PSL University, Sorbonne Universit\'e, Universit\'e de Paris, 75005 Paris, France}

\date{\today}

\begin{abstract}
We investigate and solve the weak noise theory for the semi-discrete O'Connell-Yor directed polymer. In the large deviation regime, the most probable evolution of the partition function obeys a classical non-linear system which is a non-standard discretisation of the nonlinear Schrodinger equation with mixed initial-final conditions. We show that this system is integrable and find its general solution through an inverse scattering method and a novel Fredholm determinant framework that we develop. This allows to obtain the large deviation rate function of the free energy of the
polymer model from its conserved quantities and to study its convergence to the large deviations of the Kardar-Parisi-Zhang equation. Our model also degenerates to the classical Toda chain, which further substantiates the applicability of our Fredholm framework.
\end{abstract}

\maketitle

There has been much recent progress in obtaining exact solutions for the weak noise theory (WNT) of stochastic continuous systems in 1+1 dimensions. 
This was achieved in the context of the Kardar-Parisi-Zhang (KPZ) equation 
\cite{UsWNT2021,UsWNTFlat2021,TsaiIntegrability}, introduced to describes interface growth \cite{KPZ},
and of the macroscopic fluctuation theory (MFT) 
\cite{NaftaliDNLS,krajenbrink2023crossover,grabsch2021closing,mallick2022exact,NaftaliDNLS2,GrabschBenichou2023Review},
which describes diffusive particle systems \cite{BertiniPRL2005,DerridaGershenfeld,DerridaMFTReview2007,BertiniMFT2015}.
The WNT describes the optimal configuration of height or density fields which realizes atypically large fluctuations in the presence of a weak noise
\cite{Korshunov,Baruch,lin2020short}: this is an example of a large deviation problem \cite{TouchetteReview2018}.
The progress was achieved by noticing that the nonlinear systems of equations which arise in WNT are integrable. 
For the KPZ equation, where the WNT describes the short time regime, these equations are a close cousin of the nonlinear Schrodinger equation (NLS), 
while for the MFT the connection is to the derivative nonlinear Schrodinger equation. 
Both equations are integrable using the inverse scattering method 
based on the existence of a Lax pair \cite{ZS,AblowitzKaup1974,kaup1978exact},
but solving the WNT requires handling mixed initial-final time conditions.
These connections to stochastic problems have renewed the interest in these classical integrable systems
due to the new challenge of non-standard mixed boundary conditions,
together with the possibility to solve exactly the equations using Fredholm determinants.
In fact, some of the large deviation rate functions obtained by these classical integrability methods
have also been derived \cite{le2016exact,krajenbrink2017exact,Meerson_flatST,krajenbrink2023crossover} 
from Fredholm determinant formula stemming from quantum integrability.
While complementary, the two approaches are quite distinct, and the connections between them
are not elucidated.

\begin{figure}[h!]
    \centering
    \includegraphics[scale=0.4]{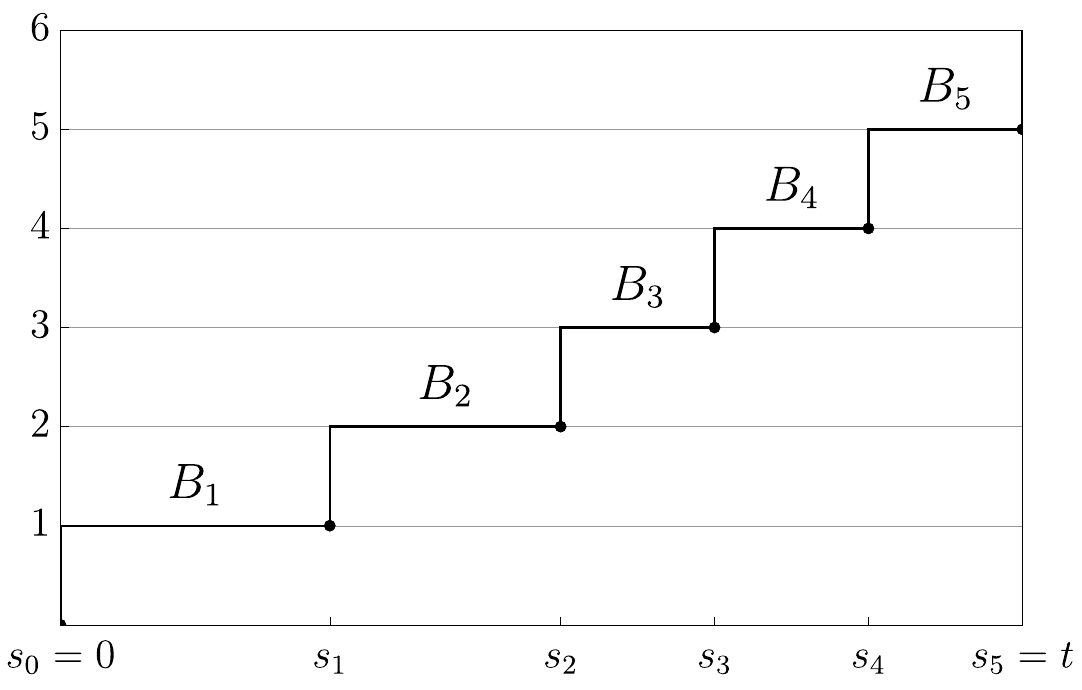}
    \caption{Example of a configuration of the O'Connell-Yor polymer, here for $N=5$, where independent Brownian live on each horizontal line.} 
    \label{fig:yo-polymer}
\end{figure}

These studies have so far remained in the scope of continuous systems and the question whether this can be extended to discrete systems is still open. In this Letter, we provide a positive answer to this question in the context of the semi-discrete O'Connell-Yor (OY) polymer
\cite{o2001brownian,O_Connell_2002,O_Connell_2012}, a discretization of the KPZ equation \cite{BorodinMacdo}.
In its point-to-point version, it is defined by the partition sum $Z_N(t)$ defined as

\be \label{ZN} 
Z_N(t) \! = \! \int_{ s_0=0<s_1<\dots<s_N=t} \hspace{-2.4cm} \rmd s_1 \dots \rmd s_{N-1} \,e^{ \sqrt{\varepsilon} \sum\limits_{j=1}^N  (B_j(s_j)-B_j(s_{j-1}))} 
\ee

The directed polymer path lives on the horizontal lines $j=1,\dots,N$ and jump upward from line $j$ to $j+1$ at 
time $s_j$, see Fig.~\ref{fig:yo-polymer}. The $B_j(s)$ are independent standard Brownian motions and represent the noise. The endpoints are fixed at $(j,s)=(1,0)$ and $(N,t)$.
For $N=1$ this is just the geometric Brownian motion  $Z_1(t) = e^{\sqrt{\varepsilon} B(t)}$
much studied in the Black-Scholes model in finance \cite{bouchaud2003theory}. 

The weak noise limit amounts to take the (inverse temperature) parameter $\varepsilon \ll 1$. Consider 
first the simplest case of the geometric Brownian motion in the weak noise, i.e. small volatility, limit. 
 Although its typical fluctuations are simple
(they are Brownian by expanding the exponential), the large deviations are 
more interesting. Indeed 
consider the observable
\be   
\overline{ e^{ - \frac{\hat{\Lambda}}{\varepsilon} Z_1(t=1)} } = \int_{\R} \frac{\rmd u}{\sqrt{2\pi \varepsilon}} e^{- \frac{u^2}{2 \varepsilon}  - \frac{\hat{\Lambda}}{\varepsilon} e^{u} } \underset{\varepsilon \to 0}{\sim }
e^{ - \frac{1}{\varepsilon} \Psi_1(\hat{\Lambda}) } 
\ee  
where overbar denotes expectation values with respect to the noise, and the 
the rate function is
\be   
\label{eq:sol-n1-lambert}
\Psi(\hat{\Lambda})  = \min_{u \in \mathbb{R}} \big( \frac{u^2}{2} + \hat{\Lambda} e^{u}\big) = \frac{W(\hat{\Lambda})^2}{2} + W(\hat{\Lambda}) 
\ee  
where $W$ is the Lambert function \cite{SM,corless1996lambertw}.
Hence it is already non-trivial for $N=1$ and describes how the rare large
fluctuations $B(t=1) = \mathcal{O}(1/\sqrt{\varepsilon})$ affect the geometric
Brownian motion. 

In this Letter, we consider such large deviation
observables for the full OY problem defined in Eq.~\eqref{ZN} with arbitrary number of lines $N$. 
This now involves a non-trivial path-dependent optimization.
We show that these large deviations are controlled by a system of deterministic nonlinear equations with mixed initial-final boundary conditions which arise from the saddle point of the dynamical action associated to the stochastic evolution of $Z_N(t)$. These equations are a discretization of the $\{P,Q \}$ system studied by us  using inverse scattering in the context of the KPZ equation in Refs.~\cite{krajenbrink2021inverse, krajenbrink2022inverse,krajenbrink2023crossover} - a close cousin of the NLS equation \cite{ZS}, and a member of the 
AKNS hierarchy \cite{AblowitzKaup1974}. Interestingly it is a nonstandard discretisation, whose nonlinearity is local on the lattice, and which is closely related to other physical models such as the discrete self-trapping dimer model \cite{kuznetsov2000quantum,enol1991alternate}, the Toda lattice 
\cite{christiansen1993integrable},
and the periodic TASEP at large time \cite{SilvaPrivate}. These equations are integrable in the sense of the existence of a Lax pair. Our contribution is the derivation of an inverse scattering theory adapted to boundary conditions which are non-local in time, together with the explicit solution of these equations in terms of direct and inverse scattering. This allows us to obtain the analytical expression of the large deviation rate function for the OY model for arbitrary $N$ as well as the optimal history of the partition function conditioned on the large deviations. We complement our study
by performing an asymptotic analysis of a determinantal formula obtained for the OY model \cite{BorodinMacdo,imamura2016determinantal}, which
agrees with our results. We also show in detail how, in the limit of
large $N$, these equations and their solutions recover the previous results for the Weak Noise Theory/short-time limit of the KPZ equation \cite{krajenbrink2021inverse}.
As an amusing byproduct, we obtain a new contour integral representation of the Lambert function.

We start by recalling the coupled stochastic equations which are obeyed by the set of all the partition functions $\{ Z_n(t) \}_{n \in \mathbb{N}}$. It reads
\begin{equation}
\label{eq:weak-noise-epsilon}
    \p_t {\sf z}_n(t) = {\sf z}_{n-1}(t)- {\sf z}_n(t)+\sqrt{\varepsilon} {\sf z}_n(t)\eta_n(t) 
\end{equation}
in Ito discretization, where the $\eta_n(t)$ are independent white noises, 
and where ${\sf z}_n(t) = e^{-(1+ \frac{\varepsilon}{2}) t} Z_n(t)$
with the convention ${\sf z}_0(t)=0$. Here
we are interested in the large deviation form of the cumulant generating function (CGF) 
\be \label{generN} 
\overline{ e^{-\frac{\Lambda}{\varepsilon} {\sf z}_N(t=1)} } \sim e^{ - \frac{1}{\varepsilon} \Psi_N(\Lambda) } 
\ee 
One of our aim is to compute the rate function $\Psi_N(\Lambda)$ explicitly for any $N$. From it one
can extract the large deviation form of the PDF of ${\cal P}_N(H)$ of ${\sf H}=\log {\sf z}_N(t=1)$
\be
{\cal P}_N(H) \sim e^{- \frac{1}{\varepsilon}  \Phi_N(H) }
\ee 
through Legendre inversion of the saddle point
\be \label{saddleprelation}
\Psi_N(\Lambda) = \min_{H \in \mathbb{R}} (\Lambda e^H + \Phi_N(H) ) 
\ee 
The calculation of the CGF is done using the path-integral representation
\be \label{representation0} 
\overline{e^{\frac{1}{\varepsilon}\int \rmd t\tiny{ \sum}_n j_n(t) z_n(t)}}= \iint \mathcal{D}\tilde{z}\mathcal{D}z e^{ -\frac{1}{\varepsilon}S[z,\Tilde{z},j]}
\ee 
in terms of the dynamical action
\bea  
\label{eq:action-yo-polymer}
&& S[z,\Tilde{z},j] =\\
&& \int_0^{\infty} \rmd t \, \sum_{n=1}^N [\tilde{z}_n (\p_t z_n-z_{n-1}+z_n )-\frac{1}{2} z_n^2 \tilde{z}_n^2  -j_n z_n]    \nn
\eea 
where the source field is here $j_n(t)=-\Lambda \delta_{n,N}\delta(t-1)$, see \cite{SM}.
In the limit $\varepsilon \ll 1$ we evaluate the r.h.s of \eqref{representation0} using 
a saddle point method which leads to the following nonlinear differential equations
\begin{equation}
    \begin{split}
        \p_t z_n&= z_{n-1}-z_n+   z_n^2 \tilde{z}_n  \\
        -\p_t \tilde{z}_n&=\tilde{z}_{n+1}-\tilde{z}_n+   z_n \tilde{z}_n^2 
    \end{split}
    \label{eq:YO-polymer-saddle}
\end{equation} 
To study the point-to-point polymer one must impose the following initial and final conditions 
\begin{equation}
\label{eq:boundary-conditions}
  z_n(0)=\delta_{n,1}, \quad   \tilde{z}_n(1)=-\Lambda\delta_{N,n} \, .
\end{equation}
Note that one also has $z_n(t)=0$ for all $n \leq 0$, and $\tilde z_n(t)=0$ for all $n \geq N+1$. Also, for these boundary conditions, we expect the symmetry $\tilde z_n(t) = - \Lambda z_{N-n+1}(1-t)$ to hold. 

It turns out that the equations \eqref{eq:YO-polymer-saddle} are integrable, as was noted in other contexts 
\cite{SilvaPrivate,merola1994novel,sklyanin2000backlund,kundu1994simple,kuznetsov2000quantum}. They
enjoy a Lax pair representation, which reads
\begin{equation}
\label{eq:scattering-problem}
        \p_t {v}_n=U_n {v}_n, \quad  {v}_{n+1}=L_n {v}_n \, ,
\end{equation}
where the Lax matrices are given explicitly by
\begin{equation}
U_n=
\begin{pmatrix}
\frac{\lambda^2-1}{2}  &   -z_{n-1} \\ 
\tilde{z}_n &  \frac{1-\lambda^2}{2}
\end{pmatrix}, \;
L_n=
\begin{pmatrix}
\frac{1}{\lambda} & \frac{z_n}{\lambda}\\
 -\frac{1}{\lambda}\tilde{z}_{n} & \lambda- \frac{1}{\lambda}z_n\tilde{z}_{n} 
\end{pmatrix} .
\label{eq:LaxPairYO-new0}
\end{equation}
Here $n$ is the lattice index and $\lambda$ is the spectral parameter and ${v}_n$ is a 2-component vector. One can check that the original system  \eqref{eq:YO-polymer-saddle} is recovered from the compatibility equation $\p_t L_n=U_{n+1}L_n -L_n U_n$.

The general strategy that we use is to first solve the direct scattering problem 
\eqref{eq:scattering-problem}. In a second stage we give the general solution of the 
inverse scattering problem and reconstruct the field $\{z_n(t),\tilde z_n(t)\}$
solution of the system \eqref{eq:YO-polymer-saddle}. We first describe mathematically the general method to solve
the system \eqref{eq:YO-polymer-saddle}, 
before applying it to the specific boundary conditions \eqref{eq:boundary-conditions} of interest for the polymer problem.
To this aim, let us denote   $ {v}_n = e^{\frac{\lambda^2-1}{2}t}\phi_n$ with $\phi_n=(\phi_1,\phi_2)^\intercal$ and $ {v}_n = e^{-\frac{\lambda^2-1}{2}t}\bar{\phi}_n$ two independent solutions of the linear Lax pair problem. Assuming that the fields $\{z_n, \tilde{z}_n \}$ vanish as $n\to \pm \infty$,
which is the case for the boundary conditions \eqref{eq:boundary-conditions},
we can choose the solutions to behave asymptotically as  $\phi_n  \simeq \lambda^{-n}(1,0)^\intercal $ and $\bar{\phi}_n \simeq \lambda^n 
  (0, -1)^\intercal$ at $n\to -\infty$. At $n\to +\infty$, each solution should be a particular linear combination of these elementary solutions.
  This allows to define the scattering amplitudes $\{a,\atilde,b,\btilde \}$ as
  \begin{equation}
  \label{eq:def-scattering-amplitude}
    \phi_n  \underset{n\to +\infty}{\simeq }
    \begin{pmatrix}
a(\lambda,t)\lambda^{-n}  \\
b(\lambda,t) \lambda^n
\end{pmatrix}, \quad 
\bar{\phi}_n \underset{n\to +\infty}{\simeq }
    \begin{pmatrix}
\btilde(\lambda,t)\lambda^{-n} \\
-\atilde(\lambda,t)\lambda^{n}
\end{pmatrix}
\end{equation}
Inserting Eqs.~\eqref{eq:def-scattering-amplitude} in the Lax equation with $U_n$ for $n\to +\infty$ yields the time-dependence of the scattering amplitudes $a(\lambda,t)=a(\lambda)$, $\atilde(\lambda,t)=\atilde(\lambda)$, $\btilde(\lambda,t)=\btilde(\lambda)e^{(\lambda^2-1)t}$ and $b(\lambda,t)=b(\lambda) e^{(1-\lambda^2)t}$. Note that the representation of the Lax matrices \eqref{eq:LaxPairYO-new0}
chosen here is particularly convenient as $\Tr(U_n)=0$ and $\Det(L_n)=1$. As a consequence, the 
Wronskian of $\{\phi_n, \bar{\phi}_n \}$ is constant in space and time \cite{SM}. This allows to 
the obtain the following normalisation relation of the scattering amplitudes \cite{footnote-ablowitz-ladik} 
$    a(\lambda)\atilde(\lambda)+b(\lambda)\btilde(\lambda)= 1 $. We expect $a(\lambda)$ to be analytic inside a contour enclosing the origin which we choose as a
circle of radius $\mathtt{R}$, i.e. for $|\lambda|<\mathtt{R}$, and 
and $\atilde(\lambda)$ to be analytic outside, i.e. for $|\lambda|>\mathtt{R}$.
Although we will use that circle notation for simplicity, it is understood below in this work 
that for $\Lambda>\Lambda^*_N$ (a large positive value defined later) 
the circle must actually be deformed into an ellipse ${\cal C}$ \cite{SM}. 

From the knowledge of the scattering amplitudes, it is possible to reconstruct explicitly the field $z_n(t)$ using a novel 
Fredholm operator formula that we now present. We first define the Fourier transforms of the ratio of scattering amplitudes called reflection coefficients as
\bea  
&&     F_t(n)=\! \oint_{\abs{\lambda}=\mathtt{R}} \! \frac{\rmd \lambda }{2\I \pi}\frac{b(\lambda,t)}{a(\lambda) } \lambda^{n-1} \\
&& \tilde{F}_t(n)=\!\oint_{\abs{\lambda}=\mathtt{R}} \!\frac{\rmd \lambda }{2\I \pi}\frac{\btilde(\lambda,t)}{\atilde(\lambda) } \lambda^{-1-n}
\eea  
where the contour integrals are taken over the $\mathtt{R}$-circle.
We then define two (space-time dependent) Hankel operators $F_{n,t}$ and $\tilde F_{n,t}$ with the following kernels
\begin{equation}
\label{eq:additive-operator-definition}
    F_{n,t}(i,j) = F_t(2n+i+j),     \tilde{F}_{n,t}(i,j) = \tilde{F}_t(2n+i+j)
\end{equation}
where the indices are positive $i,j \geq 0$. The product of two such operators $A,B$ is defined as
\begin{equation}
    (A B)(i,j)=\sum_{k>0}A(i,k)B(k,j)
\end{equation}
and we define the vector $\bra{\delta}$ with component $\delta_{i,0}$ so that $\bra{\delta}A\ket{\delta}=A(0,0)$ for any operator $A$. Then the general solution of Eq.~\eqref{eq:YO-polymer-saddle} reads
\begin{equation}
\label{eq:general-fred-sol-zn}
    z_n(t) =-\bra{\delta}\tilde{F}_{n,t}(I+F_{n,t}\tilde{F}_{n,t})^{-1}\ket{\delta}
\end{equation}
The additive structure present in Eq.~\eqref{eq:additive-operator-definition} arises from the integrability of the problem and is akin to the structure for the continuous space problem \cite{krajenbrink2020painleve, krajenbrink2021inverse,bothner2022riemann}. An analogous formula also exists for $\tilde z_n(t)$, see \cite{SM}. Note that this formula \eqref{eq:general-fred-sol-zn} can be easily evaluated numerically using a discretized version of the Bornemann algorithm \cite{krajenbrink2021inverse}.

Quite remarkably for the 
boundary conditions of interest here \eqref{eq:boundary-conditions}, the calculation of the scattering amplitudes can be performed explicitly. Indeed, by solving 
the spatial Lax equation ${\phi}_{n+1}=L_n {\phi}_n$ and ${\bar \phi}_{n+1}=L_n {\bar \phi}_n$ at $t=0$ and $t=1$, see \cite{SM}, one obtains
\begin{equation}
\begin{split}
    \btilde(\lambda)&=-\lambda^2, \quad  b(\lambda)  = \Lambda \lambda^{-2N-2}e^{\lambda^2 -1} \, .
    \end{split}
\end{equation}
The normalisation relation then implies that
\begin{equation}
\label{eq:normalisation-RH}
\begin{split}
    a(\lambda) \atilde(\lambda)&= 1+ \Lambda\lambda^{-2N}e^{\lambda^2 -1} \, .
    \end{split}
\end{equation}
To obtain the expression of $a$, $\atilde$, we solve Eq.~\eqref{eq:normalisation-RH} as a scalar Riemann-Hilbert (RH) problem. 
There are two families, or branches, of solutions of Eq.~\eqref{eq:normalisation-RH} which are relevant for our large deviation 
problem, i.e. needed to invert \eqref{saddleprelation} to obtain $\Phi_N(H)$. 

{\it Main branch, no solitons}. The first family of solution does not involve solitons and determines {\it the main branch} of
$\Psi_N(\Lambda)$. It assumes that $a(\lambda)$ has no zero for $|\lambda|< \mathtt{R}$
 and that $\atilde(\lambda)$ has no zero for $|\lambda|> \mathtt{R}$. 
 Solving the RH problem Eq.~\eqref{eq:normalisation-RH}, see \cite{SM}, one obtains
\begin{equation}
\begin{cases}
         \log \atilde(\lambda)=-\varphi(\lambda), \quad &\abs{\lambda}>\mathtt{R}\\
                 \log {a}(\lambda)=\varphi(\lambda), \quad &\abs{\lambda}<\mathtt{R}
         \end{cases}
\end{equation}
with 
\begin{equation}
\label{eq:sol-RH-scatttering}
\varphi(\lambda)=\oint_{{\cal C}} \frac{\rmd w}{2\I \pi}\frac{w}{w^2-\lambda^2}\log (1+ \Lambda w^{-2N}e^{w^2 -1})
\end{equation}
where ${\cal C}$ is a closed contour around the origin which must avoid the branch cuts of the logarithm. 
For most values of interest, $\Lambda<\Lambda_N^*$, ${\cal C}$ can be chosen as the circle $\mathtt{R}=\sqrt{N}$.
The threshold $\Lambda_N^*$ is defined in \cite{SM}, it grows very fast with $N$ (with $\Lambda_1^*=e^2$) and
plays little role below, so we stick to the circle notation. From the knowledge of the scattering amplitudes, one additionally obtains the values
taken by the conserved quantities of the problem, for details see \cite{SM}. This is achieved by Laurent or Taylor expanding 
the scattering amplitudes as
\begin{equation}
         \log \atilde(\lambda) = \sum_{n=1}^\infty \frac{\tilde{C}_n}{\lambda^{2n}}, \quad   \log {a}(\lambda) =\sum_{n=0}^\infty \lambda^{2n}C_n \, .
\end{equation}
In particular, expanding equivalently 
Eq.~\eqref{eq:sol-RH-scatttering}, we have that the value taken by the $n$-th conserved quantity is
\begin{equation}
    \tilde{C}_n = \oint_{\abs{w}=\mathtt{R}}\frac{\rmd w}{2\I \pi}w^{2n-1}\log (1+ \Lambda w^{-2N}e^{w^2 -1}) \, .
\end{equation}
Since the first conserved quantity is related to the fields $\{ z_n, \tilde{z}_n \}$ as $\tilde C_1 = -  \sum_{n=1}^N z_n(t) \tilde z_n(t)$
and is by definition time independent, it can be evaluated at $t=1$ where one has
\be   
\tilde C_1 = -  \sum_{n=1}^N z_n(t=1) \tilde z_n(t=1) = \Lambda z_N(t=1)  
\ee  
thanks to the boundary conditions \eqref{eq:boundary-conditions}. From the derivative of \eqref{saddleprelation} w.r.t. $\Lambda$ one sees that 
$\Psi_N'(\Lambda)=e^H=z_N(t=1)$. Hence the large deviation rate function is determined by the
first conserved quantity and one obtains 
\be 
\label{eq:derivative-rate-function-1}
 \Lambda \Psi_N'(\Lambda) = \oint_{\abs{w}=\mathtt{R}}\frac{\rmd w}{2\I \pi} w \log (1+  \Lambda w^{-2N}e^{w^2 -1})
\ee   
Note that for $N=1$, Eq.~\eqref{eq:derivative-rate-function-1} provides a novel integral representation of the Lambert function as $ \Lambda \Psi_1'(\Lambda)=W(\frac{\Lambda}{e})$ from Eq.~\eqref{eq:sol-n1-lambert}. A direct integration of \eqref{eq:derivative-rate-function-1} finally yields 
\be \label{eq:resultPsi}
  \Psi_N(\Lambda) = -\oint_{\abs{v}=\mathtt{R}^2}\frac{\rmd v}{2\I \pi}  \mathrm{Li}_2(- \Lambda v^{-N}e^{v -1})
\ee   
 where $\mathrm{Li}_2$ refers to the dilogarithm \cite{footnotecontour} which domain of definition restricts the validity of this formula for
 \begin{equation}
 \Lambda \geq \Lambda_c, \quad  \Lambda_c = -e^{1-N}N^N \leq 0 \, .
 \end{equation}
It allows to reconstruct the rate function $\Phi_N(H)$ with $H = \log z_N(t=1)$ 
in the range
 \begin{equation}
0 \leq z_N(t=1)  \leq  \Psi_N'(\Lambda_c)=z_N^{(c)}
 \end{equation}
and one obtains the parametric representation
\be  \label{param} 
\begin{cases}
\Phi_N(H)=\Psi_N(\Lambda) - \Lambda \Psi_N'(\Lambda) \\
H = \log \Psi_N'(\Lambda) 
\end{cases}
\ee
for $H \in (-\infty,H_c]$ as $\Lambda \in [\Lambda_c,+\infty)$.
This range contains $\Lambda=0$ which gives the typical value
$\Psi_N'(0) = e^{H_{\rm typ}} = \overline{ {\sf z}_N(t=1) }   = \frac{e^{-1}}{(N-1)!}$,
as well as the second cumulant of the partition sum
$\overline{ {\sf z}_N(t=1)^2}^c =   \frac{2^{2 N -2}}{(2 N  -1)!} e^{-2} \,  \varepsilon$, and of its logarithm
\be 
\overline{{\sf H}^2}^c =  \frac{\varepsilon}{\Phi''(H_{\rm typ})} = 
\frac{ - \varepsilon \Psi''(0)}{\Psi'(0)^2} = \frac{2^{2 N-2} (N-1)!^2}{(2 N-1)!} \, \varepsilon
\ee

{\it Second branch, with solitons}. The second family of solution involves solitons, and allows to 
obtain the rate function $\Phi_N(H)$ for the range $z_N(t=1) \geq z_N^{(c)}$, 
i.e. $H > H_c$. To see how it arises consider the right hand side of Eq.~\eqref{eq:normalisation-RH}. One can
check that for $\Lambda \in (\Lambda_c,0)$ it has four zeroes
on the real axis $\{ \pm \lambda_0, \pm \lambda_{-1} \}$ with
\begin{equation}
    \lambda_{0/-1}^2 =  -N W_{0/-1}\left(-\frac{1}{e} \big(\frac{ \Lambda}{ \Lambda_c }\big)^{\frac{1}{N}}\right) 
\end{equation}
and $|\lambda_{-1}| \geq \mathtt{R}=\sqrt{N}$ and $|\lambda_{0}|<\sqrt{N}$. Here $W_0$
and $W_{-1}$ are the two main branches of the Lambert function \cite{corless1996lambertw}.
The existence of these zeroes allows for a modified solution to the RH problem 
\eqref{eq:normalisation-RH}, where $a(\lambda)$ has two zeroes inside the
$\mathtt{R}$-circle, $\pm \lambda_0$, and $\atilde(\lambda)$ two zeroes outside, $\pm \lambda_{-1}$. The solution then reads 
\begin{equation}
\begin{cases}
         \log \atilde(\lambda)=-\varphi_{\text{soliton}}(\lambda), \quad &\abs{\lambda}>\mathtt{R}\\
                 \log {a}(\lambda)=\varphi_{\text{soliton}}(\lambda), \quad &\abs{\lambda}<\mathtt{R}
         \end{cases}
\end{equation}
with
\begin{equation}
\label{eq:sol-RH-scatttering-with-soliton}
\varphi_{\text{soliton}}(\lambda)=\varphi(\lambda)-\log\left(\frac{\lambda^2-\lambda_{-1}^2}{\lambda^2-\lambda_0^2}\right)
\end{equation}
where $\varphi(\lambda)$ is given by the same formula as in \eqref{eq:sol-RH-scatttering}.
By either Taylor or Laurent expanding the new solitonic contribution in \eqref{eq:sol-RH-scatttering-with-soliton}, one also obtains an additional (solitonic) contribution $\Delta \tilde C_n$ to the values of the conserved quantities. In particular one
obtains $\Delta \tilde C_1=\lambda_0^2-\lambda_{-1}^2$ which leads to the
following correction \cite{SM}
\begin{equation}
 \Lambda \Psi_{N,\text{soliton}}'(\Lambda)= \Lambda \Psi_N'(\Lambda)+\lambda_0^2-\lambda_{-1}^2
\end{equation}
It can be explicitly integrated over $\Lambda$ and we obtain the {\it second branch}
of the rate function $\Psi_{N} \to \Psi_{N,\text{soliton}}$ with
\begin{equation}
    \Psi_{N,\text{soliton}}(\Lambda) =  \Psi_N(\Lambda)  +  \Delta_N(\Lambda)
\end{equation}
where the solitonic contribution $\Delta_N(\Lambda)$ is

\begin{figure}[h!]
\centering
\includegraphics[scale=0.7]{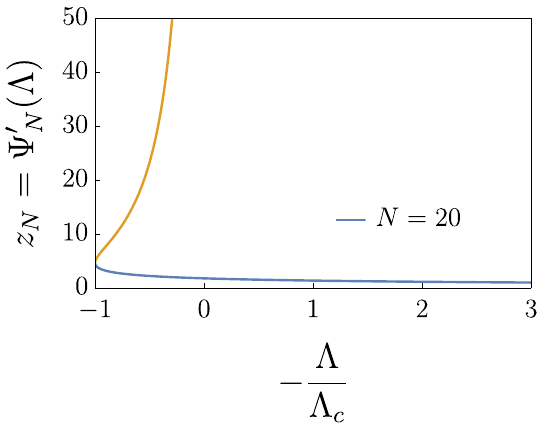}
\caption{Derivative of the rate function, $\Psi_{N}'(\Lambda)$. The main branch is in blue and the
second (solitonic) branch is in orange. The ordinate is also equal to $z_N(t=1)=e^H$ hence one can
read off the relation between $H$ and $\Lambda$ (which is not one-to-one).}
\label{fig:z}
\end{figure}

\begin{widetext}

\begin{equation} \label{DeltaSoliton} 
    \Delta_N(\Lambda) = \frac{N^2 }{2} \left(W_{-1}\left(-\frac{1}{e} \big(\frac{ \Lambda}{ \Lambda_c }\big)^{\frac{1}{N}}\right)  \left[W_{-1}\left(-\frac{1}{e} \big(\frac{ \Lambda}{ \Lambda_c }\big)^{\frac{1}{N}}\right) +2\right]-   W_{0}\left(-\frac{1}{e} \big(\frac{ \Lambda}{ \Lambda_c }\big)^{\frac{1}{N}}\right) \left[W_{0}\left(-\frac{1}{e} \big(\frac{ \Lambda}{ \Lambda_c }\big)^{\frac{1}{N}}\right) +2\right] \right)
\end{equation}
\end{widetext}
This second branch allows to reconstruct $\Phi(H)$ for $H>H_c$,
using the same parametric representation \eqref{param} where
now one replaces everywhere $\Psi_{N}(\Lambda) \to \Psi_{N,\text{soliton}}(\Lambda)$.
As $\Lambda$ increases from $\Lambda_c$ to $0$, the values
of $H$ increase from $H_c$ to $+\infty$. The two branches
of $\Psi'_{N}(\Lambda)$ are shown in Fig~\ref{fig:z} where one sees that the branches merge smoothly.

As we have obtained our solution for general $N$, it is natural to study the limit when the polymer sees a large number of lines, i.e. $N \gg 1$. It is known that in the large $N$ limit the OY polymer point-to-point partition sum converges, under the proper rescaling, to the solution of the stochastic
heat equation, i.e. to the exponential of the KPZ height, for the so-called droplet initial conditions \cite{BorodinMacdo}. 

Here we can check that Eq.~\eqref{generN} 
converges
to the corresponding equality for the KPZ equation at short time,
which was obtained in \cite{le2016exact,krajenbrink2021inverse}.
To this aim we first define the 
rescaled variables 
\begin{equation}
z=-\frac{\Lambda}{\Lambda_c}, \quad T_{\rm KPZ} = \frac{\varepsilon^2}{2N}
\end{equation}
where $T_{\rm KPZ} \ll 1$ is the time in the KPZ equation. We then expand the formula \eqref{eq:resultPsi}
for $\Psi_N(\Lambda)$ at large $N$ around the point $v=N$ on the contour,
by setting $v=N e^{q \sqrt{2/N}}$ which yields for large $N$
\be  \label{limit0} 
\frac{\Psi_N(\Lambda)}{\varepsilon}   \to   \frac{\Psi_{\rm KPZ}(z)}{\sqrt{T_{\rm KPZ}}} 
\ee   
where 
\begin{equation}
\Psi_{\rm KPZ}(z) = 
- \int_{\mathbb{R}} \frac{\rmd k}{2 \pi}   \mathrm{Li}_2(- z e^{- q^2})
\end{equation}
is the rate function for the KPZ equation with droplet initial data. Equation~\eqref{limit0}  shows
the convergence of the r.h.s of \eqref{generN} and the convergence
of the l.h.s of \eqref{generN} is obtained using  
Ref.~\cite[Section 5.4.1]{BorodinMacdo}, which allows to 
identify at large $N$
\be 
\frac{\Lambda}{\varepsilon} z_N(t=1) \to \frac{z e^{H_{\rm KPZ}}}{\sqrt{T_{\rm KPZ}}}
\ee 
where $H_{\rm KPZ}$ is the properly shifted KPZ height field,
denoted $H$ in \cite{krajenbrink2021inverse}. To control the convergence of the solitonic branch, we expand the Lambert functions $W_{0/-1}(x)$ around $x=-1/e$
in \eqref{DeltaSoliton} and obtain \cite{SM}
\be 
 \frac{ \Delta_N(\Lambda) }{\varepsilon} \simeq  \frac{\Delta_{\rm KPZ}(z) }{\sqrt{T_{\rm KPZ}}} 
 ~ , ~ \Delta_{\rm KPZ}(z) = \frac{4}{3} \left(-\log(-z)\right)^{\frac{3}{2}} 
\ee  
This shows the convergence to the corresponding solitonic branch of the KPZ equation,
obtained in \cite{krajenbrink2021inverse} (see also \cite{TsaiIntegrability}). For completeness, we have also derived the above results for $\Psi_N(\Lambda)$
from an asymptotic analysis of a determinantal representation formula for
$\overline{e^{- u Z_N(t)}}$ \cite{BorodinMacdo,imamura2016determinantal} using a first cumulant approximation,
see details in \cite{SM}.

The OY polymer is also related to the
quantum Toda lattice \cite{O_Connell_2012}. Here we
show that its weak noise theory \eqref{eq:YO-polymer-saddle} converges to 
the {\it classical} Toda lattice  \cite{toda1967vibration} in the small time limit. Indeed, using the Cole-Hopf parameterization
$z_n(t)=\alpha e^{h_n(t)+ \alpha^2 t}$, $\tilde{z}_n(t) z_n(t)=\alpha (\alpha +p_n(t))$, 
and taking $\alpha\to \infty$ we find 
the Toda dynamics in scaled time $\tau=\alpha t$ \cite{SM} 
\begin{equation}
    \p_\tau h_n = p_n  ~,~ 
    \p_\tau p_n  = e^{h_{n-1}-h_n}-e^{h_{n}-h_{n+1}} \, .
\end{equation}
Hence our results on the scattering theory and on the Fredholm determinants apply,
extending known solutions for solitons \cite{bauhardt1987fredholm}. Finally, there seems to exist a connexion between the weak noise theory of the OY polymer and an integrable spin chain which we discuss in \cite{SM}.

In conclusion we have shown that the weak noise theory of the
OY polymer is integrable for any $N$, obtained its general solution 
in terms of Fredholm determinants, and computed large deviation rate functions
from conserved quantities. The system \eqref{eq:YO-polymer-saddle} 
provides a discretization of the NLS equation with local nonlinearity
and converges at large $N$ to the weak noise of the KPZ equation.
A distinct limit is the classical Toda chain, which is related to the QR decompositions in linear algebra \cite{symes1982qr,deift1983ordinary},
hence the extension to our system may 
provide applications to linear algebra algorithms. 
Finally, this work opens the path for a general study of weak noise limit in stochastic integrable systems and their connection to classical integrability. Since semi-discrete integrable systems have found renewed interest given their connexions to random matrix theory \cite{grava2023discrete,mazzuca2023equilibrium,spohn2020generalized,spohn2021hydrodynamic,spohn2022hydrodynamic}, the model studied in this work will find additional applications. A potential outcome of this general study might lead to a broader classification of stochastic integrable models.

\begin{acknowledgments}
\paragraph{Acknowledgments.}  AK thanks Guilherme L.~F.~Silva for discussions on the existence of the Lax pair for the equations \eqref{eq:YO-polymer-saddle}. We thank Alexei Borodin for interesting discussions and collaborations on
closely related topics. We acknowledge support
from MIT-France MISTI Global Seed Funds project “Exact Solutions in Field Theories via Integrable Probability”
and the MIT math department for hospitality. PLD acknowledges support from the ANR grant ANR-17-CE30-0027-01 RaMaTraF.  
\end{acknowledgments}

\input{bibliography_orderedV1.tex}

\newpage
~
\newpage

\begin{widetext} 
\makeatletter
\renewcommand{\theequation}{S\arabic{equation}}
\renewcommand{\thefigure}{S\arabic{figure}}

\setcounter{section}{0}
\setcounter{secnumdepth}{2}

\begin{large}
\begin{center}

Supplementary Material for\\  {\it The weak noise theory of the O'Connell-Yor polymer as an integrable discretisation of the nonlinear Schrodinger equation }

\end{center}
\end{large}

We give the principal details of the calculations described in the main text of the Letter as well as additional information about the results.

{\hypersetup{linkcolor=black}
\setcounter{tocdepth}{2}
\tableofcontents
}

\newpage

Note that in the main text we use the notation ${\sf z}_n(t)$ for the OY partition sums (i.e. a random variable)
and $z_n(t)$ for the solution of the weak noise saddle-point equations. Here we will simplify
notations and use the same letter for both, and the context will make it clear which is which.

\section{Model and observable}

In this Supp.~Mat. we will consider a slight generalization of the partition sums \eqref{ZN} defined in the text, which
corresponds to adding drifts to the Brownian weights $B_j(s) \to B_j(s) + a_j s$. One thus defines for $1 \leq n \leq N$ the partition function
\be \label{ZN2} 
Z_n(t) = \int_{ s_0=0<s_1<\dots<s_n=t}  \rmd s_1 \dots \rmd s_{n-1} e^{\sqrt{\varepsilon} \sum_{j=1}^n  (B_j(s_j)-B_j(s_{j-1})) + \sum_{j=1}^n a_j (s_j-s_{j-1})} 
\ee 
It is convenient to define $z_n(t) = e^{-(1+ \frac{\varepsilon}{2}) t} Z_n(t)$, so that the first two terms read explicitly
\be  z_1(\tau) = e^{-\tau - \frac{1}{2} \varepsilon \tau} e^{\sqrt{\varepsilon} B_1(\tau) + a_1 \tau} 
\quad , \quad z_2(\tau) = e^{-\tau - \frac{1}{2} \varepsilon \tau} \int_0^\tau \rmd s e^{\sqrt{\varepsilon} B_1(s)  + a_1 s} e^{\sqrt{\varepsilon}(B_2(t)-B_2(s)) + a_2 (\tau-s) } 
\ee  
Using the rules of Ito calculus it is easy to see that these partition sums satisfy the following coupled stochastic equations
in Ito discretization for $1 \leq n \leq N$
\begin{equation}
\label{eq:weak-noise-epsilon-2}
    \p_t z_n(t) = z_{n-1}(t)- z_n(t)+\sqrt{\varepsilon} z_n(t)\eta_n(t) + a_n z_n(t) 
\end{equation}
where the $\eta_n(t)$ are standard independent
white noises, i.e. with correlators $\overline{\eta_n(t) \eta_{n'}(t')} = \delta(t-t')\delta_{nn'}$. One uses the 
convention $z_0(t)=0$ and the initial condition is $z_n(0)=\delta_{n,1}$. \\

The observable of interest is the partition function at one point $z_N(t=1)$ and its probability distribution function 
${\cal P}_N(z)$. It is useful to also introduce its logarithm $H_N= \log z_N(t=1)$ and its PDF ${\cal P}_N(H)$.
These PDF's exhibit, in the weak noise limit $\varepsilon \to 0$ the following large deviation 
principle
\be 
{\cal P}_N(z) \underset{\varepsilon \to 0^+}{\sim} e^{-\frac{1}{\varepsilon} \hat \Phi_N(z)} \quad , \quad 
{\cal P}_N(H) \underset{\varepsilon \to 0^+}{\sim} e^{-\frac{1}{\varepsilon} \Phi_N(H)} 
\ee 
To compute these rate functions, which differ only by a change of variable, $\hat \Phi_N(z)=\Phi_N(H=\log z)$
we will first compute the following generating function of the cumulants of $z_N(t=1)$
\be  \label{gener0} 
\overline{ e^{-\frac{\Lambda}{\varepsilon} z_N(t=1)} } \underset{\varepsilon \to 0^+}{\sim} e^{ - \frac{1}{\varepsilon} \Psi_N(\Lambda) } 
\ee 
and the rate function $\Psi_N(\Lambda)$. The two rate functions are related by the Legendre transform
\be 
\Psi_N(\Lambda) = \max_{z \in \mathbb{R}^+} ( \Lambda z + \hat \Phi_N(z) ) = \max_{H \in \mathbb{R} } ( \Lambda e^H + \Phi_N(H) )
\label{Legendre0} 
\ee 

\section{Dynamical field theory and saddle point equations}

We now use the standard path integral representation using source fields $j_1(t),\dots,j_N(t)$ 
\bea \label{representation} 
\overline{e^{\frac{1}{\varepsilon}\int_0^{+\infty} \rmd t\tiny{ \sum}_{n=1}^N j_n(t) z_n(t)}} &=&
\iiint \mathcal{D}\eta \mathcal{D}\tilde z \mathcal{D}z  
e^{ \int_0^{+\infty} \rmd t \, \sum_{n=1}^N [- \frac{\tilde{z}_n}{\varepsilon}  (\p_t z_n-z_{n-1}+z_n - \sqrt{\varepsilon} z_n \eta_n - a_n z_n)-\frac{1}{2} \eta_n^2  +j_n z_n] }
\\
& = & \iint \mathcal{D}\tilde{z}\mathcal{D}z e^{ -\frac{1}{\varepsilon}S[z,\Tilde{z},j]}   \label{path} 
\eea 
where $\mathcal{D}z= \prod_{n=1}^N \mathcal{D}z_n(t)$ (and similarly for $\tilde z$ and $\tilde \eta$) is the
path integral measure, in terms of the dynamical action (time dependence is implicit)
\be  
 S[z,\Tilde{z},j]  = S_0[z,\Tilde{z}] - \int_0^{+\infty} \rmd t \, \sum_{n=1}^N j_n(t) z_n(t)  \, .
\ee
The source-less action reads
\be 
S_0[z,\Tilde{z}] = \int_0^{+\infty} \rmd t \,  \sum_{n=1}^N [\tilde{z}_n (\p_t z_n-z_{n-1}+z_n - a_n z_n)-\frac{1}{2} z_n^2 \tilde{z}_n^2 ] \label{dynOY} 
\ee
where we have integrated over the noises $\eta_n(t)$. We have introduced the response field $\frac{\tilde{z}_n}{\varepsilon}$
to enforce the $N$ equations of motion. To obtain the observable in \eqref{gener0} we
need to choose the source field $j_n(t)=-\Lambda \delta_{n,N}\delta(t-1)$.\\

In the limit $\varepsilon \to 0$ the path integral \eqref{path} is dominated by the saddle point of the action $S[z,\Tilde{z},j]$. The saddle point
equations take the form of the following system of equations
\begin{equation}
    \begin{split}
        \p_t z_n&= z_{n-1}-z_n+  \frac{g}{2}  z_n^2 \tilde{z}_n  + a_n z_n \\
        -\p_t \tilde{z}_n&=\tilde{z}_{n+1}-\tilde{z}_n+ \frac{g}{2}   z_n \tilde{z}_n^2 + a_n \tilde z_n
    \end{split}
    \label{eq:YO-polymer-saddle-with-drift2}
\end{equation}
We have added for convenience the parameter $g$ for the non-linearity, but for the application here one must set $g=2$. 
Note that a priori the above saddle point equations hold for $1 \leq n \leq N$. However, for technical reasons
it is useful to extend the system \eqref{eq:YO-polymer-saddle-with-drift2} for all $n \in \mathbb{Z}$, which
we do from now on. We then consider the following initial and final conditions
\begin{equation}
 \{   z_0(t)=0, \quad z_n(0)=\delta_{n,1}, \quad \tilde{z}_{N+1}(t)=0, \quad \tilde{z}_n(1)=-\Lambda\delta_{N,n} \}\label{boundary0} 
\end{equation}
where imposing the last condition is equivalent to omitting the source term $j_n(t)=-\Lambda \delta_{n,N}\delta(t-1)$ 
in the equation of motion for $\tilde z_n$. Note that one also has $z_n(t)=0$ for all $n \leq 0$, and $\tilde z_n(t)=0$ for all $n \geq N+1$.
In practice we also take $a_n=0$ except if $n=1,\dots,N$. Finally the response field $\tilde z_n(t)$ vanishes for $t>1$ for all $n$.\\

It is useful to note the following exact symmetry of the solution of the system \eqref{eq:YO-polymer-saddle-with-drift2}
with the boundary conditions \eqref{boundary0}.
If $a_\ell = a_{N-\ell+1}$ then 
\be  
\tilde z_\ell(t) = - \Lambda z_{N-\ell+1}(1-t)  
\ee

We will obtain below the solution of this system of equations for any $N$. Once this is done one can
insert the solution into the dynamical action to obtain its value at the saddle point. Hence one has
\be 
\overline{ e^{-\frac{\Lambda}{\varepsilon} z_N(t=1)} } \underset{\varepsilon \to 0^+}{\sim}  
\iint \mathcal{D}\tilde{z}\mathcal{D}z e^{ -\frac{1}{\varepsilon} S_0^{\rm sp}[z,\Tilde{z}]} e^{-\frac{\Lambda}{\varepsilon} z_N(t=1)}
\ee 
Hence the PDF  
${\cal P}_N(z)$ of $z_N(t=1)$ is given by the optimal action $S_0^{\rm sp}[z,\Tilde{z}]$, which using
the saddle point equations simplifies into (for $g=2$)
\begin{equation} \label{optimal} 
  {\cal P}_N(z) \sim  \exp \left(- \frac{1}{2\varepsilon}\int_0^1 \rmd t \, \sum_{n=1}^N z_n^2 \tilde{z}_n^2  \right) 
\end{equation}
evaluated using the solution of \eqref{eq:YO-polymer-saddle-with-drift2} with $z_N(t=1)=z$,  this determines $\hat \Phi_N(z)$. In practice however we will obtain $z=z_N(t=1)$ as a function of
$\Lambda$ by solving \eqref{eq:YO-polymer-saddle-with-drift2}. From the Legendre
relation $z = \Psi'(\Lambda)$ obtained by taking a derivative of \eqref{Legendre0} w.r.t. $\Lambda$,
this will allow to determine $\Psi_N(\Lambda)$. The last step will be to 
obtain $\hat \Phi_N(z)$ by an inverse Legendre transform.\\

We note that the evolution equations \eqref{eq:YO-polymer-saddle-with-drift2} differ in part by the minus sign in front of the time derivative. This reflects that the evolution of $z_n$ can be seen as forward in time and the evolution of $\tilde{z}_n$ as backwards in time so that the whole problems is closer to a shooting problem rather than a dynamical problem. 

\section{Two simple cases}

\subsection{$g=0$} 

Let us consider the system \eqref{eq:YO-polymer-saddle-with-drift2} in the simple case $g=0$
with the boundary conditions \eqref{boundary0}. In this case the two equations decouple and
are simply linear. In the absence of drifts the solution is readily obtained as
\be  \label{zzt0} 
z_n(t) = e^{-t}\frac{t^{n-1}}{(n-1)!} \quad , \quad \tilde z_n(t) = - \Lambda e^{t-1} \frac{(1-t)^{N-n}}{(N-n)!}
\ee

\begin{figure}[t!]
\centering
\includegraphics[scale=0.8]{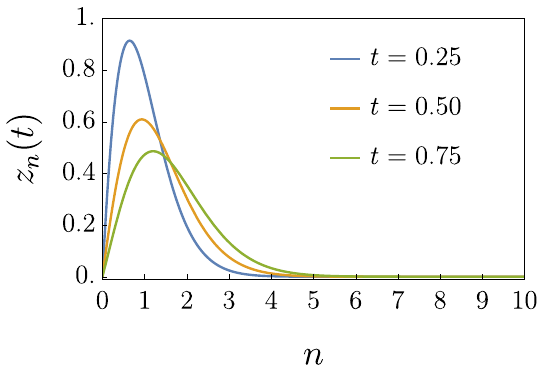}
\caption{Solution $z_n(t)$ of the problem without interaction, $g=0$, as given in Eq.~\eqref{zzt0}.}
\label{fig:....}
\end{figure}

In the presence of drifts it is easier to solve these equations using a Fourier representation
which will be useful in the following. Let us write
\be  
z_n(t) = \int_C \frac{\rmd\lambda^2}{2 \I \pi \lambda^2} \hat z_n(\lambda) e^{t (\lambda^2-1)} 
\ee  
where $C$ is a circle around $0$ in the complex plane. Inserting into
\be  
\partial_t z_n = z_{n-1} - z_{n}  + a_n z_n
\ee   
for $1 \leq n \leq N$, it gives
\be   
\hat z_n(\lambda) =\frac{\lambda^{-2}}{1 - \frac{a_n}{\lambda^2}  } \hat z_{n-1}(\lambda) 
\ee   
For $z_1(t)$ we can solve directly
\be  
\partial_t z_1 = - z_1 + a_1 z_1 \quad , \quad z_1(t) = e^{- t + a_1 t}
\ee  
Hence one has 
\be
\hat z_1(\lambda) = \frac{\lambda^2}{\lambda^2 - a_1} \quad , \quad z_1(t) = \int_C \frac{\rmd\lambda^2}{2 \I \pi \lambda^2}  \frac{\lambda^2}{\lambda^2 - a_1} e^{t (\lambda^2-1)} 
\ee   
and the contour must include $a_1$. Then we obtain for $1 \leq n \leq N$
\be  
\hat z_n(\lambda) = \frac{\lambda^{2(1-n)}}{\prod_{k=1}^{n} (1 - \frac{a_k}{\lambda^2}) }
\ee   
which gives for any $n \in \mathbb{Z}$ (using that $a_n=0$ for $n \geq N+1$) 
\be   
z_n(t) = \int_C \frac{\rmd \lambda^2}{2 \I \pi \lambda^2} \,  \frac{\lambda^{2(1-n)}}{\prod_{k=1}^{\min(n,N)} (1 - \frac{a_k}{\lambda^2}) } e^{(\lambda^2-1)t}
\ee   
which automatically vanishes for $n \geq 0$. The contour $C$ must contain all the $a_n$. Similarly one 
has $\tilde z_N(t)=- \Lambda e^{(a_N-1)(1-t)}$ and one 
obtains
\be   
\tilde z_n(t) = - \Lambda \int_C \frac{\rmd \lambda^2}{2 \I \pi \lambda^2} \,  \frac{\lambda^{2(n-N)}}{\prod_{k=1}^{\min(N-n+1,N)} (1 - \frac{a_{N-k+1}}{\lambda^2}) } 
e^{(\lambda^2-1)(1-t)}
\ee  
If we set all $a_n=0$ one can check that one recovers \eqref{zzt0}.

\subsection{$N=1$} 

For $N=1$ there are several ways to obtain the large deviation rate functions.\\

{\bf First method}. For $N=1$ the system \eqref{eq:YO-polymer-saddle-with-drift2} involves only the variables $z_1(t)$ and $\tilde z_1(t)$
and becomes
\begin{equation}
    \begin{split}
        \p_t z_1&= -z_1+ z_1^2 \tilde{z}_1  + a_1 z_1 \\
        -\p_t \tilde{z}_1&=-\tilde{z}_1+  z_1 \tilde{z}_1^2 + a_1 \tilde z_1
    \end{split}
\end{equation}
with $z_1(t=0)=1$ and $\tilde{z}_1(t=1)=- \Lambda$. The quantity $z_1(t) \tilde{z}_1(t)$ is obviously conserved
so that $z_1(t) \tilde{z}_1(t)=c_1$ where $c_1$ is a constant independent of time. Thus the system becomes
\begin{equation}
    \begin{split}
        \p_t z_1&= (c_1-1 + a_1) z_1  \\
        -\p_t \tilde{z}_1&=(c_1-1+ a_1) \tilde{z}_1
    \end{split}
\end{equation} 
and we find that $z_1(t)= e^{(c_1-1+a_1) t} $ and $\tilde z_1(t)= - \Lambda e^{(c_1-1+a_1) (1-t)} $. The constant $c_1$ is thus determined by
\be    
z_1 \tilde z_1 = c_1 = - \Lambda e^{c_1-1+a_1}   \quad , \quad c_1 = - W(\Lambda e^{a_1-1}) 
\ee   
where $W(x)$ is the Lambert function, such that $W(x) e^{W(x)}=x$. From the optimal action \eqref{optimal} we obtain,
parametrically
\be   
\hat \Phi(Z) = \frac{c_1^2}{2} \quad , \quad Z=z_1(1) = e^{c_1-1+a_1} 
\ee    
which leads to
\be  \label{S72} 
\hat \Phi(Z) = \frac{1}{2} (1-a_1+\log Z)^2
\ee

{\bf Second method}.
Alternatively going back to the original stochastic model, where the partition sum for $N=1$ is simply $z_1(t)=e^{-t} e^{\sqrt{\varepsilon} B_1(t) + a_1 t}$ 
(here beware $z_1(t)$ is not the solution of the SP equation)
one can compute directly its cumulant generating function, as sketched in the text. Indeed, denoting $u=\sqrt{\varepsilon} B(1)$ one has for small $\varepsilon$,
\be   
\overline{ e^{ - \frac{\Lambda}{\varepsilon} z_1(1)} } = \overline{ e^{ - \frac{\Lambda}{\varepsilon} e^{-1+\sqrt{\varepsilon} B(1) + a_1}} } 
\sim \int_{\R} \rmd u \, e^{- \frac{u^2}{2 \varepsilon}  - \frac{\Lambda}{\varepsilon} e^{u-1+ a_1} } \sim 
e^{ - \frac{1}{\varepsilon} \Psi(\Lambda) } 
\ee  
arising from events where the Brownian is anomalously large, i.e. $u=\sqrt{\varepsilon} B(1)=\mathcal{O}(1)$.
Here we have, for $\Lambda>0$ 
\be   
\Psi(\Lambda)  = \min_{u \in \mathbb{R}} \left( \frac{u^2}{2} + \Lambda e^{u-1+a_1} \right) 
\ee    
There is a single minimum which is reached at
\be   
u= - W(\Lambda e^{a_1-1}) 
\ee 
which is seen to equal the constant $c_1$ of the first method. This gives
\be   \label{Psi11} 
\Psi(\Lambda)  =  \frac{W(\Lambda e^{a_1-1})^2}{2} + W(\Lambda e^{a_1-1}) 
\ee   
as given in the text for $a_1=0$. This gives also 
$\Lambda \Psi'(\Lambda)=\Lambda e^{u-1+a_1}= - u =W(\Lambda e^{a_1-1})$.
Note that $u = c_1= \tilde z_1 z_1 = - \Lambda \Psi'(\Lambda)$ in agreement with the general
formula for conserved quantities (see Section~\ref{sec:conserved-quantities}).\\

Although the function $\Psi(\Lambda)$ was defined and computed for $\Lambda>0$ the formula
\eqref{Psi11} can be continued to negative $\Lambda$, on the principal branche $W=W_0$, down to $\Lambda= \Lambda_c(1) e^{a_1} = -1$. This
is in agreement with the general formula for arbitrary $N$ obtained below which
states that 
\be   
\Lambda_c(N) e^{a_1} = - e^{1-N} N^N 
\ee   

Let us now discuss the Legendre transform which relates $\hat \Phi(Z)$ and $\Psi(\Lambda)$.
One has
\be  
\Psi(\Lambda)= \min_Z ( \Lambda Z + \hat \Phi(Z) ) 
\ee  
leading to
\be \label{ZZ} 
Z = \Psi'(\Lambda) = \frac{W(\Lambda e^{a_1-1})}{\Lambda}
\ee 
The typical value of $Z$ is given by $Z_{\rm typ}=e^{a_1-1}$ using that $W_0'(0)=1$. 
Using that $W_0(-1/e)=-1$ we see that the domain $0^+<Z\leq Z_c=e^{a_1}$ corresponds to $\Lambda>\Lambda_c$.
This is the principal branch with $W=W_0$. \\

To obtain the values of $Z \in [Z_c,\infty)$ one needs to consider a second branch.
This is achieved using the second real branch of the Lambert function, $W=W_1$.
Then one has
\be 
Z :=z_1(t=1)= \Psi'(\Lambda) = \frac{W_{-1}(\Lambda e^{a_1-1})}{\Lambda}
\ee 
where $\Lambda$ increases again from $\Lambda=\Lambda_c$ back to $\Lambda=0^-$, see
Appendix~\ref{subsec:supp-mat-continuation}.\\

One obtains overall
\be
\begin{split}
& \hat \Phi(Z) = \max_{\Lambda \in [-1,+\infty) } ( - \Lambda Z + \Psi_0(\Lambda) ) \, , \qquad  0< Z \leq Z_c \\
& \hat \Phi(Z) = \min_{\Lambda \in [-1,0) } ( - \Lambda Z + \Psi_1(\Lambda) ) \, , \qquad \hspace{0.3cm} Z \geq Z_c 
\end{split}
\ee
with 
\bea 
&& \Psi_0(\Lambda) = \frac{W_0(\Lambda e^{a_1-1})^2}{2} + W_0(\Lambda e^{a_1-1}) \\
&& \Psi_1(\Lambda) = \frac{W_1(\Lambda e^{a_1-1})^2}{2} + W_1(\Lambda e^{a_1-1}) 
\eea 
One can verify that this recovers \eqref{S72} which is indeed analytic for $Z>0$.

\section{Lax pair for the problem}
The aim of this Section is to present a semi-discrete Lax pair for the system \eqref{eq:YO-polymer-saddle-with-drift2}. The zero curvature relation for a differential-difference model is defined as follows

\begin{equation}
\label{eq:Lax-compatibility-supp-mat}
    \begin{split}
        \p_t \Vec{v}_n&=U_n \Vec{v}_n\\
        \Vec{v}_{n+1}&=L_n \Vec{v}_n
    \end{split}
\end{equation}
The original system  \eqref{eq:YO-polymer-saddle-with-drift2} should be found back from the compatibility equation

\begin{equation} \label{compatibilityOY} 
\begin{split}
    \p_t L_n&=U_{n+1}L_n -L_n U_n
    \end{split}
\end{equation}
In our case, $\Vec{v}_n$ will be a two-components vector
\begin{equation}
    \vec{v}_n=(v_n^{(1)},v_n^{(2)})^\intercal 
\end{equation}
and the $2 \times 2$ matrices $U_n$, $L_n$ composing the  Lax pair read

\begin{equation}
U_n=
\begin{pmatrix}
\frac{\lambda^2-1}{2}  &   -z_{n-1} \\ 
&\\
\frac{g}{2}\tilde{z}_n &  \frac{1-\lambda^2}{2}
\end{pmatrix}, \qquad 
L_n=
\begin{pmatrix}
\frac{1}{\lambda} & \frac{z_n}{\lambda}\\
 -\frac{g}{2\lambda}\tilde{z}_{n} & \lambda- \frac{g}{2\lambda}z_n\tilde{z}_{n} - \frac{a_n}{\lambda}
\end{pmatrix}\, .
\label{eq:LaxPairYO-new}
\end{equation}
where $\lambda$ is the spectral parameter. We emphasize again here that in our problem the drifts $\{a_n \}$ are zero outside the interval $[1,N]$.\\

Quite remarkably, we note that the matrix $L_k$ admits the following factorization 
    \begin{equation}
    \label{eq:supp-mat-factorisation-lax-Ln}
        L_n = \begin{pmatrix}
1  &   0 \\ 
-\frac{g}{2}\tilde{z}_n &  1
\end{pmatrix}
\begin{pmatrix}
\frac{1}{\lambda} &   0 \\ 
0 &  \lambda - \frac{a_n}{\lambda}
\end{pmatrix}
\begin{pmatrix}
1  &   z_{n} \\ 
0 &  1
\end{pmatrix}
    \end{equation}
so that the contribution of the optimal partition function $z_n$ and the response field $\tilde{z}_n$ can be split.

\section{Definition of the scattering problem}
Assuming the fields $\{z_n, \tilde{z}_n \}$ as well as the drifts $\{a_n \}$ to decay to 0 for $n\to \pm \infty$, the asymptotic Lax matrices \eqref{eq:LaxPairYO-new} are diagonal so that we can define two sets of independent solutions asymptotically
\begin{equation}
    \phi_n  \underset{n\to -\infty }{\sim} \lambda^{-n}
    \begin{pmatrix}
1 \\
0
\end{pmatrix}, \quad 
\bar{\phi}_n \underset{n\to -\infty }{\sim} \lambda^n 
    \begin{pmatrix}
0 \\
-1
\end{pmatrix}
\end{equation}
and
\begin{equation}
    \psi_n  \underset{n\to +\infty}{\sim} \lambda^n 
    \begin{pmatrix}
0 \\
1
\end{pmatrix}, \quad 
    \bar{\psi}_n  \underset{n\to +\infty}{\sim} \lambda^{-n} 
    \begin{pmatrix}
1 \\
0
\end{pmatrix}
\end{equation}
Since there can be only two independent solutions, there exists a linear combination relating the two sets which defines the scattering amplitudes
\begin{equation}
\label{eq:def-scattering-amplitudes}
\begin{split}
    \phi_n &= a(\lambda,t) \bar{\psi}_n+b(\lambda,t)\psi_n\\
    \bar{\phi}_n &= -\atilde(\lambda,t)\psi_n + \btilde(\lambda,t)\bar{\psi}_n
    \end{split}
\end{equation}
equivalent to implying the following asymptotic conditions at $+\infty$
\begin{equation}
    \phi_n\underset{n\to +\infty}{\sim}
    \begin{pmatrix}
a(\lambda,t)\lambda^{-n}  \\
b(\lambda,t) \lambda^n
\end{pmatrix}, \quad 
\bar{\phi}_n  \underset{n\to +\infty}{\sim}
    \begin{pmatrix}
\btilde(\lambda,t)\lambda^{-n} \\
-\atilde(\lambda,t)\lambda^{n}
\end{pmatrix}
\end{equation}
In practice, the two independent solutions $ \{\vec{v}_n \}$ verifying the time evolution equation will be chosen as
\begin{equation}
\label{eq:solution-with-time}
    \vec{v}_n = e^{\frac{\lambda^2-1}{2}t}\phi_n\, \quad \text{and} \quad \vec{v}_n = e^{-\frac{\lambda^2-1}{2}t}\bar{\phi}_n
\end{equation}

\section{Time dependence of the scattering amplitudes}
Inserting the solutions \eqref{eq:solution-with-time} into the time equation of the Lax pair \eqref{eq:Lax-compatibility-supp-mat} and evaluating it at $n=+\infty$, we obtain that 

\begin{equation}
    \begin{split}
        &\p_t a(\lambda,t)=0\\
        &\p_t \atilde(\lambda,t)=0\\
        &\p_t b(\lambda,t) = (1-\lambda^2) b(\lambda,t)\\
        &\p_t \btilde(\lambda,t) = (\lambda^2-1) \btilde(\lambda,t)
    \end{split}
\end{equation}

Hence the scattering amplitudes are either time-independent (for $a(\lambda)$ and $\atilde(\lambda)$) or have a Gaussian time dependence
\begin{equation}
    \btilde(\lambda,t)=\btilde(\lambda)e^{(\lambda^2-1)t}, \quad b(\lambda,t)=b(\lambda) e^{(1-\lambda^2)t}
\end{equation}
The opposite sign in the time evolution of $b(\lambda)$ and $\btilde(\lambda)$ reflects the fact that the saddle point equations describe simultaneously forward and backward evolutions in time. This result is universal as long as the fields vanish at $\pm \infty$.

\section{Wronskian of the solution}
On top of the time evolution of the scattering amplitudes, we now determine a normalisation relation using the Wronskian of the problem. For the two solutions of the Lax problem $\{\phi_n, \bar{\phi}_n \}_{n \in \Z}$, we define the Wronskian as
\begin{equation}
    W_n =W(\phi_n,\bar{\phi}_n)= \phi_n^{(1)} \bar{\phi}_n^{(2)}-\phi_n^{(2)} \bar{\phi}_n^{(1)}
\end{equation}
From the evolutions \eqref{eq:Lax-compatibility-supp-mat}, 
We start with the time derivative and the index increment
\begin{equation}
\label{eq:ev-wronskian}
\begin{split}
    \p_t W_n &= \Tr(U_n)W_n = 0, \\
     W_{n+1}&=\Det(L_n)W_n=( 1- \frac{a_n}{\lambda^2} ) W_n
    \end{split}
\end{equation}
The Wronskian of $\{\phi_n, \bar{\phi}_n \}_{n \in \Z}$ at both infinities read
\begin{equation}
    W_{n}\underset{n\to -\infty}{\sim} -1, \quad  W_{n}\underset{n\to +\infty}{\sim} -(a\atilde+b\btilde)
\end{equation}
and we have also have from Eq.~\eqref{eq:ev-wronskian}
\be
W_{ +\infty} = \prod_{n=1}^N ( 1- \frac{a_n}{\lambda^2} ) W_{ -\infty}
\ee 
We then deduce that
\begin{equation}
\label{eq:normalisation-scattering}
    a(\lambda)\atilde(\lambda)+b(\lambda)\btilde(\lambda)= \prod_{n=1}^N ( 1- \frac{a_n}{\lambda^2} )
\end{equation}
This normalisation relation is universal. Contrary to the Ablowitz-Ladik integrable system \cite{ablowitz2004discrete, footnote-ablowitz-ladik}, the normalisation of the scattering amplitudes \eqref{eq:normalisation-scattering} does not depend on the fields $\{z_n, \tilde{z}_n \}$.

\section{Conserved quantities}  
\label{sec:conserved-quantities}

We derive in this Section the conserved quantities of the integrable system \eqref{eq:YO-polymer-saddle-with-drift2}. As we shall see, there exist an infinite amount of conserved quantities, which is standard for such models. To obtain the conserved quantities, we require three ingrediens:
\begin{enumerate}
    \item The Ricatti equation of the Lax pair;
    \item A continuity equation arising from the compatibility of the Lax equations expressed with log derivatives and the Ricatti variables and a relation between the continuity equation and the scattering amplitudes;
    \item A suitable Taylor expansion of $\log a(\lambda)$ or Laurent expansion of $\log \atilde(\lambda)$ as a function of the spectral parameter.
\end{enumerate}
\subsection{Ricatti equation}

Let us first define the Ricatti variable $\Gamma$ and its inverse $\tilde{\Gamma}$ as
\begin{equation}
    \Gamma_n = \frac{v_n^{(2)}}{v_n^{(1)}}, \quad \tilde{\Gamma}_n=\frac{1}{\Gamma_n}
\end{equation}
Dividing the two equations of the space part of the Lax pair, we obtain the following recursions for the Ricatti variable and its inverse

\begin{equation}
\label{eq:ricatti-eq-1}
   ( \Gamma_{n+1}+\tilde{z}_n)(1+z_n\Gamma_n)=\Gamma_n (\lambda^2 - a_n) 
\end{equation}
and
\begin{equation}
\label{eq:ricatti-eq-2}
   ( 1+\tilde{z}_n\tilde{\Gamma}_{n+1})(\tilde{\Gamma}_{n}+z_n)=\tilde{\Gamma}_{n+1} (\lambda^2 - a_n)
\end{equation}
As we shall see subsequently, we need to expand these equations to obtain the Taylor and Laurent series of the Ricatti variables.

\subsection{Continuity equations}
The continuity equations will be obtained as the compatibility of the dynamics of $\log v_n^{(1)}$ and $\log v_n^{(2)}$ respectively. 

\subsubsection{First continuity equation}
From the Lax pair system, we obtain the pair of equations for $\log v_n^{(1)}$
\begin{equation}
\begin{cases}
    \p_t \log v_n^{(1)}=\frac{\lambda^2-1}{2}-z_{n-1}\Gamma_n\\ 
    \log(\frac{v^{(1)}_{n+1}}{v^{(1)}_{n}})=\Delta^+ \log v^{(1)}_{n}  = \log(\frac{1}{\lambda}+\frac{z_n}{\lambda} \Gamma_n) 
    \end{cases}
\end{equation}
where we introduced the following notation for the finite difference $\Delta^+ f_n=f_{n+1}-f_n$. We rewrite the above system for convenience as 
\begin{equation}
\label{eq:continuity-eq-rescaled}
    \begin{cases}
            \p_t \log (v_n^{(1)}\lambda^{n}e^{-\frac{\lambda^2-1}{2}t})=-z_{n-1}\Gamma_n\\
            \\
            \log\left(\frac{v^{(1)}_{n+1}\lambda^{n+1}e^{-\frac{\lambda^2-1}{2}t}}{v^{(1)}_{n}\lambda^{n}e^{-\frac{\lambda^2-1}{2}t}}\right) = \Delta^+ \log\left(v^{(1)}_{n}\lambda^{n}e^{-\frac{\lambda^2-1}{2}t}\right) =\log(1+z_n\Gamma_n) 
    \end{cases}
\end{equation}
The compatibility is obtained by the commutation relation 
\begin{equation}
    \p_t \Delta^+ = \Delta^+ \p_t
\end{equation}
which yields the first compatibility equation
\begin{equation}
    \p_t \log(1+z_n \Gamma_n)  = -\Delta^+ ( z_{n-1}\Gamma_n)
\end{equation}
We therefore interpret $J_n^{(1)}= z_{n-1}\Gamma_n$ as a generalised current and $\varrho_n^{(1)}=\log(1+z_n \Gamma_n)$ as a generalised density. In particular, since the potentials $\{z_n, \tilde{z}_n \}$ vanish at infinities, we have the conservation law
\begin{equation}
    \p_t \left(\sum_{n=-\infty}^{+\infty} \varrho_n^{(1)} \right) =0
\end{equation}

With the particular choice of $\vec{v}_n = e^{\frac{\lambda^2-1}{2}t}\phi_n$ and summing the second equation of \eqref{eq:continuity-eq-rescaled} over integers in $\Z$ we obtain the expected relation between the scattering amplitude and this set of conserved charges as

\begin{equation}
\label{eq:conserved-q-scattering-a}
    \log a(\lambda) = \sum_{n=-\infty}^{+\infty} \log(1+z_n\Gamma_n)  \quad \longleftrightarrow \quad 
        a(\lambda) = \prod_{n=-\infty}^{+\infty}  (1+z_n\Gamma_n) 
    \end{equation}

\subsubsection{Second continuity equation}
We now repeat the same exercise for $\log v_n^{(2)}$ and first obtain 
\begin{equation}
\begin{cases}
    \p_t \log v_n^{(2)}=\frac{1-\lambda^2}{2}+\tilde{z}_{n}\tilde{\Gamma}_n\\
    \log(\frac{v^{(2)}_{n+1}}{v^{(2)}_{n}})=\Delta^+ \log v^{(2)}_{n}  = \log(-\tilde{z}_n \tilde{\Gamma}_n+\lambda-\frac{1}{\lambda}z_n \tilde{z}_n -\frac{a_n}{\lambda}) 
\end{cases} 
\end{equation}

which is also rewritten for convenience as

\begin{equation}
\label{eq:continuity-eq-rescaled-2}
    \begin{cases}
            \p_t \log (-v_n^{(2)}\lambda^{-n}e^{\frac{\lambda^2-1}{2}t})=\tilde{z}_{n}\tilde{\Gamma}_n\\
            \\
            \log\left(\frac{-v^{(2)}_{n+1}\lambda^{-n-1}e^{\frac{\lambda^2-1}{2}t}}{-v^{(2)}_{n}\lambda^{-n}e^{\frac{\lambda^2-1}{2}t}}\right) = \log(1-\frac{1}{\lambda^2}\tilde{z}_n (\tilde{\Gamma}_n+z_n )-\frac{a_n}{\lambda^2}) 
    \end{cases}
\end{equation}
The compatibility in this case reads
\begin{equation}
    \p_t   \log(1-\frac{1}{\lambda^2}\tilde{z}_n (\tilde{\Gamma}_n+z_n )-\frac{a_n}{\lambda^2})  = -\Delta^+ (-\tilde{z}_n\tilde{\Gamma}_n)
\end{equation}
We also interpret $J_n^{(2)} = -\tilde{z}_n\tilde{\Gamma}_n$ as a generalised current and $\varrho_n^{(2)}= \log(1-\frac{1}{\lambda^2}\tilde{z}_n (\tilde{\Gamma}_n+z_n )-\frac{a_n}{\lambda^2}) $ as a generalised density. In particular, since the potentials $\{z_n, \tilde{z}_n \}$ vanish at infinities, we have the conservation law
\begin{equation}
    \p_t \left(\sum_{n=-\infty}^{+\infty} \varrho_n^{(2)} \right)=0
\end{equation}

With the particular choice of $\vec{v}_n = e^{\frac{1-\lambda^2}{2}t}\bar{\phi}_n$ and summing the second equation of \eqref{eq:continuity-eq-rescaled-2} over integers in $\Z$ we obtain the expected relation between the scattering amplitude and this set of conserved charges as

\begin{equation}
\label{eq:conserved-q-scattering-atilde}
    \log \atilde(\lambda) = \sum_{n=-\infty}^{+\infty}  \log(1-\frac{1}{\lambda^2}\tilde{z}_n (\tilde{\Gamma}_n+z_n )-\frac{a_n}{\lambda^2})  \quad \longleftrightarrow \quad 
        \atilde(\lambda) = \prod_{n=-\infty}^{+\infty}  (1-\frac{1}{\lambda^2}\tilde{z}_n (\tilde{\Gamma}_n+z_n )-\frac{a_n}{\lambda^2}) 
    \end{equation}

\subsection{Conserved charges}
\label{subsec:supp-mat-conserved-quantities}
We now complete the determination of the conserved charges in the system by proceeding to the suitable expansion of the continuity equations. We now take all drifts equal to 0, i.e. $a_n=0$, to simplify the expressions.
\subsubsection{Taylor expansion of $a(\lambda)$}
Since $\log a(\lambda)$ is analytic close to the origin, we choose to expand Eqs.~\eqref{eq:ricatti-eq-1}-\eqref{eq:ricatti-eq-2}-\eqref{eq:conserved-q-scattering-a} in a Taylor series
\begin{equation}
    \log a(\lambda)=\sum_{\ell=0}^{\infty} \lambda^{2\ell} C_\ell, \quad \Gamma_n = \sum_{\ell=0}^{\infty} \lambda^{2\ell} \Gamma_n^{(\ell,0)}, \quad \tilde{\Gamma}_n = \sum_{\ell=0}^{\infty} \lambda^{2\ell} \tilde{\Gamma}_n^{(\ell,0)}
\end{equation}
and
\begin{equation}
\varrho_n^{(1)}=\sum_{\ell=0}^{\infty} \lambda^{2\ell} \varrho^{(1)}_{n,\ell}, \quad J_n^{(1)}=\sum_{\ell=0}^{\infty} \lambda^{2\ell} J^{(1)}_{n,\ell}, \quad C_\ell = \sum_{n \in \Z}\varrho^{(1)}_{n,\ell}
\end{equation}
Formally, there exists two solutions to the Taylor expansion of Eqs.~\eqref{eq:ricatti-eq-1}-\eqref{eq:ricatti-eq-2}:
\begin{itemize}
    \item The first one is physical as it consistent with a Taylor expansion of $\log a(\lambda)$ and it provides the expected conserved quantities;
    \item The second one is unphysical as it would impose that $\log a(\lambda)$ behaves as $\log \lambda^2$ for small spectral parameter and therefore that the solution would include a zero-mode, incompatible with the decay of $z_n, \tilde{z}_n$ at infinities. Nonetheless formally, this expansion yields additional conserved quantities, which conservation can also be checked by hand, and for completeness we will provide them. 
\end{itemize}

The first solution, the 'physical' expansion, yields for its first two terms
\begin{enumerate}
    \item For $\ell=0$
    \begin{equation}
   \varrho^{(1)}_{n,0}= \log (1-z_n \tilde{z}_{n-1}) , \quad J^{(1)}_{n,0}= -z_{n-1}\tilde{z}_{n-1}  
    \end{equation}
    \item For $\ell=1$
    \begin{equation}
   \varrho^{(1)}_{n,1}= -\frac{ z_n \tilde{z}_{n-2}}{\left( 1-z_{n-1} \tilde{z}_{n-2}\right) \left( 1-z_n \tilde{z}_{n-1}\right)}, \quad J^{(1)}_{n,1}= 1-\frac{1}{1- z_{n-1} \tilde{z}_{n-2}}
    \end{equation}
\end{enumerate}

The conserved charges are then
\begin{equation}
\begin{split}
    C_0 &= \sum_{n =-\infty}^{\infty} \log (1-z_n \tilde{z}_{n-1}) \\
    C_1 &= \sum_{n =-\infty}^{\infty}-\frac{ z_n \tilde{z}_{n-2}}{\left( 1-z_{n-1} \tilde{z}_{n-2}\right) \left( 1-z_n \tilde{z}_{n-1}\right)}
    \end{split}
\end{equation}
and one can verify using the dynamical equations for $\{ z_n, \tilde{z}_n \} \eqref{eq:YO-polymer-saddle}$  that for any solution we obtain  
\begin{equation}
 \p_t \varrho^{(1)}_{n,\ell}= J^{(1)}_{n,\ell}-J^{(1)}_{n+1,\ell}   
\end{equation}

The second solution, the 'unphysical' expansion, yields for its first two terms (we use again the same notation for simplicity)
\begin{enumerate}
    \item For $\ell=0$
    \begin{equation}
   \varrho^{(1)}_{n,0}=\log \left(\frac{z_{n+1}}{z_n \left(1-z_{n+1} \tilde{z}_n\right)}\right) , \quad J^{(1)}_{n,0}= -\frac{z_{n-1}}{z_n}
    \end{equation}
    \item For $\ell=1$
    \begin{equation}
   \varrho^{(1)}_{n,1}=\frac{\frac{z_{n+2}}{z_{n+2} \tilde{z}_{n+1}-1}+\frac{z_{n+1}^2}{z_n}}{z_{n+1} \left(z_{n+1} \tilde{z}_n-1\right)}, \quad J^{(1)}_{n,1}=-\frac{z_{n-1} z_{n+1}}{z_n^2 \left(z_{n+1} \tilde{z}_n-1\right)}
    \end{equation}
\end{enumerate}

\subsubsection{Laurent expansion of $\atilde(\lambda)$}

Since $\log \atilde(\lambda)$ is analytic for large $\abs{\lambda}$, we choose to expand Eqs.~\eqref{eq:ricatti-eq-1}-\eqref{eq:ricatti-eq-2}-\eqref{eq:conserved-q-scattering-atilde} as  Laurent series
\begin{equation}
    \log \atilde(\lambda)=\sum_{\ell=1}^{\infty} \frac{\tilde{C}_\ell}{\lambda^{2\ell} }, \quad \Gamma_n = \sum_{\ell=1}^{\infty} \frac{\Gamma_n^{(\ell,\infty)}}{\lambda^{2\ell} }, \quad \tilde{\Gamma}_n = \sum_{\ell=1}^{\infty} \frac{\tilde{\Gamma}_n^{(\ell,\infty)}}{\lambda^{2\ell} }
\end{equation}
and
\begin{equation}
\varrho_n^{(2)}=\sum_{\ell=1}^{\infty} \frac{\varrho^{(2)}_{n,\ell}}{\lambda^{2\ell} }, \quad J_n^{(2)}=\sum_{\ell=1}^{\infty} \frac{J^{(2)}_{n,\ell}}{\lambda^{2\ell} }, \quad \tilde{C}_\ell = \sum_{n \in \Z}\varrho^{(2)}_{n,\ell}
\end{equation}
Formally, there exists two solutions to the Laurent expansion of Eqs.~\eqref{eq:ricatti-eq-1}-\eqref{eq:ricatti-eq-2}:
\begin{itemize}
    \item The first one is physical as it consistent with a Laurent expansion of $\log \atilde(\lambda)$ and it provides the expected conserved quantities;
    \item The second one is unphysical as it would impose that $\log \atilde(\lambda)$ behaves as $\log \lambda^2$ for large spectral parameter whereas we seek a solution for which $\log \atilde(\lambda)\to 0$ for $\abs{\lambda}\to \infty$. Nonetheless, formally this expansion yields additional conserved quantities, which conservation can also be checked by hand, and for completeness we will provide them. 
\end{itemize}

The first solution, the 'physical' expansion, yields for its first two terms

\begin{enumerate}
    \item For $\ell=1$
    \begin{equation}
    \label{eq:conserved-quantities-laurent-suppmat}
   \varrho^{(2)}_{n,1}=-z_n \tilde{z}_n  , \quad J^{(2)}_{n,1}=-z_{n-1}\tilde{z}_n   
    \end{equation}
    \item For $\ell=2$
    \begin{equation}
   \varrho^{(2)}_{n,2}= -\frac{1}{2} \tilde{z}_n \left(z_n^2 \tilde{z}_n+2 z_{n-1}\right), \quad J^{(2)}_{n,2}=-\tilde{z}_n \left(z_{n-1}^2 \tilde{z}_{n-1}+z_{n-2}\right)
    \end{equation}
\end{enumerate}

The conserved charges are then
\begin{equation}
\begin{split}
    \tilde{C}_1 &=-\sum_{n =-\infty}^{\infty}z_n \tilde{z}_n   \\
    \tilde{C}_2 &=- \sum_{n =-\infty}^{\infty}\frac{1}{2} \tilde{z}_n \left(z_n^2 \tilde{z}_n+2 z_{n-1}\right)
    \end{split}
\end{equation}
and one can verify using the dynamical equations for $\{ z_n, \tilde{z}_n \} \eqref{eq:YO-polymer-saddle}$  that for any solution we obtain  
\begin{equation}
 \p_t \varrho^{(2)}_{n,\ell}= J^{(2)}_{n,\ell}-J^{(2)}_{n+1,\ell}   
\end{equation}

The second solution, the 'unphysical' expansion, yields for its first two terms (we use again the same notation for simplicity)

\begin{enumerate}
    \item For $\ell=1$
    \begin{equation}
   \varrho^{(2)}_{n,1}=\log \left(\frac{ \tilde{z}_{n+1}}{\tilde{z}_n}\right) , \quad J^{(2)}_{n,1}=z_n \tilde{z}_n+\frac{\tilde{z}_{n+1}}{\tilde{z}_n}
    \end{equation}
    \item For $\ell=2$
    \begin{equation}
   \varrho^{(2)}_{n,2}=z_{n+1} \tilde{z}_{n+1}-\frac{\tilde{z}_{n+1}}{\tilde{z}_n}+\frac{\tilde{z}_{n+2}}{\tilde{z}_{n+1}} , \quad J^{(2)}_{n,2}=\frac{z_{n+1} \tilde{z}_{n+1}^2}{\tilde{z}_n}-\frac{\tilde{z}_{n+1}^2}{\tilde{z}_n^2}+\frac{\tilde{z}_{n+2}}{\tilde{z}_n}
    \end{equation}
\end{enumerate}

\subsection{Discussion on Generalized Gibbs Ensembles and other flows generated by the conserved charges}

We first comment on the fact that the quantity $(z_n \tilde{z}_n)$ is at the same a local density and a local current as
\begin{equation}
   - z_n \tilde{z}_n=J_{n,0}^{(1)}=\varrho_{n,1}^{(2)}
\end{equation}

Such equality also arises in other models and has \textit{"surprising consequences"}, see Ref.~\cite[Sec.~2.1]{spohn2021hydrodynamic} in the context of the Toda lattice. It would be interesting in the present context to analyse the consequence of this identity.\\

What is more, from the knowledge of the conserved quantities, it is possible to formally define a Generalized Gibbs measure on the fields $z_n,\tilde{z}_n$ that is conserved by the dynamics \eqref{eq:YO-polymer-saddle}. Such construction was done for a variety of semi-discrete integrable models in recent works, see Refs.~\cite{grava2023discrete,mazzuca2023equilibrium,spohn2020generalized,spohn2021hydrodynamic,spohn2022hydrodynamic}. To this aim, we select a subset of conserved quantities, e.g. $C_0, \tilde{C}_1, \tilde{C}_2$, and a set of conjugated "temperature" (in this example $\beta_0, \beta_1, \beta_2$) and construct the measure
\begin{equation}
\label{eq:measure-GGE}
    \prod_{n=1}^N \rmd z_n \rmd \tilde{z}_n \exp (-\beta_0 C_0 -\beta_1 \tilde{C}_1-\beta_2 \tilde{C}_2)
\end{equation}
If the initial conditions of the OY system are chosen randomly according to the measure \eqref{eq:measure-GGE}, we then expect this distribution of $z_n, \tilde{z}_n$ to be stationary as the dynamics will preserve $C_0, \tilde{C}_1, \tilde{C}_2$. Such measure can a priori be extended to arbitrary number of conserved quantities and for the usual semi-discrete integrable models, this measure was related to measures appearing in Random Matrix Theory \cite{grava2023discrete,mazzuca2023equilibrium,spohn2020generalized,spohn2021hydrodynamic,spohn2022hydrodynamic}, we expect such a connexion to hold in this case as well.\\

Finally, we address a last comment about the conserved charges which is the flow induced by these. We can recast the original OY system \eqref{eq:YO-polymer-saddle} as 
\begin{equation}
\label{eq:supp-mat-hamiltonian-flow}
    \begin{split}
        \p_t z_n&= \frac{\delta}{\delta \tilde{z}_n} (\tilde{C}_1-\tilde{C}_2)\\ 
        -\p_t \tilde{z}_n&= \frac{\delta }{\delta z_n}(\tilde{C}_1-\tilde{C}_2)
    \end{split}
\end{equation}
Similarly to Ref.~\cite{merola1994novel}, one could define other dynamics by replacing in \eqref{eq:supp-mat-hamiltonian-flow} the term $\tilde{C}_1-\tilde{C}_2$ by any linear combination of conserved quantities. The differential system induced would remain integrable with the same space Lax matrix $L_n$ in \eqref{eq:LaxPairYO-new} but the time Lax matrix $U_n$ would have to be modified.\\

The Hamiltonian system \eqref{eq:supp-mat-hamiltonian-flow} can also be rewritten in a symplectic form by introducing the suitable Poisson bracket. To this aim, introduce the field gradient
\begin{equation}
    \nabla_n =  
    \begin{pmatrix}
        \dfrac{\delta}{\delta z_n} \\ 
        \dfrac{\delta}{\delta \tilde{z}_n} \, ,
    \end{pmatrix}
\end{equation}
and rewrite \eqref{eq:supp-mat-hamiltonian-flow} as 
\begin{equation}
     \p_t  \begin{pmatrix}
      z_n \\ \tilde{z}_n
    \end{pmatrix} =
    \begin{pmatrix}
        0 & 1 \\
        -1 & 0
    \end{pmatrix}
         \nabla_n     (\tilde{C}_1-\tilde{C}_2) \, .
\end{equation}
Then by introducing the Poisson bracket 
\begin{equation}
\label{eq:sup-mat-canonical-poisson}
    \langle F,G \rangle = \sum_{n=-\infty}^{+\infty} \nabla_n (F)     \begin{pmatrix}
        0 & 1 \\
        -1 & 0
    \end{pmatrix} \nabla_n(G) \, ,
\end{equation}
we have that if $I[z,\tilde{z}]$ is a functional of $z,\tilde{z}$, its dynamic is given by
\begin{equation}
    \p_t I[z,\tilde{z}]=\langle I[z,\tilde{z}],\tilde{C}_1-\tilde{C}_2 \rangle \, .
\end{equation}
In the present case, the canonical Poisson bracket \eqref{eq:sup-mat-canonical-poisson} is the natural structure behind the dynamics but other Poisson brackets could appear in different semi-discrete integrable models as is the case for continuous systems where the Poisson bracket is not the same for the nonlinear Schrodinger equation \cite{ablowitz2004discrete} and the derivative nonlinear Schrodinger equation \cite{kaup1978exact}.

\section{Solution of the scattering problem for arbitrary $g$ }

We now obtain the solution of the scattering problem associated to the non-linear system for arbitrary $g$, for the 
specific initial and final conditions considered in this paper, that we now recall
\begin{equation}
    z_n(t=0)=\delta_{n,1} \quad , \quad  \tilde{z}_n(t=1)=-\Lambda\delta_{N,n} \label{boundary00} 
\end{equation}
and we recall that $z_{n \leq 0}(t)=0$ and $\tilde{z}_{n \geq N+1}(t)=0$. 
The scattering problem is defined by the recursion, for $n \in \mathbb{Z}$
\be 
\label{eq:scattcomp}
             \phi_{n+1} = \begin{pmatrix}
  \phi^{(1)}_{n+1} \\
 \phi^{(2)}_{n+1} 
\end{pmatrix}
             =
              \begin{pmatrix}
\frac{1}{\lambda} & \frac{z_n}{\lambda}\\
 -\frac{g}{2\lambda}\tilde{z}_{n} & \lambda- \frac{g}{2\lambda}z_n\tilde{z}_{n} - \frac{a_n}{\lambda}
\end{pmatrix}
  \begin{pmatrix}
  \phi^{(1)}_{n} \\
 \phi^{(2)}_{n} 
\end{pmatrix}
\ee 
together with the same equation for $\bar \phi_n$, with the following boundary conditions at $n \to \pm \infty$ 
\bea \label{bc01} 
 &&    \phi_n  \sim \lambda^{-n}
    \begin{pmatrix}
1 \\
0
\end{pmatrix}, \quad 
\bar{\phi}_n \sim \lambda^n 
    \begin{pmatrix}
0 \\
-1
\end{pmatrix}, \quad n \to -\infty \\
&&  \label{bc02}  
    \phi_n  \sim 
    \begin{pmatrix}
a(\lambda)\lambda^{-n}  \\
b(\lambda) e^{(1-\lambda^2) t} \lambda^n
\end{pmatrix}, \quad 
\bar{\phi}_n \sim 
    \begin{pmatrix}
\btilde(\lambda) e^{(\lambda^2-1) t} \lambda^{-n} \\
-\atilde(\lambda)\lambda^{n}
\end{pmatrix}, \quad n \to +\infty 
\eea

We will now consider the scattering problem successively at the initial time $t=0$ and at the final time $t=1$.
In each case we consider it first for $\phi$ and then for $\bar \phi$.

\subsection{Scattering problem at $t=0$}

\subsubsection{A.1 For $ \phi$}

One must solve, from \eqref{eq:scattcomp} and the initial condition \eqref{boundary00} 
\begin{equation}
\begin{split}
& \phi^{(1)}_{n+1}=\frac{1}{\lambda}\phi^{(1)}_{n} + \frac{1}{\lambda} \delta_{n,1} \phi^{(2)}_{n} \\
& \phi^{(2)}_{n+1}= -\frac{g}{2\lambda }\tilde{z}_n\phi^{(1)}_{n}+ (\lambda -\frac{g}{2\lambda }\tilde{z}_n \delta_{n,1} - \frac{a_n}{\lambda}) \phi^{(2)}_{n} \label{pair0} 
\end{split}
\end{equation}
Using \eqref{bc01}, the first equation implies that
\begin{equation}
    \begin{split}
        \phi_n^{(1)}&=\lambda^{-n}, \qquad n\leq 1\\
    \phi_n^{(1)}&=\lambda^{-n}+\lambda^{1-n}\phi_1^{(2)}, \qquad n\geq 2\\
    \end{split}
\end{equation}
In particular we obtain from \eqref{bc02} that
\begin{equation} \label{aphi1} 
    a(\lambda) = 1+\lambda \phi_1^{(2)}
\end{equation}
The second equation in \eqref{pair0} then leads to (we recall that $a_n=0$ for $n<1$ and for $n>N$)
\begin{equation}
\begin{split}
    \phi^{(2)}_{n+1}&= -\frac{g}{2 }\tilde{z}_n \lambda^{-n-1}+ \lambda \phi^{(2)}_{n} , \quad n<1\\
    \phi^{(2)}_{n+1}&= -\frac{g}{2 }\tilde{z}_n \lambda^{-n-1}(1+\lambda \phi_1^{(2)})+ \lambda (1- \frac{a_n}{\lambda^2})\phi^{(2)}_{n} , \quad n>1\\
    \end{split}
\end{equation}
which implies using \eqref{bc01} 
\bea \label{soluphi01} 
&&     \phi^{(2)}_{n}=\sum_{k=-\infty}^{n-1}-\frac{g}{2}\tilde{z}_k  \lambda^{n-2-2k} , \quad n\leq 1\\
&&     \phi^{(2)}_{2} = -\frac{g}{2\lambda^2}\tilde{z}_1+ (\lambda -\frac{g}{2\lambda }\tilde{z}_1(t=0) { - \frac{a_1}{\lambda}} ) \phi^{(2)}_{1} , \quad n=2 \\
&&  \phi^{(2)}_{n}=\sum_{k=2}^{n-1}\bigg( -\frac{g}{2}\tilde{z}_k  \lambda^{n-2-2k}(1+\lambda \phi_1^{(2)}) 
\prod_{\ell=k+1}^{n-1} (1- \frac{a_\ell}{\lambda^2}) \bigg)
+\lambda^{n-2} \prod_{\ell=2}^{n-1} (1- \frac{a_\ell}{\lambda^2}) \phi^{(2)}_{2} , \quad n\geq 3  \label{soluphi03} 
\eea

We can simplify the $n=2$ term as
\begin{equation}
\begin{split}
    \phi^{(2)}_{2}& = -\frac{g}{2\lambda^2}\tilde{z}_1(t=0)+ (\lambda -\frac{g}{2\lambda }\tilde{z}_1(t=0) {- \frac{a_1}{\lambda} }) \frac{a(\lambda)-1}{\lambda} \\
    &= a(\lambda)(1 -\frac{g}{2\lambda^2 }\tilde{z}_1(t=0) {- \frac{a_1}{\lambda^2} } )  -1 + {\frac{a_1}{\lambda^2} } \\ 
    \end{split}
\end{equation}
We finally obtain $a(\lambda)$ from \eqref{aphi1} with $\phi^{(2)}_{1}$ given in \eqref{soluphi01}, and $b(\lambda)$
from the asymptotic behavior at $n \to +\infty$ of \eqref{soluphi03} 
\bea  
a(\lambda) &=& 1- \frac{g}{2}\sum_{k=-\infty}^{0}\tilde{z}_k(t=0)  \lambda^{-2k} \label{resa} \\
b(\lambda) & =&  \frac{a(\lambda)}{\lambda^2}\sum_{k=1}^{\infty}\bigg( -\frac{g}{2}\tilde{z}_k(t=0)  \lambda^{-2k} \prod_{\ell=k+1}^{N} (1- \frac{a_\ell}{\lambda^2}) \bigg) 
    + \frac{a(\lambda)-1}{\lambda^2} \prod_{\ell=1}^{N} (1- \frac{a_\ell}{\lambda^2}) \label{resb} 
\eea

\subsubsection{A.2 For $ \bar \phi$}

The vector $\bar \phi$ satisfies the same equation as $\phi$ but with different boundary conditions
at infinity, see \eqref{bc01}, \eqref{bc02}
\begin{equation} \label{eqphib0} 
\begin{split}
& \bar \phi^{(1)}_{n+1}=\frac{1}{\lambda}\bar\phi^{(1)}_{n} + \frac{1}{\lambda} \delta_{n,1} \bar\phi^{(2)}_{n} \\
& \bar\phi^{(2)}_{n+1}= -\frac{g}{2\lambda }\tilde{z}_n \bar\phi^{(1)}_{n}+ (\lambda -\frac{g}{2\lambda }\tilde{z}_n \delta_{n,1} {- \frac{a_n}{\lambda}})\bar \phi^{(2)}_{n} 
\end{split}
\end{equation}
Using \eqref{bc01} the first equation gives
\begin{equation}
\begin{split}
    &\bar \phi^{(1)}_{n}=0, \quad  n\leq 1\\
    &\bar \phi^{(1)}_{n}=\frac{\bar{\phi}_1^{(2)}}{\lambda^{n-1}}, \quad n\geq 2\\
    \end{split}
\end{equation}
which from \eqref{bc02} implies that 
\begin{equation}
    \btilde(\lambda)=\lambda \bar{\phi}_1^{(2)}
\end{equation}
Using \eqref{bc01} the second equation in \eqref{eqphib0} can be solved as
 \begin{equation}
 \begin{split}
   \bar \phi^{(2)}_{n}&=-\lambda^n,  \quad n\leq1\\
   \bar \phi^{(2)}_{2}&=-(\lambda-\frac{g}{2\lambda }\tilde{z}_1(t=0) {- \frac{a_1}{\lambda}} )\lambda,  \quad n=2 \\
    \bar\phi^{(2)}_{n}&=\sum_{k=2}^{n-1} \big( \frac{g}{2}\tilde{z}_k  \lambda^{n-2k} 
 \prod_{\ell=k+1}^{n-1} (1- \frac{a_\ell}{\lambda^2}) \bigg) 
 +\lambda^{n-2} \prod_{\ell=2}^{n-1} (1- \frac{a_\ell}{\lambda^2}) \bar \phi^{(2)}_{2} , \quad n\geq 3 
   \end{split}
 \end{equation}
 From the $n \to +\infty$ asymptotics using \eqref{bc02} we thus obtain
 \be  
 \atilde(\lambda) = -\sum_{k=2}^{\infty} \bigg( \frac{g}{2}\tilde{z}_k  \lambda^{-2k} \prod_{\ell=k+1}^{N} (1- \frac{a_\ell}{\lambda^2}) \bigg) 
 -\lambda^{-2} \prod_{\ell=2}^{N} (1- \frac{a_\ell}{\lambda^2}) \bar \phi^{(2)}_{2}
   \ee   
which, upon the replacement of the value of 
$\bar \phi^{(2)}_{1}=-\lambda$ 
and of $\bar \phi^{(2)}_{2}$ leads to

     \bea 
    \atilde(\lambda)&=&  \prod_{\ell=1}^{N} (1- \frac{a_\ell}{\lambda^2})  - \sum_{k=1}^{\infty} \bigg( \frac{g}{2}\tilde{z}_k(t=0)  \lambda^{-2k} \prod_{\ell=k+1}^{N} (1- \frac{a_\ell}{\lambda^2}) \bigg) \label{resat}  \\      
     \btilde(\lambda)&=& -\lambda^2  \label{resbt} 
        \eea

At this stage it is interesting to verify the normalization relation, by inserting \eqref{resat} and \eqref{resb}

\bea  
&& a(\lambda) \atilde(\lambda) +b(\lambda) \btilde(\lambda) =  a(\lambda) \prod_{\ell=1}^{N} (1- \frac{a_\ell}{\lambda^2})  - a(\lambda) \sum_{k=1}^{\infty} \bigg( \frac{g}{2}\tilde{z}_k(t=0)  \lambda^{-2k} \prod_{\ell=k+1}^{N} (1- \frac{a_\ell}{\lambda^2}) \bigg) \\
&& -
a(\lambda) \sum_{k=1}^{\infty}\bigg( -\frac{g}{2}\tilde{z}_k(t=0)  \lambda^{-2k} \prod_{\ell=k+1}^{N} (1- \frac{a_\ell}{\lambda^2}) \bigg) 
    + (1-a(\lambda))  \prod_{\ell=1}^{N} (1- \frac{a_\ell}{\lambda^2}) \\
    && = \prod_{\ell=1}^{N} (1- \frac{a_\ell}{\lambda^2})
\eea  
which provides a non trivial check of our calculations. 

\subsection{Scattering problem at $t=1$}

\subsubsection{B.1 For $ \phi$}

Using that $\tilde z_n(t=1)=-\Lambda \delta_{N,\ell}$ one must solve
\begin{equation} \label{ee} 
\begin{split}
& \phi^{(1)}_{n+1}=\frac{1}{\lambda}\phi^{(1)}_{n} + \frac{1}{\lambda}z_n \phi^{(2)}_{n} \\
& \phi^{(2)}_{n+1}= \frac{g\Lambda}{2\lambda }\delta_{n,N}\phi^{(1)}_{n}+ (\lambda +\frac{g\Lambda}{2\lambda }z_n \delta_{n,N} {- \frac{a_n}{\lambda}}) \phi^{(2)}_{n} 
\end{split}
\end{equation}
Let  us solve first the second equation, using the boundary condition \eqref{bc01}. with $t=1$ One finds
(recalling that $a_n=0$ for $n \geq N+1$)
\begin{equation}
\begin{split}
    \phi_n^{(2)}&=0, \quad n \leq N\\
    \phi_n^{(2)}&=\frac{g\Lambda}{2 }\lambda^{n-N-2}\phi^{(1)}_{N}, \quad n \geq N+1\\
    \end{split}
\end{equation}
Hence from \eqref{bc02} we obtain 
\begin{equation}
    b(\lambda) e^{1-\lambda^2} = \frac{g\Lambda}{2 }\lambda^{-N-2}\phi^{(1)}_{N}
\end{equation}
We can now solve the first equation in \eqref{ee} and obtain
\begin{equation}
\begin{split}
    \phi^{(1)}_n &= \lambda^{-n},\quad n\leq N+1\\
    \phi^{(1)}_{n} &= \lambda^{-n}+\sum_{k=N+1}^{n-1}  z_k(t=1) \frac{g\Lambda}{2 }\lambda^{2k-n-N-2}\phi^{(1)}_{N} ,\quad n\geq N+2\\
\end{split}
\end{equation}
From the asymptotics for $n \to +\infty$ and \eqref{bc02}, inserting $\phi^{(1)}_{N}=\lambda^{-1}$ one finds
\bea  
    a(\lambda )&=&   1+\frac{g\Lambda}{2 }\sum_{k=N+1}^{\infty}  z_k(t=1) \lambda^{2k-2N-2} \label{resanew} \\
    b(\lambda) e^{1-\lambda^2}  & = & \frac{g\Lambda}{2 }\lambda^{-2N-2} \label{resbnew} 
    \eea  
    
    \subsubsection{B.2 For $ \bar \phi$}

One must solve for $ \bar \phi$
\begin{equation} \label{ff} 
\begin{split}
& \bar \phi^{(1)}_{n+1}=\frac{1}{\lambda}\bar \phi^{(1)}_{n} + \frac{1}{\lambda}z_n \bar \phi^{(2)}_{n} \\
& \bar \phi^{(2)}_{n+1}= \frac{g\Lambda}{2\lambda }\delta_{n,N}\bar \phi^{(1)}_{n}+ (\lambda +\frac{g\Lambda}{2\lambda }z_n \delta_{n,N} {- \frac{a_n}{\lambda}}) \bar \phi^{(2)}_{n} 
\end{split}
\end{equation}
Again we start with the second equation, using the asymptotics in \eqref{bc01}. 
We find
\begin{equation}
\begin{split}
    \bar \phi^{(2)}_n &= - \lambda^n { \prod_{\ell=1}^{n-1} (1 - \frac{a_\ell}{\lambda^2})}  , \quad n\leq N\\
    \bar \phi^{(2)}_n&= \lambda^{n-N-1} (  \frac{g\Lambda}{2\lambda }\bar \phi^{(1)}_{N}- (\lambda +\frac{g\Lambda}{2\lambda }z_N {- \frac{a_N}{\lambda}} ) \lambda^N { \prod_{\ell=1}^{N-1} (1 - \frac{a_\ell}{\lambda^2})} ), \quad n\geq N+1\\
    \end{split}
\end{equation}
From the $n \to +\infty$ asymptotics and \eqref{bc02} one finds
\begin{equation} \label{aaaa} 
    \atilde(\lambda)= -\lambda^{-N-1} (  \frac{g\Lambda}{2\lambda }\bar \phi^{(1)}_{N}- (\lambda +\frac{g\Lambda}{2\lambda }z_N {- \frac{a_N}{\lambda}} ) \lambda^N { \prod_{\ell=1}^{N-1} (1 - \frac{a_\ell}{\lambda^2})} )
\end{equation}
The solution for $\bar \phi^{(2)}_n$ can then be inserted in the first equation in \eqref{ff}. It reads
\begin{equation}
    \begin{split}
         \bar \phi^{(1)}_{n+1}&=\frac{1}{\lambda}\bar \phi^{(1)}_{n} - \lambda^{n-1}z_n { \prod_{\ell=1}^{n-1} (1 - \frac{a_\ell}{\lambda^2})} , \quad n\leq N\\
        \bar \phi^{(1)}_{n+1}&=\frac{1}{\lambda}\bar \phi^{(1)}_{n} + z_n \lambda^{n-N-2} (  \frac{g\Lambda}{2\lambda }\bar \phi^{(1)}_{N}- (\lambda +\frac{g\Lambda}{2\lambda }z_N {- \frac{a_N}{\lambda}} ) \lambda^N { \prod_{\ell=1}^{N-1} (1 - \frac{a_\ell}{\lambda^2})} ) , \quad n\geq N+1\\
    \end{split}
\end{equation}
Taking into account the boundary condition \eqref{bc01}, its solution is found as
\begin{equation}
    \begin{split}
        \bar \phi^{(1)}_{n} &=- \sum_{k=-\infty}^{n-1} z_k \lambda^{2k-n} { \prod_{\ell=1}^{k-1} (1 - \frac{a_\ell}{\lambda^2})} , \quad n \leq N+1\\
        \bar \phi^{(1)}_{n} &=\lambda^{-n+N+1}\bar \phi^{(1)}_{N+1}  + (  \frac{g\Lambda}{2\lambda }\bar \phi^{(1)}_{N}- (\lambda +\frac{g\Lambda}{2\lambda }z_N {- \frac{a_N}{\lambda}}) \lambda^N { \prod_{\ell=1}^{N-1} (1 - \frac{a_\ell}{\lambda^2})} )\sum_{k=N+1}^{n-1} z_k \lambda^{2k-n-N-1} , \quad n \geq N+2
    \end{split}
\end{equation}
From the asymptotics at $n \to +\infty$ and \eqref{bc02} one finds
\begin{equation} \label{bbbb} 
    \btilde(\lambda) e^{\lambda^2-1} = \lambda^{N+1}\bar \phi^{(1)}_{N+1}  + (  \frac{g\Lambda}{2\lambda }\bar \phi^{(1)}_{N}- (\lambda +\frac{g\Lambda}{2\lambda }z_N {- \frac{a_N}{\lambda}}) \lambda^N { \prod_{\ell=1}^{N-1} (1 - \frac{a_\ell}{\lambda^2})} )\sum_{k=N+1}^{\infty} z_k \lambda^{2k-N-1} 
\end{equation}
Inserting the value of $\bar \phi^{(1)}_{N}$ and of $\bar \phi^{(1)}_{N+1}$  in \eqref{aaaa} and 
in \eqref{bbbb} we obtain
\bea
        \atilde(\lambda) &=& { \prod_{\ell=1}^{N} (1 - \frac{a_\ell}{\lambda^2})} +\frac{g\Lambda}{2 }\sum_{k=-\infty}^{N} \bigg( z_k(t=1) \lambda^{2k-2N-2} { \prod_{\ell=1}^{k-1} (1 - \frac{a_\ell}{\lambda^2})} \bigg) \label{resatnew}  \\
          \btilde(\lambda)  e^{\lambda^2-1} &=& - \sum_{k=-\infty}^{N} z_k(t=1) \lambda^{2k} { \prod_{\ell=1}^{k-1} (1 - \frac{a_\ell}{\lambda^2})}   -    \atilde(\lambda)  \sum_{k=N+1}^{\infty} z_k(t=1) \lambda^{2k} \label{resbtnew} 
    \eea

\subsection{Summary: result for the scattering amplitudes}

In summary, the scattering amplitudes $b(\lambda)$ and $\btilde(\lambda)$ have been completely
determined. They read
\begin{equation}
\begin{split}
    \btilde(\lambda)&=-\lambda^2\\
    b(\lambda) & = \frac{g\Lambda}{2}\lambda^{-2N-2}e^{\lambda^2 -1} 
        \end{split}
\end{equation}
By contrast, for each of the amplitudes $a(\lambda)$ and $\atilde(\lambda)$ we
has obtained only two relations to the two unknown set of variables
$z_n(t=1)$ and $\tilde z_n(t=0)$. These relations read
\begin{equation}
\begin{split}
    a(\lambda)&= 1+\frac{g\Lambda}{2 }\sum_{n=N+1}^{\infty}  z_n(t=1) \lambda^{2n-2N-2}\\
    &= 1- \frac{g}{2}\sum_{n=-\infty}^{0}\tilde{z}_n(t=0)  \lambda^{-2n}
\end{split}
\end{equation}
A priori we have obtained each relation (each line) separately. However one can 
check that thanks to the symmetry 
\be  \label{sym0} 
\{ \tilde z_\ell(t) = - \Lambda z_{N-\ell+1}(1-t)  \quad , \quad a_\ell = a_{N-\ell+1} \}
\ee    
these two expressions are identical. Similarly we 
have found
\be    
\begin{split}
\atilde(\lambda) &= { \prod_{\ell=1}^{N} (1 - \frac{a_\ell}{\lambda^2})} +\frac{g\Lambda}{2 }\sum_{k=-\infty}^{N} \bigg( z_k(t=1) \lambda^{2k-2N-2} { \prod_{\ell=1}^{k-1} (1 - \frac{a_\ell}{\lambda^2})} \bigg)\\   
&=  \prod_{\ell=1}^{N} (1- \frac{a_\ell}{\lambda^2})  - \sum_{k=1}^{\infty} \bigg( \frac{g}{2}\tilde{z}_k(t=0)  \lambda^{-2k} \prod_{\ell=k+1}^{N} (1- \frac{a_\ell}{\lambda^2}) \bigg)
\end{split}
\ee   
and one can check that the two lines are again identical thanks
to the symmetry \eqref{sym0}. \\

Although the expressions for $a(\lambda)$ and $\atilde(\lambda)$ seem complicated
the product $a(\lambda) \atilde(\lambda)$ has a simple expression, thanks to
the normalization relation (which we have checked to hold) 
 \be 
    a(\lambda) \atilde(\lambda)= 1- b(\lambda) \btilde(\lambda) 
    = \prod_{\ell=1}^{N} (1 - \frac{a_\ell}{\lambda^2}) + \frac{g\Lambda}{2}\lambda^{-2N}e^{\lambda^2 -1} 
 \ee 

This equation is amenable to a solution using scalar Riemann-Hilbert methods.

\section{Solution of the Riemann-Hilbert problem for the scattering amplitudes}

Here we restrict to the case with zero drift $a_\ell=0$ and we set the value of the coupling constant $g=2$. We need to solve 
\be 
    a(\lambda) \atilde(\lambda)= G(\lambda) 
    = 1 + \Lambda \lambda^{-2N} e^{\lambda^2 -1} 
 \ee 

Let us first study the function $G(\lambda)$ (which is a function of $\lambda^2$). It has an infinity of zeroes in the complex plane
(they are represented by red dots in the Figures~\ref{fig:completplot0p5}, \ref{fig:completplot-0p5}, \ref{fig:completplot-5} and \ref{fig:completplot100000}). To obtain their analytical expressions one writes 
\be
\lambda^{-2N} e^{\lambda^2} = - \frac{e}{\Lambda} 
\ee 
Taking the power $1/N$ and multiplying by $N$ we obtain
\be 
\frac{N}{\lambda^2} e^{\frac{\lambda^2}{N} } = N \left(- \frac{e}{\Lambda}\right)^{1/N} e^{- 2 \I \pi \frac{n}{N}}
\ee
with $n=0,\dots,N-1$ and for any $z=|z| e^{i \theta}$, we denote $z^{1/N}=|z|^{1/N} e^{\I \theta/N}$.\\

Now, since the solution of $e^x/x=y$ is $x=- W_k(-1/y)$, where $W_k$ is any of the branches of 
the Lambert function with $k \in \mathbb{Z}$ \cite{corless1996lambertw}, we obtain that the zeroes of $G(\lambda)$ can be written as $\pm \lambda_{k,n}$, where
\be \label{zeroesformula}
\lambda_{k,n}^2 = - N W_k( - e^{-1} (\frac{\Lambda}{\Lambda_c})^{1/N} e^{2 \I \pi \frac{n}{N}} ) , \quad k \in \mathbb{Z}, \quad  n=0,\dots,N-1
\ee 
We have further defined
\be 
\Lambda_c = - e^{1-N} N^N < 0 \, .
\ee 
Below we will always denote $\lambda_k=\lambda_{k,0}$. Finally, let us note that
$G(\lambda)$ has a pole of order $2N$ at the origin $\lambda=0$. 
\\ 

\begin{figure}[t!]
    \centering
\includegraphics[scale=0.7]{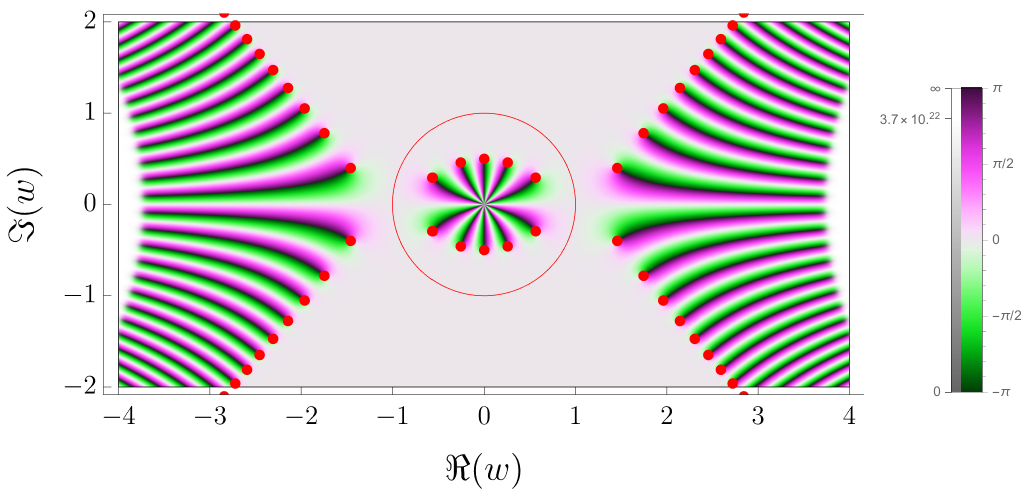}
    \caption{Representation {of $\arg(G)$, i.e. the phase (shown by a color code)} 
    of the rescaled function  $w\mapsto 1+\frac{\Lambda}{|\Lambda_c|} w^{-2N}e^{N(w^2-1)}=G(\lambda=w \sqrt{N})$ for $N=5$ and 
    $\frac{\Lambda}{|\Lambda_c|}=0.5$ in the complex plane. The plot was made using the \texttt{ComplexPlot} function of Mathematica with the \texttt{GreenPinkTones} color function. 
    The red dots represent the zeroes of the function $G$ and we have represented the circle of radius $|w|=1$ which
    corresponds to contour ${\cal C}$ (equal to the circle $|\lambda|=\mathtt{R}=\sqrt{N}$ mentionned in the text. The black lines
    can be interpreted as the positions of branch cuts in $\log G $ and we see that the contour ${\cal C}$ does not cross them.}
    \label{fig:completplot0p5}
\end{figure}

\begin{figure}[t!]
    \centering
\includegraphics[scale=0.7]{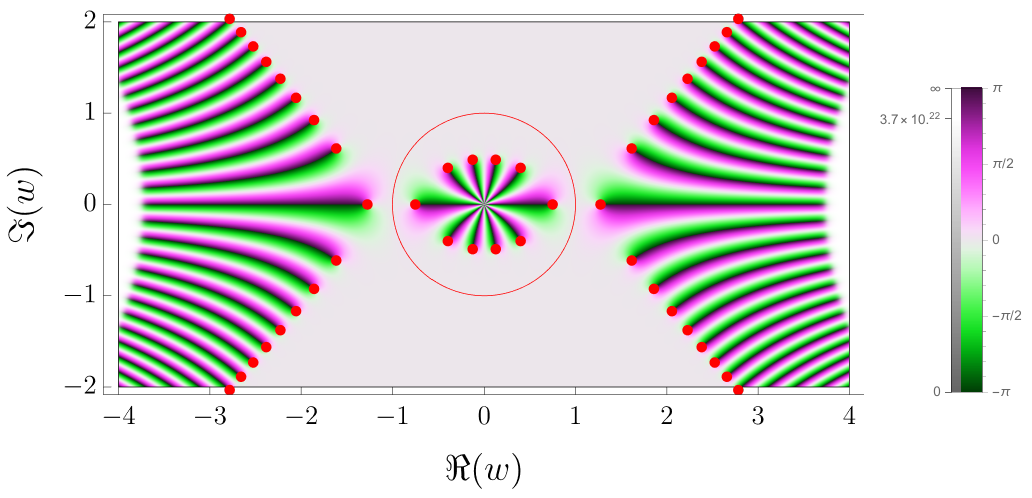}
    \caption{{Same as Fig.~\ref{fig:completplot0p5} with $\frac{\Lambda}{|\Lambda_c|}=-0.5$. Note the
    four zeroes on the real axis which correspond to (in the order of increasing real parts)
    to $\{ - \lambda_{-1},-\lambda_0,\lambda_0, \lambda_{-1} \}$. This feature happens
    for any $\frac{\Lambda}{|\Lambda_c|} \in ]0,1[$. These zeroes are related to
    the soliton rapidities, for the solution where solitons are present, see discussion in the text.
    As $\frac{\Lambda}{|\Lambda_c|} \to -1^+$ these zeroes get closer and merge pairwise.
    For $\frac{\Lambda}{|\Lambda_c|} < -1$ the zeroes go along a curve in the plane, see Fig.~\ref{fig:completplot-5}.}
    }
    \label{fig:completplot-0p5}
\end{figure}

\begin{figure}[t!]
    \centering
\includegraphics[scale=0.7]{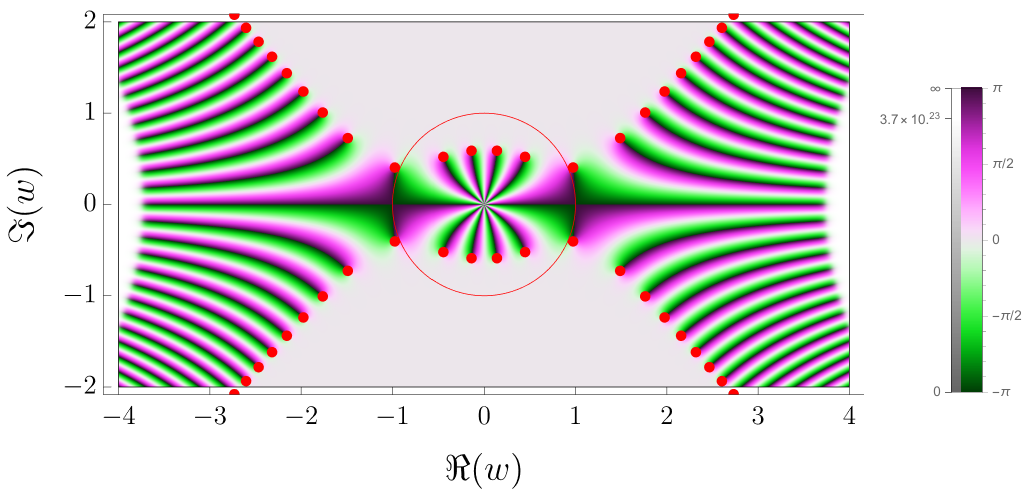}
    \caption{Same as Fig.~\ref{fig:completplot0p5} with $\frac{\Lambda}{|\Lambda_c|}=-5$. The horizontal branch cuts originating from the origin and from infinity have merged at $\frac{\Lambda}{|\Lambda_c|}=-1$ and expand in lower and upper half planes. We will not study in details 
    the case $\frac{\Lambda}{|\Lambda_c|}<-1$ here since we do not need it for our large deviation problem.
    }
    \label{fig:completplot-5}
\end{figure}

\begin{figure}[t!]
    \centering
\includegraphics[scale=0.7]{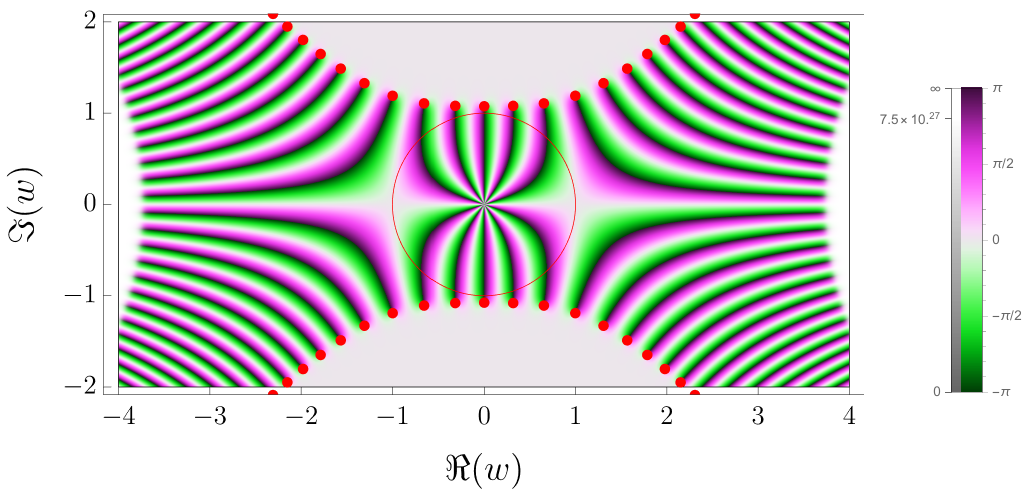}
    \caption{Same as Fig.~\ref{fig:completplot0p5} with $\frac{\Lambda}{|\Lambda_c|}=100000$. As discussed in the text, in the special
    case of large positive $\Lambda$, one needs to deform the circular contour (in an obvious way) to avoid the branch cuts in $\log G$.
    }
    \label{fig:completplot100000}
\end{figure}

\begin{figure}[t!]
    \centering
\includegraphics[scale=0.7]{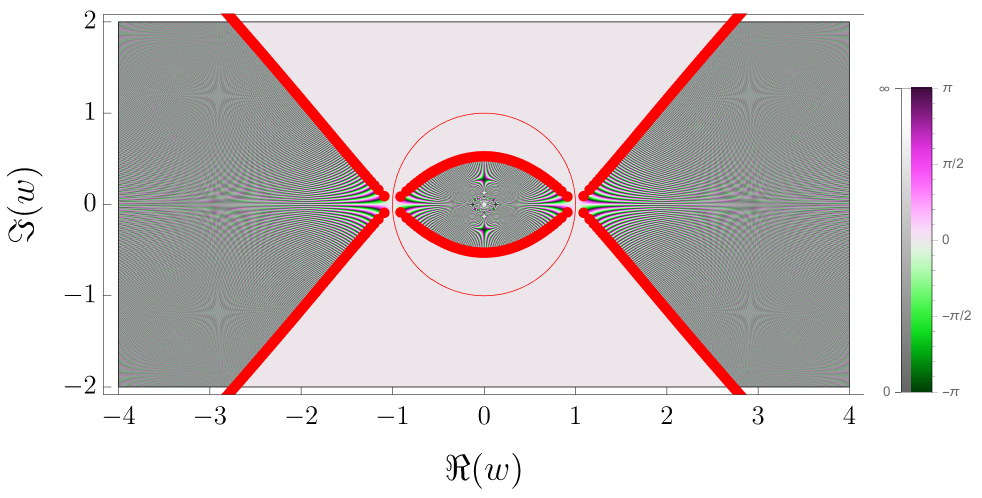}
    \caption{{Same as Fig.~\ref{fig:completplot0p5} with $\frac{\Lambda}{|\Lambda_c|}=1$ and $N=100$. }
    }
    \label{fig:completplot1N100}
\end{figure}

\begin{figure}[t!]
    \centering
\includegraphics[scale=0.66]{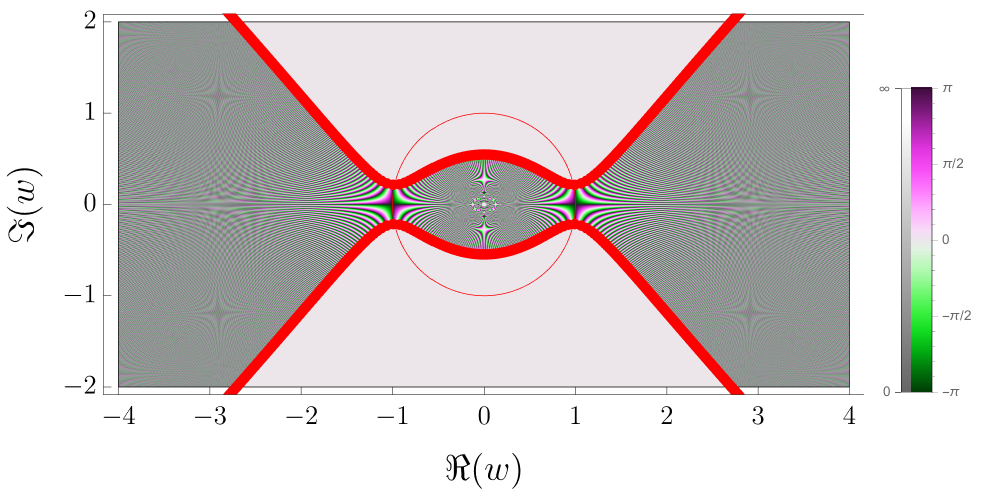}
    \caption{{Same as Fig.~\ref{fig:completplot0p5} with $\frac{\Lambda}{|\Lambda_c|}=-10000$ and $N=100$. }
    }
    \label{fig:completplot-10000N100}
\end{figure}

Now recall that we expect $a(\lambda)$ to be analytic inside a contour ${\cal C}$ (we will
call ${\cal D}$ the interior of ${\cal C}$) and  $\atilde(\lambda)$ to be analytic outside the contour 
${\cal C}$ in the complementary domain ${\cal D}^c$. In practice, in most cases
one can take the contour ${\cal C}$ to be a circle of radius $\mathtt{R}=\sqrt{N}$,
in which case $a(\lambda)$ is analytic for for $|\lambda|<\mathtt{R}$
and $\atilde(\lambda)$ is analytic for $|\lambda|>\mathtt{R}$.
In some special case however (see below) we will need to deform the circle.\\

From the knowledge of the zeroes of $G(\lambda)$, one can then determine the solution for $a(\lambda)$ and $\atilde(\lambda)$. There are two types of solutions:

{\bf Solution without soliton}. 
Let us first assume that $a(\lambda)$ has no zeroes for $|\lambda|<\mathtt{R}$
and that $\atilde(\lambda)$ has no zeroes for $|\lambda|>\mathtt{R}$, or more generally for $\lambda \in {\cal D}$
and $\lambda \in {\cal D}^c$ respectively.
From Cauchy's theorem, and taking into account that both functions are even functions of $\lambda$
one has, for $\lambda \in {\cal D}$
\begin{equation}
    \begin{split}
        \log {a}(\lambda) =&\oint_{{\cal C}}\frac{\rmd w}{2\I \pi}\frac{w}{w^2-\lambda^2}\log {a}(w)\\
        0=&\oint_{{\cal C}}\frac{\rmd w}{2\I \pi}\frac{w}{w^2-\lambda^2}\log \atilde(w), \quad \lambda \in {\cal D}
    \end{split}
\end{equation}
where in the second equality we have closed the contour at infinity (assuming that $\atilde(\lambda)$
goes to unity for $|\lambda| \to +\infty$). Similarly one has, for $\lambda \in {\cal D}^c$
\begin{equation}
    \begin{split}
        \log \tilde {a}(\lambda) =&-\oint_{{\cal C}}\frac{\rmd w}{2\I \pi}\frac{w}{w^2-\lambda^2}\log \tilde {a}(w)\\
        0=&\oint_{{\cal C}}\frac{\rmd w}{2\I \pi}\frac{w}{w^2-\lambda^2}\log {a}(w), \quad \lambda \in {\cal D}^c 
    \end{split}
\end{equation}

Subtracting these equations, we find 
\begin{equation}
    \begin{split}
         \log \atilde(\lambda) =&- \oint_{{\cal C}}\frac{\rmd w}{2\I \pi}\frac{w}{w^2-\lambda^2}\log a(w)\atilde(w) =  - \varphi(\lambda) \quad , \quad \lambda \in {\cal D}^c 
         \\
 \log {a}(\lambda) =& \, \varphi(\lambda) 
         \quad , \quad \lambda \in {\cal D} 
    \end{split}
\end{equation}
where everywhere we denote
\be  \label{phidef0} 
\varphi(\lambda) = \oint_{{\cal C}} \frac{\rmd w}{2\I \pi}\frac{w}{w^2-\lambda^2}\log (1 + \Lambda w^{-2N}e^{w^2 -1}) \quad , \quad 
\lambda \notin  {\cal C} \, .
\ee

Note that since the number of zeroes of $G(\lambda)$ is infinite, 
$\atilde(\lambda)$ has zeroes inside the contour ${\cal C}$ (i.e. in ${\cal D}$), and
$a(\lambda)$ has an infinite number of zeroes outside the contour ${\cal C}$ (i.e. in ${\cal D}^c$). 
The positions of these zeroes are shown as red dots in the Figures~\ref{fig:completplot0p5}, \ref{fig:completplot-0p5}, \ref{fig:completplot-5}, \ref{fig:completplot100000} for $N=5$.
These zeroes lead to branch cuts in the 
function $\log G(\lambda)$ as can be seen on these figures. We have checked
that for $\frac{\Lambda}{\Lambda_c} \in (-1,+\infty[$, which is the region
that we need for our large deviation problem, 
there is a choice of the contour ${\cal C}$ so that no branch cut is
crossed when integrating $\log G(\lambda)$ over the
contour ${\cal C}$. There exists a threshold value $\Lambda^*(N)$
such that for $\Lambda < \Lambda^*(N)$ the contour ${\cal C}$ can be
chosen to be the circle of radius $\mathtt{R}=\sqrt{N}$ (represented
as a circle of radius unity in the rescaled variables in the figures). 
For $\Lambda > \Lambda^*(N)$ the contour can be chosen as a deformed circle
as can be seen from Fig.~\ref{fig:completplot100000}. The value 
$\Lambda^*(N)$ increases very fast with $N$ and corresponds to
the first time that a zero initially inside the circle hits the circle
as $\Lambda$ increases
(we did not attempt to find its analytical value, except for
$N=1$ where one has $\Lambda^*(1)=e^2$.\\

Finally, we also plotted the zeroes for $N=100$ in Figs.~\ref{fig:completplot1N100} and \ref{fig:completplot-10000N100}.
One can see that they fall on some limit curves that we did not attempt to describe or interpret. \\

{\bf Solution with a soliton}. From the considerations discussed in the text, we know that there should be
another branch of solutions with solitons. Indeed for $\Lambda \in [\Lambda_c,0)$
there are two pairs of real zeroes of $G(\lambda)$, 
$\{\pm \lambda_{0}, \pm \lambda_{-1} \}$ (we recall that
we denote $\lambda_k=\lambda_{k,0}$) as can be seen in Fig.~\ref{fig:completplot-0p5}. Furthermore one has 
$0<\lambda_{0} < \mathtt{R}=\sqrt{N}$ and 
$\mathtt{R}=\sqrt{N} < \lambda_{-1}$. Thus in the interval $\Lambda \in [\Lambda_c,0[$
one can consider a solution such that 
$a(\lambda)$ has two zeroes at $\lambda = \pm \lambda_{0}$ 
inside the $\mathtt{R}$-circle,
and $\atilde(\lambda)$ has two zeroes at $\lambda = \pm \lambda_{1}$,
outside the $\mathtt{R}$-circle. It is then natural to redefine
    \be
    a(\lambda)= \alpha(\lambda)\frac{\lambda^2-\lambda_0^2}{\lambda^2-\lambda_{-1}^2} \quad , \quad  \atilde(\lambda)=\tilde{\alpha}(\lambda)\frac{\lambda^2-\lambda_{-1}^2}{\lambda^2-\lambda_0^2}
   \ee
Since the product $a(\lambda) \atilde(\lambda)= \alpha(\lambda) \tilde \alpha(\lambda)=G(\lambda)$
and $\tilde \alpha(\lambda)$ also tend to one at infinity, we can apply the same
manipulations as above to $\alpha(\lambda)$ and $\tilde \alpha(\lambda)$. The solution thus reads
\bea \label{finala} 
a(\lambda) = \frac{\lambda^2-\lambda_0^2}{\lambda^2-\lambda_{-1}^2} \times
\begin{cases}   e^{\varphi(\lambda)} \quad &|\lambda| < \mathtt{R} = \sqrt{N} \\
 G(\lambda)^{1/2} e^{\tilde \varphi(\lambda)}  
\quad &|\lambda| = \mathtt{R} \\
 G(\lambda) e^{\varphi(\lambda)} 
\quad &|\lambda| > \mathtt{R} 
\end{cases} 
\eea 
\bea \label{finalat} 
\atilde(\lambda) = \frac{\lambda^2-\lambda_{-1}^2}{\lambda^2-\lambda_0^2} \times \begin{cases}   G(\lambda) e^{- \varphi(\lambda)} \quad &|\lambda| < \mathtt{R} = \sqrt{N} \\
 G(\lambda)^{1/2} e^{- \tilde \varphi(\lambda)}  
\quad &|\lambda| = \mathtt{R} \\
  e^{- \varphi(\lambda)} 
\quad &|\lambda| > \mathtt{R} 
\end{cases} 
\eea 
where $\varphi(\lambda)$ was defined in \eqref{phidef0} and 
\be 
\tilde \varphi(\lambda) = \dashint_{\abs{w}=\mathtt{R}}\frac{\rmd w}{2\I \pi}\frac{w}{w^2-\lambda^2}\log (1 + \Lambda w^{-2N}e^{w^2 -1}) 
\quad 
|\lambda| = \mathtt{R} 
\ee 
is given as a principal value. The discussion above about the contour and branch cut applies identically.
This is quite analogous to the situation for the KPZ equation (which as we show is
reached for $N \to +\infty$) as found in \cite{krajenbrink2021inverse} and confirmed by the rigorous recent work of Ref.~\cite{TsaiIntegrability}.

\section{Calculation of the rate functions}

\subsection{Values of the conserved quantities} 

From Section \ref{sec:conserved-quantities} we can now obtain the values taken by
the conserved quantities $\{ \tilde C_n, C_n \}$ from the coefficients of the Laurent or Taylor series of
the scattering amplitudes. One writes
\bea 
&& \log \atilde(\lambda) = - \varphi(\lambda) = \sum_{n=1}^\infty \frac{\tilde{C}_n}{\lambda^{2n}} \quad , \quad |\lambda| > \mathtt{R} \\
&& \log a(\lambda) =  \varphi(\lambda) = \sum_{n=0}^\infty \lambda^{2n}C_n   \quad , \quad |\lambda| < \mathtt{R}  
\eea 
Recalling the definition \eqref{phidef0} of $\varphi(\lambda)$,
this leads to 
\bea   
&&     C_n  = \oint_{\abs{w}=\mathtt{R}}\frac{\rmd w}{2\I \pi}\frac{1}{w^{2n+1}}\log (1+ \Lambda w^{-2N}e^{w^2 -1}), \quad n\in [0,\infty]\\
&&     \tilde{C}_n = \oint_{\abs{w}=\mathtt{R}}\frac{\rmd w}{2\I \pi}w^{2n-1}\log (1+ \Lambda w^{-2N}e^{w^2 -1}), \quad n\in [1,\infty]
\label{soluCtilde} 
\eea  
This is the result in the absence of soliton. In the presence of soliton there is an additional additive contribution
so that the values of the conserved quantities are now $C_n + \Delta C_n$ and $\tilde C_n + \Delta \tilde C_n$
respectively, where $C_n,\tilde C_n$ are still given by the above expressions and, from \eqref{finalat} 
for $|\lambda| > \mathtt{R}$ 
\bea 
 \log(\frac{\lambda^2-\lambda_{-1}^2}{\lambda^2-\lambda_0^2}) =  \sum_{n=1}^\infty \frac{1}{\lambda^{2n}}\frac{\lambda_0^{2n}-\lambda_{-1}^{2n}}{n}
 = \sum_{n=1}^\infty \frac{\Delta \tilde{C}_n}{\lambda^{2n}}
\eea 
Similarly from \eqref{finala} 
for $|\lambda| < \mathtt{R}$ one obtains 
\bea 
 \log(\frac{\lambda^2-\lambda_{0}^2}{\lambda^2-\lambda_{-1}^2}) =  \log(\frac{\lambda_{0}^2}{\lambda_{-1}^2})
+  \sum_{n=1}^\infty \lambda^{2n}\frac{\lambda_{-1}^{-2n}-\lambda_0^{-2n}}{n}
 = \Delta C_0+ \sum_{n=1}^\infty \lambda^{2 n} \Delta C_n 
\eea  
This leads to 
\be 
\label{change} 
\begin{split}
 \Delta C_0 &= \log(\lambda_{0}^2)-\log(\lambda_{-1}^2), \\
 \Delta C_n &= \frac{\lambda_{-1}^{-2n}-\lambda_0^{-2n}}{n} \quad \text{for} \quad n \geq 1 \, , \\
 \Delta \tilde{C}_n &= \frac{\lambda_0^{2n}-\lambda_{-1}^{2n}}{n} \quad \hspace{0.4cm} \text{for} \quad n \geq 1 \, . 
\end{split}
\ee   

\subsection{Rate function $\Psi(\Lambda)$ : main branch, no soliton} 

Let us now recall the first conserved quantity from Section~\ref{sec:conserved-quantities}.
With the value of the coupling constant set to $g=2$, it reads
\be   
\tilde C_1 = - \frac{g}{2} \sum_{n=-\infty}^{+\infty} z_n(t) \tilde z_n(t)
\ee  
and it is independent of time $t$. Let us set and evaluate it at time $t=1$ using the
boundary condition $z_n(1)=- \Lambda \delta_{n,N}$. We obtain the relation
\be   
\tilde C_1 =  \Lambda z_N(t=1) 
\ee 
By definition $z_N(t=1)=e^{H_N}$ is our observable and by taking a derivative of Eq.~\eqref{Legendre0} w.r.t. the Legendre parameter $\Lambda$ we see that 
\be 
\Psi'(\Lambda) = z_N(t=1)  
\ee 
Consider here the case without soliton. 
Hence, using \eqref{soluCtilde} for $n=1$ we obtain
\be \label{ll} 
 \Lambda \Psi'(\Lambda) = \oint_{\abs{w}=\mathtt{R}} \frac{\rmd w}{2\I \pi} w \log (1+ \Lambda w^{-2N}e^{w^2 -1})
= \oint_{\abs{v}=\mathtt{R}^2}\frac{\rmd v}{2\I \pi}  \log (1+ \Lambda v^{-N}e^{v -1})
\ee   
where in the second equality there is no additional factor of $2$ since the circle is run twice. 
We recall that we can choose $\mathtt{R}=\sqrt{N}$.

\begin{remark} For $N=1$ we know from the direct solution, see e.g. \eqref{ZZ}, that 
    \begin{equation}
    \Lambda \Psi'(\Lambda)  = W_0(\frac{\Lambda}{e})     
    \end{equation}
   Hence this provides an integral representation of the Lambert function (which to our knowledge is novel).
   \end{remark}

Since $\Psi(0)=0$, we can integrate \eqref{ll} and obtain
\be 
\begin{split}
  \Psi(\Lambda) &= -\oint_{\abs{w}=\mathtt{R} }\frac{\rmd w}{2\I \pi} w \mathrm{Li}_2(- \Lambda w^{-2N}e^{w^2 -1})
  = -\oint_{\abs{v}=\mathtt{R}^2}\frac{\rmd v}{2\I \pi}  \mathrm{Li}_2(- \Lambda v^{-N}e^{v -1})\\
\end{split}
\ee   
\\

{\bf Series expansion of $\Psi_N(\Lambda)$ and cumulants of $z_N(t=1)$}. Recall the contour representation of the $n$-th Hermite polynomial
\be  
H_n(x) = n!   \oint \frac{\rmd z}{2 \I \pi} \frac{e^{2 x z-z^2}}{z^{n+1}} 
\ee   
where the contour encloses the origin.
Expanding the logarithm in Eq.~\eqref{ll} as a series, this leads to
\bea    
   \Lambda \Psi'(\Lambda) && =  \sum_{p \geq 1}  \frac{(-1)^{p+1}}{p} \oint_{\abs{w}=\mathtt{R}}\frac{\rmd w}{2\I \pi w}  
\Lambda^p e^{-p} 
w^{-2 N p+ 2}e^{pw^2 } \\
&& = \sum_{p \geq 1}  \frac{(-1)^{p+1}}{p}  
(\Lambda e^{-1})^p (-p)^{N p-1} 
 \frac{ H_{2 N p-2}(0) }{(2 N p-2)!} \\
   && = \sum_{p \geq 1}  \frac{(-1)^{p+1}}{p^2}  
(\Lambda e^{-1} p^N )^p 
 \frac{1} {(N p-1)!} 
\eea    
where we have used the value of the Hermite polynomial of even indices at the origin
\be   
\frac{H_{2N}(0)}{(2 N)!} = \frac{\cos(\pi N)}{N!} 
\ee 
Integrating once w.r.t $\lambda$ we obtain 
\be 
\Psi_N(\Lambda)  = \sum_{p \geq 1} b_p (\Lambda e^{-1})^p \quad , \quad b_p =(-1)^{p+1} \frac{p^{N p-3}}{(N p -1)!} 
\ee 
One finds that the radius of convergence of this series in $\Lambda$ is precisely $|\Lambda_c|$ 
defined above $\Lambda_c = -e^{1-N}N^N$, which is consistent with the behavior of the
solutions. \\

Let us note that the typical value 
is given by the coefficient for $p=1$ i.e. one has
\be  
\Psi_N'(0) = e^{H_{\rm typ}} = \overline{ z_N(t=1) }  = b_1 e^{-1} = \frac{1}{e (N-1)!} 
\ee   
which coincides with the prediction from the Poisson kernel. More generally, since $\Psi_N(\Lambda)$ is the generating function of the
cumulants of $z_N(t=1)$, from its definition \eqref{gener0}, one has, to leading order in $\varepsilon\ll 1$
\be 
\Psi_N(\Lambda) =  \sum_{p \geq 1} (-1)^{p+1} \frac{\Lambda^p }{p!} \varepsilon^{1-p} \overline{ z_N(t=1)^p }^c 
\ee 
which implies that 
\be 
 \overline{ z_N(t=1)^p }^c = (-1)^{p+1} p! \varepsilon^{p-1} b_p e^{-p} = 
p!  \frac{p^{N p-3}}{(N p -1)!} e^{-p} \,  \varepsilon^{p-1} 
\ee 
In particular the variance reads
\be 
\overline{ z_N(t=1)^2}^c =   \frac{2^{2 N -2}}{(2 N  -1)!} e^{-2} \,  \varepsilon  
\ee 
One can easily check the first two cumulants for $N=1$. 
Let us recall that $z_N(t)=e^{- (1 + \frac{\varepsilon}{2})t} Z_N(t)$. For $N=1$ one
has $z_1(t)=e^{- (1 + \frac{\varepsilon}{2})t + \sqrt{\varepsilon} B(t)} $
and one finds $\overline{ z_1 } \simeq e^{-1}$
and $\overline{ z_1^2 }^c = e^{-2} \varepsilon$ in agreement with the above formula.

Finally, one can obtain the cumulants of $H=\log z_N(1)$ from the derivatives of the rate function $\Phi(H)$ 
(see e.g. in \cite[Sec.~4.2.5 of the Supp. Mat.]{krajenbrink2017exact}).
Here they scale as $\overline{H^q}^c \sim \varepsilon^{1-q}$. The variance reads 
\be 
\overline{H^2}^c =  \frac{\varepsilon}{\Phi''(H_{\rm typ})} = - \varepsilon
\frac{ \Psi''(0)}{\Psi'(0)^2} = \frac{2^{2 N-2} (N-1)!^2}{(2 N-1)!} \, \varepsilon
\ee

\subsection{Rate function $\Psi(\Lambda)$ : second branch, with a soliton} 
\label{subsec:supp-mat-continuation}
In presence of a soliton, from \eqref{change} we see that the value of the conserved quantity $\tilde C_1$
is changed by
\be 
\Delta \tilde{C}_1 = \lambda_0^{2}-\lambda_{-1}^{2} 
\ee 
On the second (solitonic) branch, the rate function $\Psi$ thus becomes 
$\Psi_N(\Lambda) = \Psi_N^{\rm main}(\Lambda) + \Delta_N(\Lambda)$
with 
\be \label{soludeltaprime} 
\Lambda \Delta_N'(\Lambda) = \Delta \tilde{C}_1 = N W_{-1}(-e^{-1} (\frac{\Lambda}{\Lambda_c})^{1/N} ) - 
N W_{0}(-e^{-1} (\frac{\Lambda}{\Lambda_c})^{1/N} )
\ee 
where we have replaced $\lambda_{0/-1}=\lambda_{0/-1,0}$ by 
their explicit values from \eqref{zeroesformula}. 
Upon integration one finds that 
\begin{equation}  \label{soludeltaSM} 
    \Delta_N(\Lambda) = \frac{N^2 }{2} W_{-1}(-\frac{1}{e}(\frac{\Lambda}{\Lambda_c})^{1/N}) \left(W_{-1}(-\frac{1}{e}(\frac{\Lambda}{\Lambda_c})^{1/N})+2\right)- \frac{N^2 }{2}  W_{0}(-\frac{1}{e}(\frac{\Lambda}{\Lambda_c})^{1/N}) \left(W_{0}(-\frac{1}{e}(\frac{\Lambda}{\Lambda_c})^{1/N})+2\right)
\end{equation}
which vanishes as it should at $\Lambda=\Lambda_c$ which is the point where solitons are spontaneously generated.

\section{Convergence to KPZ}

\subsection{Convergence of the large deviation rate function}
We will now show that the large deviation result for the OY polymer
\be \label{eqeq} 
\overline{ \exp \left(-\frac{\Lambda}{\varepsilon} z_N(\tau=1) \right) } \sim 
 \exp\left(  -\frac{1}{\varepsilon} \Psi_N(\Lambda) \right) 
\ee 
converges to the similar result for the short time KPZ equation as $N \to +\infty$ under
a proper rescaling. We will consider successively the right hand side of \eqref{eqeq} (the large deviation form)
and its left hand side (the observable) and show that each side converges to the corresponding
one for the KPZ equation.\\

Consider the r.h.s. of \eqref{eqeq}. Let us consider first the main branch. One recalls that
\be  \label{psipsi} 
 \Psi(\Lambda) = -\oint_{\abs{v}=N}\frac{\rmd v}{2\I \pi}  \mathrm{Li}_2(- \Lambda v^{-N}e^{v -1})\\
\ee 
Note that since $\Lambda^*_N$ grows very fast with $N$ we need only to
consider the case $\Lambda < \Lambda^*_N$ where the contour is a circle.
We parametrize the circle of radius $N$ as $v=N e^{\I k}$, and for
large $N$ we expand around $k=0$ as
\be  
v- N \log v = N - N \log N - N \frac{k^2}{2} + \mathcal{O}(N k^3) 
\ee   
Inserting this expansion into \eqref{psipsi} we obtain 
\bea  
    \Psi(\Lambda)
&&  \simeq - N 
\int_{\mathbb{R}} \frac{\rmd k}{2 \pi}  e^{\I k} \mathrm{Li}_2(- \Lambda  e^{N -1 - N \log N} e^{- N k^2/2}) \\
&& \simeq - \sqrt{2 N } 
\int_{\mathbb{R}} \frac{\rmd q}{2 \pi}   \mathrm{Li}_2(\frac{\Lambda}{\Lambda_c}  e^{- q^2})
\eea  
where we recall that $\Lambda_c = - e^{1-N} N^N$. We can then make the connection on the right hand
side of the equation \eqref{eqeq} with the known result for the short time KPZ 
equation with droplet initial conditions \cite{le2016exact,krajenbrink2021inverse} as
\be  
\exp\left(- \frac{\Psi_N(\Lambda)}{\varepsilon}  \right) \to \exp\left(- \frac{1}{\sqrt{T_{\rm KPZ}} }\Psi_{\rm KPZ}(z)\right)
\quad , \quad 
\Psi_{\rm KPZ}(z) = 
- \int_{\mathbb{R}} \frac{\rmd k}{2 \pi}   \mathrm{Li}_2(- z e^{- q^2})
\ee  
with the identification
\be   
z = - \frac{\Lambda}{\Lambda_c}    \quad , \quad \frac{\sqrt{2 N}}{\varepsilon} = \frac{1}{\sqrt{T_{\rm KPZ}}}
\ee  
Hence the "KPZ time" $T_{\rm KPZ}$ must be identified with the weak noise parameter $\varepsilon^2/(2 N)$.\\ 

In the case of the second (solitonic) branch we need to study the large $N$ limit of $\Delta_N(\Lambda)$
given by Eq.~\eqref{soludeltaSM}. We recall the expansions of the Lambert functions for $0<x<1$ and $N \to +\infty$
\bea  
&& W_0(- e^{-1} x^{1/N}) = -1 + \sqrt{\frac{2}{N}} \sqrt{\log(1/x)} + \frac{2}{3 N} \log x + \mathcal{O}(\frac{1}{N^{3/2}}) \\
&& W_{-1}(- e^{-1} x^{1/N}) = -1 - \sqrt{ \frac{2}{N} } \sqrt{\log(1/x)} + \frac{2}{3 N} \log x  + \mathcal{O}(\frac{1}{N^{3/2}}) 
\eea  
From theses expansions we obtain from \eqref{soludeltaprime} 
\be 
\Lambda \Delta_N'(\Lambda) = - 2 \sqrt{2 N} \sqrt{\log(\Lambda_c/\Lambda)} + \mathcal{O}(1/N^{1/2}) 
\ee 
Integrating over $\Lambda$ one finds, at large $N$
\be 
 \Delta_N(\Lambda) =   \sqrt{2 N} \frac{4}{3} (\log(\Lambda_c/\Lambda))^{3/2} + \mathcal{O}(1/N^{1/2}) 
\ee 
which has precisely the form 
\be 
 \frac{ \Delta_N(\Lambda) }{\varepsilon} \simeq  \frac{1}{\sqrt{T_{\rm KPZ}}} \Delta_{\rm KPZ}(z) 
 \quad , \quad \Delta_{\rm KPZ}(z) = \frac{4}{3} (-\log(-z))^{3/2} 
\ee  
which agrees with the result for the KPZ equation in Refs.~\cite{le2016exact,krajenbrink2021inverse}. \\

Now we need to consider the left hand side of \eqref{eqeq}, and identify the observables in the two models.
For this we will use the result from Ref.~\cite[Section 5.4.1 Formula (5.9)]{BorodinMacdo}, which
we first state in the notations of that paper. There is it shown that the OY
partition sum denoted there $e^{F_N(T)}$ converges for $N \to +\infty$ to the 
continuum directed polymer partition sum $Z(X,T)$ (in the notations of \cite{BorodinMacdo}) as
\be  
e^{F_N(\sqrt{T N} + X)} \simeq  e^{ N + \frac{\sqrt{N T}+X}{2} + X \sqrt{\frac{N}{T}} + \frac{N}{2} \log (T N) - N \log N } Z(X,T) 
\ee    
On the other hand the partition sums defined here are related to those in that paper via 
\be 
z_N(t)= e^{- t (1 + \frac{\varepsilon}{2} )} Z_N(t)  \quad , \quad Z_N(t) = \varepsilon^{1-N} e^{F_N(\varepsilon t)}  
\ee 
We can choose $X=0$, and we obtain the convergence
\be 
z_N(t=1)= e^{- (1 + \frac{\varepsilon}{2} )} \varepsilon^{1-N} e^{F_N(\varepsilon)} \simeq   \varepsilon
e^{ N -1 - N \log N } Z(0,T) = - \frac{\varepsilon}{\Lambda_c} Z(0,T)
\ee 
where we have identified $\varepsilon = \sqrt{T N}$, i.e. we set $T=\varepsilon^2/N$.  
Let us now recall that in our previous work \cite{krajenbrink2021inverse} we define the
continuum directed polymer partition $Z_{\rm KPZ}(X, T_{\rm KPZ})$ as the solution of the SHE
\be 
\partial_t Z_{\rm KPZ} = \partial_x^2 Z_{\rm KPZ} + \sqrt{2} \eta(x,t) Z_{\rm KPZ}
\ee 
where $\eta(x,t)$ is the standard (unit variance) space-time white noise. Comparing with 
Ref.~\cite{BorodinMacdo} we see that we need to write
\be
Z(X,T) = Z_{\rm KPZ}(X, T_{\rm KPZ})   \quad , \quad 2 T_{\rm KPZ}= T 
\ee 
Thus we have the large $N$ behavior
\be 
z_N(t=1) \simeq - \frac{\varepsilon}{\Lambda_c} Z_{\rm KPZ}(0, T_{\rm KPZ}=\frac{\varepsilon^2}{2 N})
\ee 
Hence the l.h.s of \eqref{eqeq} becomes 
\be 
 \overline{  \exp \left(-\frac{\Lambda}{\varepsilon} z_N(t=1) \right) }  
\simeq \overline{ \exp \left(- z  Z_{\rm KPZ}(0,T_{\rm KPZ})  \right) } 
    = \overline{ \exp \left(-\frac{ze^{H(T_{\rm KPZ})}}{ \sqrt{T_{\rm KPZ}}    } \right) } 
\ee 
where we recall that $z=-\Lambda/\Lambda_c$ and that for the droplet solution
of the KPZ solution we have defined $H(T_{\rm KPZ}) = \log Z_{\rm KPZ}(0,T_{\rm KPZ}) 
+ \frac{1}{2} \log T_{\rm KPZ} $ in Ref.~\cite{krajenbrink2021inverse}.

\subsection{Convergence of the dynamical system \eqref{eq:YO-polymer-saddle} to the nonlinear Schrodinger equation}

In this Section, we investigate how to take the limit of the WNT of the O'Connell-Yor polymer \eqref{eq:YO-polymer-saddle} to the WNT equations of the Kardar-Parisi-Zhang equation directly. To this aim, we define a scaling variable $x_0$, which can be interpreted as a lattice length, and and introduce the rescaled space and time as
\begin{equation}
    X=x_0(t-n), \quad T=\frac{x_0^2}{2} t
\end{equation}
We then define the partition function $Z$ as
\begin{equation}
\label{eq:partition-func-limit-kpz-wnt}
    z_n(t)=Z(x_0t-x_0n,\frac{x_0^2}{2} t)  \quad \longleftrightarrow \quad Z(X,T)=z_{\frac{2T}{x_0^2}-\frac{X}{x_0}}(\frac{2T}{x_0^2})
\end{equation}
and similarly for $\tilde{z}_n$ which we relate to a response field $\tilde{Z}$. Definition~\eqref{eq:partition-func-limit-kpz-wnt} implies the differential relations 
\begin{equation}
\label{eq:convergence-operator-2}
\begin{split}
    \p_t z_n(t)&=x_0\p_X Z+\frac{x_0^2}{2} \p_T Z\\
    z_{n+1}-z_n&=-x_0\p_X Z+\frac{x_0^2}{2}\p_X^2 Z+\mathcal{O}(x_0^3)\\
    z_{n-1}-z_n&=x_0\p_X Z+\frac{x_0^2}{2}\p_X^2 Z+\mathcal{O}(x_0^3)
    \end{split}
\end{equation}
where we have assumed a small $x_0$ expansion in the last two equations.  In order to keep only one dominant term in the expansion, we consider the following combination
\begin{equation}
\label{eq:convergence-operator-1}
    \p_t z_n + z_{n+1}-z_n = \frac{x_0^2}{2} (\p_T Z+\p_X^2 Z)+\mathcal{O}(x_0^3)
\end{equation}

Upon convergence, we find that the original dynamical system \eqref{eq:YO-polymer-saddle} reads at leading order
\begin{equation}
    \begin{split}
        x_0^2\p_T Z&=x_0^2\p_X^2 Z+2Z^2 \tilde{Z}\\
        -x_0^2\p_T \tilde{Z}&=x_0^2\p_X^2 \tilde{Z}+2Z \tilde{Z}^2\\
    \end{split}
\end{equation}
To obtain a final system independent on $x_0$, we finally rescale the fields $Z$ and $\tilde{Z}$ as
\begin{equation}
\label{eq:scaling-limit-PQ}
        Q_{\rm KPZ}(X,T)=\frac{1}{x_0}Z(X,T), \qquad   P_{\rm KPZ}(X,T) =\frac{1}{x_0}\tilde{Z}(X,T)
\end{equation}
to obtain the $\{ P,Q \} $ system studied in the context of the WNT of the KPZ equation \cite{krajenbrink2021inverse}.

\subsection{Convergence of the Lax pair \eqref{eq:LaxPairYO-new0} to the Lax pair of the nonlinear Schrodinger equation}

Recalling the driftless Lax pair for the O'Connell-Yor system

\begin{equation}
\label{eq:sup-mat-lax-def-discrete}
        \p_t {v}_n=U_n {v}_n, \quad {v}_{n+1}=L_n {v}_n, \quad 
U_n=
\begin{pmatrix}
\frac{\lambda^2-1}{2}  &   -z_{n-1} \\ 
\tilde{z}_n &  \frac{1-\lambda^2}{2}
\end{pmatrix}, \quad
L_n=
\begin{pmatrix}
\frac{1}{\lambda} & \frac{z_n}{\lambda}\\
 -\frac{1}{\lambda}\tilde{z}_{n} & \lambda- \frac{1}{\lambda}z_n\tilde{z}_{n} \, .
\end{pmatrix} 
\end{equation}
From \eqref{eq:convergence-operator-1} and \eqref{eq:convergence-operator-2} we observe that in order to study the convergence of the Lax pair, we need to consider the following combination
\begin{equation}
\label{eq:finite-diff-time-deriv-lax}
  \p_t v_n + v_{n+1}-v_n= ( U_n + L_n - I_2) v_n
\end{equation}
as well as the first order in $x_0$ of the equations $\p_t v_n=U_n v_n$ and $v_{n+1}-v_n=(L_n-I_2)v_n$ where $I_2$ is the identity matrix. Indeed, in the continuum we want to transform the Lax pair \eqref{eq:sup-mat-lax-def-discrete}
into a continuous version 
\begin{equation}
\label{eq:sup-mat-lax-def-continuous}
    \p_X V(X,T)=U_1 V(X,T), \quad \p_T V(X,T) = U_2 V(X,T) 
\end{equation}
Equation \eqref{eq:convergence-operator-1} together with Eq.~\eqref{eq:finite-diff-time-deriv-lax} provides a convergence to the operator $(\p_T + \p_X^2) V$ which in terms of the continuous Lax matrices \eqref{eq:sup-mat-lax-def-continuous} reads $(U_2+U_1^2 +\p_X U_1)V$.\\

To study the convergence of the Lax pair, we choose the following scaling at leading order (see also \eqref{eq:convergence-operator-2})
\begin{equation}
\begin{split}
    &\lambda(x_0) = e^{\I \frac{\lambda}{2}x_0},\\
    & \tilde{z}_n(t)=x_0 \tilde{Z}(X,T), \quad \tilde{z}_{n-1}(t)=x_0 \tilde{Z}(X,T)+x_0^2 \p_X \tilde{Z}(X,T)+ \frac{x_0^3}{2}\p^2_X \tilde{Z}(X,T) \\
    & z_n(t) = x_0 Z(X,T), \quad z_{n-1}(t)=x_0 Z(X,T)+x_0^2 \p_X Z(X,T)+ \frac{x_0^3}{2}\p^2_X Z(X,T)
    \end{split}
\end{equation}

Under this scaling, a straightforward replacement gives
\begin{equation}
\begin{split}
    L_n &\to I_2 -x_0 U_1 + \frac{x_0^2}{2}(U_1^2 +\p_X U_1) +x_0^2  (U_2-W)+ \mathcal{O}(x_0^3)\\
    U_n &\to x_0 U_1 + x_0^2 W+ \mathcal{O}(x_0^3)
\end{split}
\end{equation}
where
\begin{equation}
\label{eq:lax-matrix-PQ1}
    U_1=
    \begin{pmatrix}
       \frac{\I \lambda}{2} &-Z(X,T) \\
        \tilde{Z}(X,T)& -\frac{\I \lambda}{2}\\
    \end{pmatrix}
\end{equation}
\begin{equation}
\label{eq:lax-matrix-PQ2}
    U_2=
    \begin{pmatrix}
 \frac{1}{2} Z(X,T) \tilde{Z}(X,T)-\frac{\lambda ^2}{4} & -\frac{1}{2} \p_X Z(X,T)-\frac{\I \lambda  }{2} Z(X,T) \\
 -\frac{1}{2} \p_X\tilde{Z}(X,T)+\frac{\I \lambda}{2}    \tilde{Z}(X,T) &  \frac{\lambda}{4} ^2-\frac{1}{2} Z(X,T) \tilde{Z}(X,T) \\
    \end{pmatrix}
\end{equation}
\begin{equation}
    W=
    \begin{pmatrix}
 -\frac{\lambda ^2}{4} & -\p_X Z(X,T) \\
 0 & \frac{\lambda ^2}{4} \\
    \end{pmatrix}
\end{equation}
We recognize in Eqs.~\eqref{eq:lax-matrix-PQ1}-\eqref{eq:lax-matrix-PQ2} the matrices $U_1$ and $U_2$ of the continuous nonlinear Schrodinger equation obtained in the weak noise theory of the KPZ equation \cite{krajenbrink2021inverse}. In particular we see the convergence
\begin{equation}
    U_n+L_n-I_2 \to x_0^2(U_2+U_1^2 +\p_X U_1)
\end{equation}
as expected. This concludes the convergence of the Lax pairs of the discrete problem towards the continuous problem.

\section{Triangular representation of the solution to the Lax problem}

Here we set $g=2$ and the drifts $\{a_\ell\}=0$ and we recall the asymptotics
\begin{equation}
    \phi_n  \sim \lambda^{-n}
    \begin{pmatrix}
1 \\
0
\end{pmatrix}, \quad 
\bar{\phi}_n \sim \lambda^n
    \begin{pmatrix}
0 \\
-1
\end{pmatrix}, \quad n\to -\infty
\end{equation}
and
\begin{equation}
    \psi_n  \sim \lambda^n
    \begin{pmatrix}
0 \\
1
\end{pmatrix}, \quad 
    \bar{\psi}_n \sim \lambda^{-n} 
    \begin{pmatrix}
1 \\
0
\end{pmatrix}, \quad 
 n\to +\infty
\end{equation}
and
\begin{equation}
    \phi_n  \sim 
    \begin{pmatrix}
a(\lambda,t)\lambda^{-n}  \\
b(\lambda,t) \lambda^n
\end{pmatrix}, \quad 
\bar{\phi}_n \sim 
    \begin{pmatrix}
\btilde(\lambda,t)\lambda^{-n} \\
-\atilde(\lambda,t)\lambda^{n}
\end{pmatrix}, \quad n\to +\infty
\end{equation}

We have that $\lambda^n \phi_n $ and $\lambda^{-n} \psi_n $ are analytic inside the circle $|\lambda|<\mathtt{R}$
and that $\lambda^{-n} \bar \phi_n $ and $\lambda^{n} \bar \psi_n $ are analytic outside of the circle $|\lambda|>\mathtt{R}$. 
(for simplicity we consider the case $\Lambda<\Lambda^*_N$ but the same manipulations extent to
$\Lambda>\Lambda^*_N$, the contour being a deformed circle).
\\

One introduces the triangular representations
\be
\label{eq:triang-repn-psi}
\psi_n = \lambda^n \sum_{p \geq 0} \lambda^{p}
    \begin{pmatrix}
K_1(n,n+p) \\
K_2(n,n+p)
\end{pmatrix}=\sum_{p \geq 0} \lambda^{n+p} K(n,n+p), \quad \phi_n = \lambda^{-n} \sum_{p \geq 0} \lambda^{p}
    \begin{pmatrix}
M_1(n,n+p) \\
M_2(n,n+p)
\end{pmatrix}, \quad \abs{\lambda}<\mathtt{R}
\ee
\be
\label{eq:triang-repn-psi-bar}
\bar \psi_n = \lambda^{-n} \sum_{p \geq 0} \lambda^{-p}
    \begin{pmatrix}
\bar K_1(n,n+p) \\
\bar K_2(n,n+p)
\end{pmatrix}=\sum_{p \geq 0} \lambda^{-n-p} \bar{K}(n,n+p),   \quad \bar{\phi}_n =   \lambda^{n} \sum_{p \geq 0} \lambda^{-p}
    \begin{pmatrix}
\bar M_1(n,n+p) \\
\bar M_2(n,n+p)
\end{pmatrix}, \quad \abs{\lambda}>\mathtt{R}
\ee

The goal is to find the values of the first coefficients in the triangular representation. 
Let us recall the Lax matrix where we took the drifts $\{a_\ell \}$ equal to 0.
\begin{equation}
L_n=
\begin{pmatrix}
\frac{1}{\lambda} & \frac{z_n}{\lambda}\\
 -\frac{1}{\lambda}\tilde{z}_{n} & \lambda- \frac{1}{\lambda}z_n\tilde{z}_{n} 
\end{pmatrix} .
\label{eq:LaxPairYO-new01}
\end{equation}

\subsection{Triangular decomposition of $\bar \psi_n$}

Injecting into $\bar \psi_{n+1} = L_n \bar \psi_n$ one finds the recursion relations for $p \geq 0$
\bea
&& \bar K_2(n,n)=0 \quad , \quad \bar K_2(n,n+1) = 0 \\
&& \bar K_1(n+1,n+1+p)=\bar K_1(n,n+p) + z_n \bar K_2(n,n+p) \\
&& \bar K_2(n+1,n+1+p)=-  \tilde z_n \bar K_1(n,n+p) + \bar K_2(n,n+2+p) -  \tilde z_n z_n \bar K_2(n,n+p) 
\eea

In particular, this implies using $p=0$ 
\bea 
&& \bar K_1(n,n)= c_0 = 1 \quad , \quad  \tilde z_n = \frac{ \bar K_2(n,n+2) }{\bar K_1(n,n)} = \bar K_2(n,n+2)
\eea 
where $c_0=1$ comes from the asymptotics at large $n$. Using $p=1$ one has
\bea 
\bar K_1(n,n+1)= c_1 = 0 \quad , \quad \bar K_2(n,n+3) = 0 
\eea 
After further examination for all $p$ one obtains
\be 
\begin{split}
&\bar K_1(n,n+2 k +1) = c_{2k+1}= 0 \\
&\bar K_2(n,n+2 k +1) = 0
\end{split}
\ee 
where all constants $c_{2k+1}$ are determined by the asymptotic $n\to \infty$ condition. We summarize our findings with the following expansion as well as Table~\ref{table:triangular-barpsi}
\begin{equation}
\label{eq:asymptotic-psibar-n}
    \bar{\psi}_n =  \lambda^{-n}   \begin{pmatrix}
 1 \\
 0
\end{pmatrix}+
\lambda^{-n-2}   
\begin{pmatrix}
 -\sum_{\ell=n}^{\infty} z_\ell \tilde{z}_{\ell}  \\
\tilde{z}_n
\end{pmatrix}
+\mathcal{O}(\lambda^{-n-4}) \, .
\end{equation}

\begin{table}[t!]
    \centering
    \begin{tabular}{p{3cm} p{5cm} p{4cm}}
    \hline
    \hline\\[-0.5ex]
    index  & $ \bar K_1$& $ \bar K_2$ \\[2ex]
    \hline\\[-0.5ex]
      $(n,n)$  & $ \bar K_1(n,n)=1$& $ \bar K_2(n,n)=0$ \\
       $(n,n+1)$  & $ \bar K_1(n,n+1)=0$& $ \bar K_2(n,n+1)=0$ \\
       $(n,n+2)$  &  $\bar{K}_1(n,n+2)=-\sum_{\ell=n}^{\infty} z_\ell \tilde{z}_{\ell} $ & $ \bar K_2(n,n+2)= \tilde z_n $ \\
        $(n,n+2k+1)$ & $\bar K_1(n,n+2 k +1) = 0 $ & $\bar K_2(n,n+2 k +1) = 0$\\[1ex]
        \hline
    \hline
    \end{tabular}
    \caption{Summary of the results for $\bar K_1$ and $\bar K_2$}
    \label{table:triangular-barpsi}
\end{table}

\subsection{Triangular decomposition of $ \psi_n$}

Injecting into $ \psi_{n+1} = L_n  \psi_n$ one finds the recursion relations for $p \geq 0$
\be
\begin{split}
& K_1(n,n) + z_n K_2(n,n) = 0 \\
& K_1(n,n+1) + z_n K_2(n,n+1) = 0 \\
& K_1(n+1,n+1+p) = K_1(n,n+p+2) + z_n K_2(n,n+p+2) \\
& K_2(n+1,n+1+p) = -  \tilde z_n ( K_1(n,n+p+2) + z_n K_2(n,n+p+2) ) + K_2(n,n+p) 
\end{split}
\ee
In particular, this implies using  $p=0$ 
\begin{equation}
    \begin{split}
        K_1(n+1,n+1) &= K_1(n,n+2) + z_n K_2(n,n+2)\\
        K_2(n+1,n+1) &= - \tilde z_n ( K_1(n,n+2) + z_n K_2(n,n+2) ) + K_2(n,n)\\
        &= -  \tilde z_n K_1(n+1,n+1) + K_2(n,n)\\
        &= \tilde z_n z_{n+1}K_2(n+1,n+1) + K_2(n,n)
    \end{split}
\end{equation}
With $p=1$ we further obtain the cancellation of all odd terms in the triangular decomposition. Hence we obtain the following result
\begin{equation}
    {\psi}_n = \lambda^{n}    \prod_{\ell=n}^{\infty}(1-\tilde{z}_\ell z_{\ell+1}) \begin{pmatrix}
 -z_n \\
 1
\end{pmatrix}+
\mathcal{O}(\lambda^{n+2})   
\end{equation}
also summarized in Table~\ref{table:triangular-psi}.
\begin{table}[t!]
    \centering
    \begin{tabular}{p{2cm} p{5cm} p{5cm}}
    \hline
    \hline\\[-0.5ex]
    index  & $  K_1$& $  K_2$\\[2ex]
    \hline\\[-0.5ex]
      $(n,n)$  & $  K_1(n,n)=-z_n \prod_{\ell=n}^{\infty}(1-\tilde{z}_\ell z_{\ell+1})$& $  K_2(n,n)=\prod_{\ell=n}^{\infty}(1-\tilde{z}_\ell z_{\ell+1})$ \\
        $(n,n+2p+1)$ & $ K_1(n,n+2 p +1) = 0 $ & $ K_2(n,n+2 p +1) = 0$\\[1ex]
        \hline
    \hline
    \end{tabular}
    \caption{Summary of the results for $K_1$ and $K_2$} 
    \label{table:triangular-psi}
\end{table}

\subsection{Triangular decomposition of $\bar \phi_n$}

We also give for completeness the results for $\bar \phi_n$ and $\phi_n$ although
we will not use them below.   Injecting into $\bar{\phi}_{n+1}=L_n\bar{\phi}_n$ one finds the asymptotic expansion
\begin{equation}
\label{eq:asymptotic-value-barphin}
    \bar{\phi}_n =  \lambda^{n}   \begin{pmatrix}
 0 \\
 -1
\end{pmatrix}+
\lambda^{n-2}   
\begin{pmatrix}
-z_{n-1}  \\
\sum_{\ell=-\infty}^{n-1}\frac{g}{2}z_\ell \tilde{z}_{\ell} 
\end{pmatrix}
+\mathcal{O}(\lambda^{n-4}) \, .
\end{equation}

\subsection{Triangular decomposition of $ \phi_n$}

Injecting into ${\phi}_{n+1}=L_n{\phi}_n$ one finds the asymptotic expansion
\begin{equation}
    {\phi}_n = \lambda^{-n}    \prod_{\ell=-\infty}^{n-1}(1-\frac{g}{2}\tilde{z}_{\ell-1} z_{\ell}) 
    \begin{pmatrix}
 1 \\
 -\frac{g}{2}\tilde{z}_{n-1}
\end{pmatrix}+
\mathcal{O}(\lambda^{-n+2})   \, .
\end{equation}

\subsection{Additional series expansion for the Lax problem}
As in Ref.~\cite{merola1994novel} we provide an additional discrete integral representation of the solutions to the Lax problem. Grouping the solution of the Lax problem into a $2\times 2$ matrix $\Psi_n = \{\phi_n, \bar{\phi}_n \}$, we first define a rescaled matrix $W_n$ so that
\begin{equation}
    \Psi_n = \lambda^{-n \sigma_3} \tilde{W}_n , \quad  \sigma_3 = \begin{pmatrix}
1  &   0 \\ 
0 &  -1
\end{pmatrix}, \quad
\lambda^{-n \sigma_3} = \left(
\begin{array}{cc}
 \lambda ^{-n} & 0 \\
 0 & \lambda ^n \\
\end{array}
\right)
\end{equation}

whose evolution and asymptotic value are given by 
\begin{equation}
\label{eq:evolution-wtilde}
    \tilde{W}_{\ell+1}=\lambda^{(\ell+1)\sigma_3}L_\ell \lambda^{-\ell \sigma_3}\tilde{W}_\ell, \quad \tilde{W}_{+\infty}=     \begin{pmatrix}
a  &   \btilde \\ 
b &  -\atilde
\end{pmatrix}
\end{equation}

Rewriting the matrix $\tilde{W}_n$ using a telescopic form and using Eq.~\eqref{eq:evolution-wtilde}, we first obtain that
\begin{equation}
\begin{split}
    \tilde{W}_n &= \tilde{W}_{+\infty}-\sum_{\ell=n}^{\infty} (\tilde{W}_{\ell+1}-\tilde{W}_\ell)\\
     &=        
     \begin{pmatrix}
a  &   \btilde \\ 
b &  -\atilde
\end{pmatrix}
-\sum_{\ell=n}^{\infty} (\lambda^{(\ell+1)\sigma_3}L_\ell \lambda^{-\ell \sigma_3}-\mathds{1}_2)\tilde{W}_\ell\\
\end{split}
\end{equation}
Going back to the $\Psi_n= \{\phi_n, \bar{\phi}_n \}$ space, we have
\begin{equation}
\begin{split}
    \Psi_n      &=        \lambda^{-n \sigma_3} 
     \begin{pmatrix}
a  &   \btilde \\ 
b &  -\atilde
\end{pmatrix}
-  \sum_{\ell=n}^{\infty} \lambda^{(\ell-n) \sigma_3}(\lambda^{\sigma_3}L_\ell -\mathds{1}_2)\Psi_\ell\\
 &=        \lambda^{-n \sigma_3} 
     \begin{pmatrix}
a  &   \btilde \\ 
b &  -\atilde
\end{pmatrix}
-  \sum_{\ell=n}^{\infty} \lambda^{(\ell+1-n) \sigma_3-1}
     \begin{pmatrix}
0  &   z_\ell \\ 
-\tilde{z}_\ell &  - z_\ell \tilde{z}_\ell
\end{pmatrix}
\Psi_\ell\\
 &=        \lambda^{-n \sigma_3} 
     \begin{pmatrix}
a  &   \btilde \\ 
b &  -\atilde
\end{pmatrix}
-  \sum_{\ell=n}^{\infty}
     \begin{pmatrix}
 \lambda^{\ell-n}  &   0 \\ 
0 &   \lambda^{-\ell+n-2} 
\end{pmatrix}
    \begin{pmatrix}
0  &   z_\ell \\ 
-\tilde{z}_\ell &  - z_\ell \tilde{z}_\ell
\end{pmatrix}
\Psi_\ell\\
\end{split}
\end{equation}
Hence we can obtain the recursion for $\phi_n$ and $\bar{\phi}_n$ by inspection of the columns as
\begin{equation}
    \phi_n =
        \begin{pmatrix}
a(\lambda)\lambda^{-n}  \\
b(\lambda) \lambda^n
\end{pmatrix}
-  \sum_{\ell=n}^{\infty}
     \begin{pmatrix}
 \lambda^{\ell-n}  &   0 \\ 
0 &   \lambda^{-\ell+n-2} 
\end{pmatrix}
    \begin{pmatrix}
0  &   z_\ell \\ 
-\tilde{z}_\ell &  - z_\ell \tilde{z}_\ell
\end{pmatrix}
\phi_\ell
\end{equation}
and 
\begin{equation}
    \bar{\phi}_n =
        \begin{pmatrix}
\btilde(\lambda)\lambda^{-n}  \\
-\atilde(\lambda) \lambda^n
\end{pmatrix}
-  \sum_{\ell=n}^{\infty}
     \begin{pmatrix}
 \lambda^{\ell-n}  &   0 \\ 
0 &   \lambda^{-\ell+n-2} 
\end{pmatrix}
    \begin{pmatrix}
0  &   z_\ell \\ 
-\tilde{z}_\ell &  - z_\ell \tilde{z}_\ell
\end{pmatrix}
\bar{\phi}_\ell
\end{equation}
Since $\psi_n$ and $\bar{\psi}_n$ are linear combinations of $\phi_n$ and $\bar{\phi}_n$, we further obtain that
\begin{equation}
\label{eq:discrete-integral-psi-n}
    \psi_n = 
          \lambda^n  \begin{pmatrix}
0 \\
1
\end{pmatrix}
-  \sum_{\ell=n}^{\infty}
     \begin{pmatrix}
 \lambda^{\ell-n}  &   0 \\ 
0 &   \lambda^{-\ell+n-2} 
\end{pmatrix}
    \begin{pmatrix}
0  &   z_\ell \\ 
-\tilde{z}_\ell &  - z_\ell \tilde{z}_\ell
\end{pmatrix}
\psi_\ell
\end{equation}

and
\begin{equation}
    \bar{\psi}_n = 
          \lambda^{-n}  \begin{pmatrix}
1 \\
0
\end{pmatrix}
-  \sum_{\ell=n}^{\infty}
     \begin{pmatrix}
 \lambda^{\ell-n}  &   0 \\ 
0 &   \lambda^{-\ell+n-2} 
\end{pmatrix}
    \begin{pmatrix}
0  &   z_\ell \\ 
-\tilde{z}_\ell &  - z_\ell \tilde{z}_\ell
\end{pmatrix}
\bar{\psi}_\ell
\end{equation}

Note that to have only higher order terms in the right hand side of Eq.~\eqref{eq:discrete-integral-psi-n}, i.e. $\psi_\ell$ with $\ell>n$, it is possible to rewrite the series as

    \begin{equation}
    \begin{split}
    \lambda^{-n}\psi_n &= 
            \begin{pmatrix}
0 \\
1
\end{pmatrix}
-  \sum_{\ell=n}^{\infty}
     \begin{pmatrix}
 \lambda^{2\ell-2n}  &   0 \\ 
0 &   \lambda^{-2} 
\end{pmatrix}
    \begin{pmatrix}
0  &   z_\ell \\ 
-\tilde{z}_\ell &  - z_\ell \tilde{z}_\ell
\end{pmatrix}
 \lambda^{-\ell} \psi_\ell\\
 &= 
            \begin{pmatrix}
0 \\
1
\end{pmatrix}
-  \sum_{\ell=n}^{\infty}\lambda
     \begin{pmatrix}
 \lambda^{2\ell-2n}  &   0 \\ 
0 &   \lambda^{-2} 
\end{pmatrix}
    \begin{pmatrix}
0  &   z_\ell \\ 
-\tilde{z}_\ell &  - z_\ell \tilde{z}_\ell
\end{pmatrix}
  L_{\ell}^{-1}\lambda^{-(\ell+1)} \psi_{\ell+1}\\
  &= 
            \begin{pmatrix}
0 \\
1
\end{pmatrix}
-  \sum_{\ell=n}^{\infty}
     \begin{pmatrix}
 \lambda^{2\ell-2n}  &   0 \\ 
0 &   1 
\end{pmatrix}
    \begin{pmatrix}
 z_\ell \tilde{z}_\ell  &   z_\ell \\ 
-\tilde{z}_\ell &  0
\end{pmatrix}
  \lambda^{-(\ell+1)} \psi_{\ell+1}\\
 \end{split}
\end{equation}
These series representations have been used in Ref.~\cite{merola1994novel} to obtain analyticity results on the solution to the Lax problem. Note further that we also have the matrix factorisation \, .

\begin{equation}
        \begin{pmatrix}
0  &   z_\ell \\ 
-\tilde{z}_\ell &  - z_\ell \tilde{z}_\ell  
\end{pmatrix}
=
        \begin{pmatrix}
1  &   0 \\ 
-\tilde{z}_\ell &  1
\end{pmatrix}
\begin{pmatrix}
1  &   z_\ell \\ 
0 &  1
\end{pmatrix} - \mathds{1}_2
\end{equation}

\section{Gelfand-Levitan-Marchenko equations}\label{sec:GLM-eq-solution}

\subsection{Notations and intermediate objects}

To obtain the explicit solutions $\{z_n, \tilde{z}_n \}$ as a function of the scattering data, one needs to obtain the Gelfand-Levitan-Marchenko equations of the problem. To this aim, we start by rewriting the linear relation between $\phi, \bar{\phi}$ and $\psi, \bar{\psi}$ and we further insert the triangular representation of $\psi_n$ and $\bar{\psi}_n$ obtained in Eqs.~\eqref{eq:triang-repn-psi}--\eqref{eq:triang-repn-psi-bar}.

\begin{equation}
\label{eq:def-scattering-amplitudes-GLM}
\begin{split}
  \frac{1}{a(\lambda) }  \phi_k &= \bar{\psi}_k+\frac{b(\lambda,t)}{a(\lambda) }\psi_k\\
    \frac{1}{\atilde(\lambda)}\bar{\phi}_k &= -\psi_k + \frac{\btilde(\lambda,t)}{\atilde(\lambda)}\bar{\psi}_k
    \end{split}
\end{equation}

The properties we will require are that

\begin{itemize}
    \item $\lambda^n \phi_n $, $\lambda^{-n} \psi_n $ and $a(\lambda)$ are analytic inside the circle $|\lambda|<\mathtt{R}$ \, ,
    \item  $\lambda^{-n} \bar \phi_n $, $\lambda^{n} \bar \psi_n $ and $\atilde(\lambda)$ are analytic outside of the circle $|\lambda|>\mathtt{R}$.
\end{itemize}

The analyticity properties of the various function will be used to proceed to a contour integration of Eq.~\eqref{eq:def-scattering-amplitudes-GLM} using that
    \begin{enumerate}
        \item For a function analytic inside the circle of radius $\mathtt{R}$, we have that
        \begin{equation}
                \label{eq:rule-integration1}\oint_{|\lambda|=\mathtt{R}} \frac{\rmd \lambda}{2\I \pi}  \frac{1}{\lambda -\xi_1} f(\lambda) =  f(\xi_1)\Theta(\, \abs{\xi_1}<\mathtt{R})
        \end{equation}
        \item For a function analytic outside the circle of radius $\mathtt{R}$, we have that
              \begin{equation}
                \label{eq:rule-integration2}
                \oint_{|\lambda|=\mathtt{R}} \frac{\rmd \lambda}{2\I \pi}  \frac{1}{\lambda -\xi_1} f(\lambda) = \lim_{\lambda\to +\infty}  f(\lambda)- f(\xi_1)\Theta(\, \abs{\xi_1}>\mathtt{R})
        \end{equation}
    \end{enumerate}

where $\Theta$ denotes the Heaviside function. Furthermore, we use a contour integral representation of the Kronecker delta function
\begin{equation}
\label{eq:kronecker-contour}
    \delta_{m,n} = \oint_{|\lambda|=\mathtt{R}} \frac{\rmd \lambda}{2\I \pi } \lambda^{m-n-1} \, .
\end{equation}
We define the Fourier transform on the circle of the reflection coefficients. If the scattering data $\{ a, \atilde \}$ do not have any zero, it reads

\begin{equation}
\label{eq:def-fourier-scattering}
\begin{split}
    F(n)&:= \oint_{|\lambda|=\mathtt{R}}\frac{\rmd \lambda }{2\I \pi}\lambda^{n-1}\frac{b(\lambda,t)}{a(\lambda) } = \oint_{|\lambda|=\mathtt{R}} \frac{\rmd \lambda }{2\I \pi}\lambda^{n-1}\frac{b(\lambda)}{a(\lambda) } e^{(1-\lambda^2)t} \\
    \tilde{F}(n)&:= \oint_{|\lambda|=\mathtt{R}} \frac{\rmd \lambda }{2\I \pi}\lambda^{-n-1}\frac{\btilde(\lambda,t)}{\atilde(\lambda) } = \oint_{|\lambda|=\mathtt{R}} \frac{\rmd \lambda }{2\I \pi}\lambda^{-n-1}\frac{\btilde(\lambda)}{\atilde(\lambda) } e^{(\lambda^2-1)t} 
\end{split}
\end{equation}

In the case of the presence of soliton, i.e. the scattering data having zeroes as in the current problem, we first proceed to the replacement
    \be
    a(\lambda)\to  a(\lambda)\frac{\lambda^2-\lambda_0^2}{\lambda^2-\lambda_{-1}^2} \quad , \quad  \atilde(\lambda)\to \atilde(\lambda)\frac{\lambda^2-\lambda_{-1}^2}{\lambda^2-\lambda_0^2}
   \ee

If the scattering data have simple zeroes, e.g. $a( \pm\lambda_0)=0$ and $\atilde(\pm \lambda_{-1})=0$, then the expressions of $F(n)$ and $\tilde{F}(n)$ are modified to take into account the respective poles as 

\begin{equation}
\begin{split}
    F(n)&:=   \oint_{|\lambda|=\mathtt{R}} \frac{\rmd \lambda }{2\I \pi}\lambda^{n-1}\frac{b(\lambda)}{a(\lambda) } \frac{\lambda^2-\lambda_{-1}^2}{\lambda^2-\lambda_0^2} e^{(1-\lambda^2)t}+ \frac{\lambda_0^2-\lambda_{-1}^2}{2\lambda_0}\left[\lambda^{n-1}\frac{b(\lambda)}{a(\lambda) }  e^{(1-\lambda^2)t}\right]_{-\lambda_0}^{\lambda_0} \\
    \tilde{F}(n)& := \oint_{|\lambda|=\mathtt{R}} \frac{\rmd \lambda }{2\I \pi}\lambda^{-n-1}\frac{\btilde(\lambda)}{\atilde(\lambda) } \frac{\lambda^2-\lambda_0^2}{\lambda^2-\lambda_{-1}^2}e^{(\lambda^2-1)t}+ \frac{\lambda_{-1}^2-\lambda_0^2}{2\lambda_{-1}}\left[\lambda^{-n-1}\frac{\btilde(\lambda)}{\atilde(\lambda) }e^{(\lambda^2-1)t}\right]_{-\lambda_{-1}}^{\lambda_{-1}} 
\end{split}
\end{equation}

where the notation $[f(\lambda)]_a^b$ stands for $f(b)-f(a)$.\\

Inserting the precise value of the scattering data, i.e.
\begin{equation}
    \btilde(\lambda)=-\lambda^2, \; b(\lambda)=\Lambda \lambda^{-2N-2}e^{\lambda^2-1}, \; a(\lambda)=e^{\varphi(\lambda)}, \; \atilde(\lambda)=e^{-\varphi(\lambda)}
\end{equation}
we then obtain in the absence of soliton
    \begin{equation}
    \begin{split}
    F(n) &= \Lambda \oint_{|\lambda|=\mathtt{R}}  \frac{\rmd \lambda }{2\I \pi \lambda}\lambda^{n-2-2N} e^{(\lambda^2-1)(1-t)-\varphi(\lambda)} \\
    \tilde{F}(n)&=-\oint_{|\lambda|=\mathtt{R}}  \frac{\rmd \lambda }{2\I \pi \lambda}\lambda^{-(n-2)}  e^{(\lambda^2-1)t+\varphi(\lambda)} \end{split}
\end{equation}
and in the presence of solitons
  \begin{equation}
    \begin{split}
    F(n) &= \Lambda \oint_{|\lambda|=\mathtt{R}}  \frac{\rmd \lambda }{2\I \pi \lambda}\lambda^{n-2-2N} \frac{\lambda^2-\lambda_{-1}^2}{\lambda^2-\lambda_0^2}e^{(\lambda^2-1)(1-t)-\varphi(\lambda)}+ \Lambda\frac{\lambda_0^2-\lambda_{-1}^2}{2\lambda_0}\left[\lambda^{n-3-2N}e^{(\lambda^2-1)(1-t)-\varphi(\lambda)}\right]_{-\lambda_0}^{\lambda_0} \\
    \tilde{F}(n)&=-\oint_{|\lambda|=\mathtt{R}}  \frac{\rmd \lambda }{2\I \pi \lambda}\lambda^{-(n-2)}  \frac{\lambda^2-\lambda_{0}^2}{\lambda^2-\lambda_{-1}^2}e^{(\lambda^2-1)t+\varphi(\lambda)} - \frac{\lambda_{-1}^2-\lambda_0^2}{2\lambda_{-1}}\left[\lambda^{-n+1}e^{(\lambda^2-1)t+\varphi(\lambda)}\right]_{-\lambda_{-1}}^{\lambda_{-1}} 
    \end{split}
\end{equation}

\begin{remark}
Note that the two Fourier transforms verify the discrete linear evolution equations
    \begin{equation}
    \label{eq:time-deriv-fourier-reflec-coeff}
    \begin{split}
        \p_t F(2n) &= F(2n)-F(2(n+1))\\
        \p_t \tilde{F}(2n) &= \tilde{F}(2(n-1))-\tilde{F}(2n)\\
    \end{split}
    \end{equation}
    \end{remark}

\subsection{Derivation of the Gelfand-Levitan-Marchenko equations}

\subsubsection{First equation}

We start by considering the first equation of \eqref{eq:def-scattering-amplitudes-GLM}, multiply it by $\lambda^n/(\lambda-\xi_1)$ for $\xi_1\in \C$ and integrate it over $\lambda$.

\begin{equation}
    \oint_{|\lambda|=\mathtt{R}}   \frac{\rmd \lambda}{2\I \pi}  \frac{\lambda^n}{\lambda -\xi_1}\frac{ \phi_n}{a(\lambda) }  = \oint_{|\lambda|=\mathtt{R}}   \frac{\rmd \lambda}{2\I \pi}\frac{\lambda^n}{\lambda -\xi_1} \bar{\psi}_n+ \oint_{|\lambda|=\mathtt{R}}   \frac{\rmd \lambda}{2\I \pi}\frac{\lambda^n}{\lambda -\xi_1} \frac{b(\lambda,t)}{a(\lambda) }\psi_n
\end{equation}
We first assume that we are in the case without soliton, i.e. no zero of $a(\lambda)$ and $\atilde(\lambda)$ in their domain of analyticity. Using the rules \eqref{eq:rule-integration1} and \eqref{eq:rule-integration2}, we obtain 
\begin{equation}
\label{eq:GLM-derivation-1}
    \xi_1^n \frac{\phi_n(\xi_1)}{a(\xi_1)} \Theta(\, \abs{\xi_1}< 1)-\lim_{\abs{\lambda}\to \infty}\lambda^n \bar{\psi}_n = -\xi_1^n \bar{\psi}_n(\xi_1)\Theta(\, \abs{\xi_1}> 1)+\oint_{|\lambda|=\mathtt{R}}    \frac{\rmd \lambda}{2\I \pi}\frac{\lambda^n}{\lambda -\xi_1} \frac{b(\lambda,t)}{a(\lambda) }\psi_n
\end{equation}
We then denote the following limit
\begin{equation}
    \begin{split}
        \mathcal{I}_1 &=  \lim_{\abs{\lambda}\to \infty}\lambda^n \bar{\psi}_n\\
    \end{split}
\end{equation}
Taking $\abs{\xi_1}=\mathtt{R}^+$, Eq.~\eqref{eq:GLM-derivation-1} reads
\begin{equation}
    -\mathcal{I}_1  = -\xi_1^n \bar{\psi}_n(\xi_1)+\oint_{|\lambda|=\mathtt{R}}     \frac{\rmd \lambda}{2\I \pi}\frac{\lambda^n}{\lambda -\xi_1} \frac{b(\lambda,t)}{a(\lambda) }\psi_n
\end{equation}
Multiplying the equation by $\xi_1^{m-n-1}$ with $m\geq n$ and integrating $\xi_1$ over the circle of radius $\mathtt{R}$, we further obtain

\begin{equation}
\begin{split}
    -\oint_{|\xi_1|=\mathtt{R}}     \frac{\rmd \xi_1}{2\I \pi}\xi_1^{m-n-1}\mathcal{I}_1  = -\oint_{|\xi_1|=\mathtt{R}}     \frac{\rmd \xi_1}{2\I \pi}\xi_1^{m-1}\bar{\psi}_n(\xi_1)+\oint_{|\xi_1|=\mathtt{R}}    \frac{\rmd \xi_1}{2\I \pi}\xi_1^{m-n-1}\oint_{|\lambda|=\mathtt{R}}    \frac{\rmd \lambda}{2\I \pi}\frac{\lambda^n}{\lambda -\xi_1} \frac{b(\lambda,t)}{a(\lambda) }\psi_n
\end{split}
\end{equation}
Making use of the contour integral representation of the Kronecker function \eqref{eq:kronecker-contour} and expanding the term $1/\lambda -\xi_1$ as a series, we have

\begin{equation}
\begin{split}
    -\delta_{m,n}\mathcal{I}_1  &= -\oint_{|\xi_1|=\mathtt{R}}     \frac{\rmd \xi_1}{2\I \pi}\xi_1^{m-1} \bar{\psi}_n(\xi_1)-\sum_{\ell=0}^\infty\oint_{|\xi_1|=\mathtt{R}}      \frac{\rmd \xi_1}{2\I \pi}\oint_{|\lambda|=\mathtt{R}}      \frac{\rmd \lambda}{2\I \pi}\xi_1^{m-n-1}\frac{\lambda^{n+\ell}}{\xi_1^{\ell+1}} \frac{b(\lambda,t)}{a(\lambda) }\psi_n\\
     &= -\oint_{|\xi_1|=\mathtt{R}}      \frac{\rmd \xi_1}{2\I \pi}\xi_1^{m-1}\bar{\psi}_n(\xi_1)-\sum_{\ell=0}^\infty  \delta_{m,n+\ell+1} \oint_{|\lambda|=\mathtt{R}}      \frac{\rmd \lambda}{2\I \pi}\lambda^{n+\ell} \frac{b(\lambda,t)}{a(\lambda) }\psi_n\\
       &= -\oint_{|\xi_1|=\mathtt{R}}      \frac{\rmd \xi_1}{2\I \pi}\xi_1^{m-1}\bar{\psi}_n(\xi_1)-(1-\delta_{m,n}) \oint_{|\lambda|=\mathtt{R}}      \frac{\rmd \lambda}{2\I \pi}\lambda^{m-1} \frac{b(\lambda,t)}{a(\lambda) }\psi_n\\
\end{split}
\end{equation}

Inserting the triangular representation of $\psi_n$, $\bar{\psi}_n$ from Eqs.~\eqref{eq:triang-repn-psi}-\eqref{eq:triang-repn-psi-bar} with the Fourier transform of the scattering data \eqref{eq:def-fourier-scattering} we obtain for all $m \geq n$
\begin{equation}
\begin{split}
    \delta_{m,n}\mathcal{I}_1
       &= \bar{K}(n,m)+(1-\delta_{m,n})\sum_{p\geq 0}K(n,n+2p) F(m+n+2p)\\
\end{split}
\end{equation}

Choosing $m=n$, this implies that
    \begin{equation}
    \label{eq:glm-diagonal1}
        \mathcal{I}_1 = \bar{K}(n,n) = \begin{pmatrix}
 1 \\
 0
\end{pmatrix}
    \end{equation}
The value of $\mathcal{I}_1$ can be read in Eq.~\eqref{eq:asymptotic-psibar-n} and the one of $\bar{K}(n,n)$ in Table~\ref{table:triangular-barpsi}.  
    Furthermore, choosing $m>n$ and separating the $p=0$ term out of the sum, we obtain the first Gelfand-Levitan-Marchenko equation as
    \begin{equation}
    \label{eq:GLM-eq-1}
        \bar{K}(n,m)+K(n,n)F(m+n)+\sum_{p> 0}K(n,n+2p) F(m+n+2p)=0, \quad m>n
    \end{equation}

\subsubsection{Second equation}

We repeat the procedure to obtain the second GLM equation. To this aim, we consider the second equation of \eqref{eq:def-scattering-amplitudes-GLM}, multiply it by $\lambda^{-n}/(\lambda -\xi_2)$ for $\xi_2 \in \C$ and integrate it over $\lambda$

\begin{equation}
     \oint_{|\lambda|=\mathtt{R}} \frac{\rmd \lambda}{2\I \pi}  \frac{\lambda^{-n}}{\lambda -\xi_2}    \frac{\bar{\phi}_n}{\atilde(\lambda)} = -   \oint_{|\lambda|=\mathtt{R}} \frac{\rmd \lambda}{2\I \pi}  \frac{\lambda^{-n}}{\lambda -\xi_2}   \psi_n +   \oint_{|\lambda|=\mathtt{R}}\frac{\rmd \lambda}{2\I \pi}  \frac{\lambda^{-n}}{\lambda -\xi_2}    \frac{\btilde(\lambda,t)}{\atilde(\lambda)}\tilde{\psi}_n
\end{equation}

We first assume that we are in the case without soliton, i.e. no zero of $a(\lambda)$ and $\atilde(\lambda)$ in their domain of analyticity. Using the rules \eqref{eq:rule-integration1} and \eqref{eq:rule-integration2}, we obtain 

\begin{equation}
-\xi_2^{-n}\frac{\bar{\phi}_n(\xi_2)}{\atilde(\xi_2)}\Theta(\, \abs{\xi_2}>1)+\lim_{\abs{\lambda}\to +\infty}\lambda^{-n} \frac{\bar{\phi}_n}{\atilde(\lambda)} = -  \xi_2^{-n}  \psi_n(\xi_2)\Theta(\, \abs{\xi_2}<1) +   \oint_{|\lambda|=\mathtt{R}} \frac{\rmd \lambda}{2\I \pi}  \frac{\lambda^{-n}}{\lambda -\xi_2}    \frac{\btilde(\lambda,t)}{\atilde(\lambda)}\bar{\psi}_n
\end{equation}

We then denote the following limit
\begin{equation}
    \begin{split}
        \mathcal{I}_2 &=\lim_{\abs{\lambda}\to +\infty}\lambda^{-n} \frac{\bar{\phi}_n}{\atilde(\lambda)}
    \end{split}
\end{equation}
We now take $\abs{\xi_2}=\mathtt{R}^-$ so that

\begin{equation}
\mathcal{I}_2  = -  \xi_2^{-n}  \psi_n(\xi_2) +   \oint_{|\lambda|=\mathtt{R}} \frac{\rmd \lambda}{2\I \pi}  \frac{\lambda^{-n}}{\lambda -\xi_2}    \frac{\btilde(\lambda,t)}{\atilde(\lambda)}\bar{\psi}_n
\end{equation}
Multiplying the equation by $\xi_2^{n-m-1}$ with $m\geq n$ and integrating $\xi_2$ over the circle of radius $\mathtt{R}$, we further obtain

\begin{equation}
\oint_{|\xi_2|=\mathtt{R}} \frac{\rmd \xi_2}{2\I \pi}\xi_2^{n-m-1}\mathcal{I}_2  = - \oint_{|\xi_2|=\mathtt{R}} \frac{\rmd \xi_2}{2\I \pi}\xi_2^{-m-1} \psi_n(\xi_2) +  \oint_{|\xi_2|=\mathtt{R}} \frac{\rmd \xi_2}{2\I \pi}\xi_2^{n-m-1} \oint_{|\lambda|=\mathtt{R}} \frac{\rmd \lambda}{2\I \pi}  \frac{\lambda^{-n}}{\lambda -\xi_2}    \frac{\btilde(\lambda,t)}{\atilde(\lambda)}\bar{\psi}_n
\end{equation}
equivalent to 

\begin{equation}
\begin{split}
\delta_{m,n}\mathcal{I}_2  &= - \oint_{|\xi_2|=\mathtt{R}} \frac{\rmd \xi_2}{2\I \pi}\xi_2^{-m-1} \psi_n(\xi_2) +  \oint_{|\xi_2|=\mathtt{R}} \frac{\rmd \xi_2}{2\I \pi}\xi_2^{n-m-1} \oint_{|\lambda|=\mathtt{R}} \frac{\rmd \lambda}{2\I \pi}  \frac{\lambda^{-n}}{\lambda -\xi_2}    \frac{\btilde(\lambda,t)}{\atilde(\lambda)}\bar{\psi}_n\\
&= - \oint_{|\xi_2|=\mathtt{R}} \frac{\rmd \xi_2}{2\I \pi}\xi_2^{-m-1} \psi_n(\xi_2) + \sum_{\ell=0}^{\infty} \oint_{|\xi_2|=\mathtt{R}} \frac{\rmd \xi_2}{2\I \pi}\xi_2^{\ell+n-m-1} \oint_{|\lambda|=\mathtt{R}} \frac{\rmd \lambda}{2\I \pi}  \lambda^{-n-\ell-1}   \frac{\btilde(\lambda,t)}{\atilde(\lambda)}\bar{\psi}_n\\
&= - \oint_{|\xi_2|=\mathtt{R}} \frac{\rmd \xi_2}{2\I \pi}\xi_2^{-m-1} \psi_n(\xi_2) +   \oint_{|\lambda|=\mathtt{R}} \frac{\rmd \lambda}{2\I \pi}  \lambda^{-m-1}   \frac{\btilde(\lambda,t)}{\atilde(\lambda)}\bar{\psi}_n\\
\end{split}
\end{equation}

Inserting the triangular representation of $\psi_n$, $\bar{\psi}_n$ from Eqs.~\eqref{eq:triang-repn-psi}-\eqref{eq:triang-repn-psi-bar} with the Fourier transform of the scattering data \eqref{eq:def-fourier-scattering} we obtain for all $m \geq n$

\begin{equation}
\begin{split}
\delta_{m,n}\mathcal{I}_2  &= -K(n,m)+\sum_{p\geq 0}\bar{K}(n,n+2p)\tilde{F}(m+n+2p)
\end{split}
\end{equation}

    Choosing $n=m$, this implies that
    \begin{equation}
    \label{eq:glm-diagonal2}
\begin{split}
\mathcal{I}_2  &= -K(n,n)+\sum_{p\geq 0}\bar{K}(n,n+2p)\tilde{F}(2n+2p) = \begin{pmatrix}
 0 \\
 -1
\end{pmatrix}
\end{split}
\end{equation}
The value of $\mathcal{I}_2$ can be read from Eq.~\eqref{eq:asymptotic-value-barphin} along side $\lim_{\lambda \to \infty} \atilde(\lambda)=1$ obtained for instance from Eq.~\eqref{finalat}.
    Furthermore, choosing $m>n$ and separating the $p=0$ term out of the sum, we obtain the second Gelfand-Levitan-Marchenko equation as
\begin{equation}
\label{eq:GLM-eq-2}
\begin{split}
 K(n,m)-\bar{K}(n,n)\tilde{F}(m+n)-\sum_{p> 0}\bar{K}(n,n+2p)\tilde{F}(m+n+2p)=0, \quad m>n
\end{split}
\end{equation}

\subsection{Operator valued equations}
The GLM equations obtained in Eqs.~\eqref{eq:GLM-eq-1}--\eqref{eq:GLM-eq-2} involved summations over explicit indices. We rewrite in this Section these equations using operators and invert these to obtain explicitly the solution $\{ z_n, \tilde{z}_n\}$ of the original system \eqref{eq:YO-polymer-saddle}.\\

Let us define four operators $\{F_n, \tilde{F}_n, K_n, \bar{K}_n \}$ indexed by $n\in \Z$ with the following kernels
\begin{equation}
\begin{split}
    F_n(i,j) &= F(2n+i+j)\\
    \tilde{F}_n(i,j) &= \tilde{F}(2n+i+j)\\
    K_n(i,j)&=K(i+n,n+j) \\
    \bar{K}_n(i,j)&=\bar{K}(i+n,n+j)\\
    \end{split}
\end{equation}
for $i,j \in \N$ and we recall that the operators $K_n$ and $\bar{K}_n$ are upper-triangular operators. Using the four kernels introduced and choosing $m=n+2j$ for $j>0$, we rewrite the two GLM equations \eqref{eq:GLM-eq-1}--\eqref{eq:GLM-eq-2} as
\begin{equation}
\label{eq:GLM-kernel-indices}
\begin{split} 
\bar{K}_n(0,2j)+K_n(0,0)F_n(0,2j)+\sum_{p> 0}K_n(0,2p) F_n(2p,2j)&=0\\
    K_n(0,2j)-\bar{K}_n(0,0)\tilde{F}_n(0,2j) - \sum_{p> 0}\bar{K}_n(0,2p)\tilde{F}_n(2p,2j)&=0
    \end{split}
\end{equation}
These two equations suggest to define an operator product as 
\begin{equation}
\label{eq:def-product-discrete-operator}
    (\mathcal{O}_1\mathcal{O}_2)(i,j)=\sum_{p>0}\mathcal{O}_1(i,2p)\mathcal{O}_2(2p+j)
\end{equation}
and to define the left (resp. right) projectors to zero $\bra{\delta}$ (resp. $\ket{\delta}$), so that for any operator $\mathcal{O}$ with kernel $\mathcal{O}(i,j)$, we have
\begin{equation}
    (\bra{\delta}\mathcal{O})(j)=\mathcal{O}(0,j), \quad (\mathcal{O}\ket{\delta})(i)=\mathcal{O}(i,0), \quad \bra{\delta} \mathcal{O}\ket{\delta}=\mathcal{O}(0,0)
\end{equation}
Equipped with the operator product and the projector, we rewrite the diagonal contribution of the GLM equations, i.e. Eqs.~\eqref{eq:glm-diagonal1} and \eqref{eq:glm-diagonal2},  as
        \begin{equation}
        \label{eq:diagonal-GLM-equations}
        \begin{split}
\mathcal{I}_1 &= \bra{\delta}\bar{K}_n\ket{\delta} = \begin{pmatrix}
 1 \\
 0
\end{pmatrix}\\
\mathcal{I}_2  &= -\bra{\delta}K_n \ket{\delta}+\bra{\delta}\bar{K}_n\ket{\delta}\bra{\delta}\tilde{F}_n\ket{\delta}+\bra{\delta}\bar{K}_n\tilde{F}_n\ket{\delta}= \begin{pmatrix}
 0 \\
 -1
\end{pmatrix}
        \end{split}
    \end{equation}
where the diagonal value of the kernel of $K_n$ reads from Table~\ref{table:triangular-psi}
        \begin{equation}
    \bra{\delta} K_n \ket{\delta}=  \prod_{\ell=n}^{\infty}(1-\frac{g}{2}\tilde{z}_\ell z_{\ell+1}) \begin{pmatrix}
 -z_n \\
 1
\end{pmatrix}
    \end{equation}
The non-diagonal GLM equations \eqref{eq:GLM-kernel-indices} now read
\begin{equation}
\label{eq:glm-operator-valued-1}
    \bra{\delta}\bar{K}_n + \bra{\delta}K_n \ket{\delta}\bra{\delta} F_n + \bra{\delta}K_n F_n = 0
\end{equation}
and
\begin{equation}
\label{eq:glm-operator-valued-2}
    \bra{\delta}K_n-\bra{\delta}\bar{K}_n \ket{\delta}\bra{\delta} \tilde{F}_n - \bra{\delta}\bar{K}_n \tilde{F}_n = 0
\end{equation}

\subsection{Inversion of the GLM equations and explicit expressions of $z_n, \tilde{z}_n$}
We invert the GLM equations to obtain the kernels of $K_n$ and $\bar{K}_n$ solely in terms of the ones of $F_n$ and $\tilde{F}_n$. This in turn is used to obtain the explicit expressions of $z_n, \tilde{z}_n$ as a function of the operators $F_n$ and  $\tilde{F}_n$. 

\subsubsection{Inversion of the GLM equation}

In the first approach, we separate all the diagonals terms in \eqref{eq:glm-operator-valued-1} and \eqref{eq:glm-operator-valued-2} from the rest and respectively replace the values of $\bra{\delta} K_n$ or $\bra{\delta} \bar{K}_n$ in each equation. From this, one obtains

  \begin{equation}
  \label{eq:glm-operator-valued-inversed-1}
        \bra{\delta}\bar{K}_n=-\bra{\delta}K_n\ket{\delta} \bra{\delta}F_n\frac{1}{1+\tilde{F}_nF_n}-\bra{\delta}\bar{K}_n\ket{\delta}\bra{\delta}\frac{\tilde{F}_nF_n}{1+\tilde{F}_nF_n}
    \end{equation}

    \begin{equation}
    \label{eq:glm-operator-valued-inversed-2}
\bra{\delta}K_n=\bra{\delta}\bar{K}_n\ket{\delta} \bra{\delta}\tilde{F}_n\frac{1}{1+F_n\tilde{F}_n}-\bra{\delta}K_n\ket{\delta}\bra{\delta}\frac{F_n\tilde{F}_n}{1+F_n\tilde{F}_n}
    \end{equation}
    where in both equations the right index of the kernels have to be strictly greater than $n$, meaning that e.g. the operator $(\bra{\delta}K_n)(j)=K(n,n+j)$ can only be evaluated for $j>0$.\\

As a corollary, we can multiply Eq.~\eqref{eq:dual-inversion-1} to the right by $\tilde{F}_n$ to evaluate the following product, useful for the diagonal contribution of the GLM equations.

       \begin{equation}
       \label{eq:operator-product-1}
       \begin{split}
        \bra{\delta}\bar{K}_n\tilde{F}_n\ket{\delta}&=-\bra{\delta}K_n\ket{\delta} \bra{\delta}F_n\frac{1}{1+\tilde{F}_nF_n}\tilde{F}_n\ket{\delta}-\bra{\delta}\bar{K}_n\ket{\delta}\bra{\delta}\frac{\tilde{F}_nF_n}{1+\tilde{F}_nF_n}\tilde{F}_n\ket{\delta}\\
         &= -\bra{\delta}\bar{K}_n\ket{\delta}\bra{\delta}\tilde{F}_n\ket{\delta}-\bra{\delta}K_n\ket{\delta} \bra{\delta}\frac{F_n\tilde{F}_n}{1+F_n\tilde{F}_n}\ket{\delta}+\bra{\delta}\bar{K}_n\ket{\delta}\bra{\delta}\frac{1}{1+\tilde{F}_nF_n}\tilde{F}_n\ket{\delta}\\
        \end{split}
    \end{equation}

\subsubsection{Exact solution for $z_n$}\label{subsubsec:exact-sol-zn}

We will now prove the following result

\begin{equation}
\label{eq:exact-expression-zn}
    z_n  = -\bra{\delta}\tilde{F}_n\frac{1}{1+F_n\tilde{F}_n}\ket{\delta}
\end{equation}

To prove this, we will use Eqs.~\eqref{eq:diagonal-GLM-equations} and \eqref{eq:operator-product-1}, indeed

\begin{equation}
    \begin{split}
         \begin{pmatrix}
 0 \\
 -1
 \end{pmatrix}&=-\bra{\delta} K_n \ket{\delta}+\bra{\delta} \bar{K}_n\ket{\delta}\bra{\delta} \tilde{F}_n\ket{\delta}+\bra{\delta} \bar{K}_n\tilde{F}_n\ket{\delta}\\
 &=-\bra{\delta} K_n \ket{\delta}\left(1+\bra{\delta}\frac{F_n\tilde{F}_n}{1+F_n\tilde{F}_n}\ket{\delta}\right)+\bra{\delta}\bar{K}_n\ket{\delta}\bra{\delta}\frac{1}{1+\tilde{F}_nF_n}\tilde{F}_n\ket{\delta}\\
 &=- \prod_{\ell=n}^{\infty}(1-\tilde{z}_\ell z_{\ell+1}) \begin{pmatrix}
 -z_n \\
 1
\end{pmatrix}
\left(1+\bra{\delta}\frac{F_n\tilde{F}_n}{1+F_n\tilde{F}_n}\ket{\delta}\right)+
 \begin{pmatrix}
 1 \\
 0
\end{pmatrix}\bra{\delta}\frac{1}{1+\tilde{F}_nF_n}\tilde{F}_n\ket{\delta}\\
    \end{split}
\end{equation}

Isolating the second component of the above vector we find that

\begin{equation}
\label{eq:Fredholm-normalisation-1}
   \prod_{\ell=n}^{\infty}(1-\tilde{z}_\ell z_{\ell+1})   \left(1+\bra{\delta}\frac{F_n\tilde{F}_n}{1+F_n\tilde{F}_n}\ket{\delta}\right)=1
\end{equation}

Then isolating the first component of the above vector and using \eqref{eq:Fredholm-normalisation-1} we find the desired result which is

\begin{equation}
    z_n = -\bra{\delta}\tilde{F}_n\frac{1}{1+F_n\tilde{F}_n}\ket{\delta} 
\end{equation}

We present two other representations of the exact expression of $z_n$ which will be useful later on in Section~\ref{subsec:sanity-check-yo-system}. Anticipating the result of Eq.~\eqref{eq:finite-difference-discrete-kernel} stating that $F_{n+1}\tilde{F}_{n+1}= F_{n} \tilde{F}_{n}- F_{n+1}\ket{\delta}\bra{\delta} \tilde{F}_{n+1}$, we consider the finite difference of the resolvant and use the Sherman-Morrison formula

\begin{equation}
\label{eq:Shermann-Morrison-resolvant1}
\begin{split}
    \frac{1}{1+F_{n+1}\tilde{F}_{n+1}}    &=\frac{1}{1+F_{n} \tilde{F}_{n}}+\frac{1}{1-\bra{\delta}\tilde{F}_{n+1}\frac{1}{1+F_{n}\tilde{F}_{n}}{F}_{n+1}\ket{\delta}}\frac{1}{1+F_{n} \tilde{F}_{n}}F_{n+1}\ket{\delta}\bra{\delta} \tilde{F}_{n+1}\frac{1}{1+F_{n} \tilde{F}_{n}}\\ 
\end{split}
\end{equation}

We multiply \eqref{eq:Shermann-Morrison-resolvant1} by $\tilde{F}_{n+1}$ to the left and by $\ket{\delta}$ to the right and obtain

\begin{equation}
\begin{split}
    \bra{\delta}\tilde{F}_{n+1}\frac{1}{1+F_{n}\tilde{F}_{n}} \ket{\delta}   
           & =(1-\bra{\delta} \tilde{F}_{n+1}\frac{1}{1+ F_n \tilde{F}_n}F_{n+1}\ket{\delta})\bra{\delta}\tilde{F}_{n+1}\frac{1}{1+F_{n+1} \tilde{F}_{n+1}}\ket{\delta}   \\ 
\end{split}
\end{equation}

This leads to the second expression for $z_n$
\begin{equation}
\label{eq:exact-expression-zn-2e}
    z_n = - \frac{\bra{\delta}\tilde{F}_{n}\frac{1}{1+F_{n-1}\tilde{F}_{n-1}} \ket{\delta}   }{1-\bra{\delta} \tilde{F}_{n}\frac{1}{1+ F_{n-1} \tilde{F}_{n-1}}F_{n}\ket{\delta}}
\end{equation}

A third expression can be obtained using that $ F_{n} \tilde{F}_{n}=F_{n+1}\tilde{F}_{n+1}+ F_{n+1}\ket{\delta}\bra{\delta} \tilde{F}_{n+1}$ and the Sherman-Morrison formula

\begin{equation}
\label{eq:Shermann-Morrison-resolvant2}
\begin{split}
    \frac{1}{1+F_{n}\tilde{F}_{n}}    &=\frac{1}{1+F_{n+1} \tilde{F}_{n+1}}-\frac{1}{1+\bra{\delta}\tilde{F}_{n+1}\frac{1}{1+F_{n+1}\tilde{F}_{n+1}}{F}_{n+1}\ket{\delta}}\frac{1}{1+F_{n+1} \tilde{F}_{n+1}}F_{n+1}\ket{\delta}\bra{\delta} \tilde{F}_{n+1}\frac{1}{1+F_{n+1} \tilde{F}_{n+1}}\\ 
\end{split}
\end{equation}

We multiply \eqref{eq:Shermann-Morrison-resolvant2} by $\tilde{F}_{n+1}$ to the left and by $\ket{\delta}$ to the right and obtain

\begin{equation}
\begin{split}
    \bra{\delta}\tilde{F}_{n+1}\frac{1}{1+F_{n}\tilde{F}_{n}} \ket{\delta}   
      & =\frac{\bra{\delta}\tilde{F}_{n+1}\frac{1}{1+F_{n+1} \tilde{F}_{n+1}}\ket{\delta}}{1+\bra{\delta}\tilde{F}_{n+1}\frac{1}{1+F_{n+1}\tilde{F}_{n+1}}{F}_{n+1}\ket{\delta}}    \\ 
\end{split}
\end{equation}
This leads to the third expression for $z_n$

    \begin{equation}
    \label{eq:exact-expression-zn-3e}
         z_n = -\bra{\delta}\tilde{F}_n\frac{1}{1+F_{n-1}\tilde{F}_{n-1}}\ket{\delta} (1+\bra{\delta}\tilde{F}_{n}\frac{1}{1+F_{n}\tilde{F}_{n}}{F}_{n}\ket{\delta})
    \end{equation}

\subsubsection{Exact solution for $\tilde{z}_n$}

We continue by taking the right index of Eq.~\eqref{eq:glm-operator-valued-inversed-1} equal to $m=n+2$.

\begin{equation}
    \begin{split}
        \begin{pmatrix}
 -\sum_{\ell=n}^{\infty}z_\ell \tilde{z}_{\ell}  \\
\tilde{z}_n
\end{pmatrix}&=- \prod_{\ell=n}^{\infty}(1-\tilde{z}_\ell z_{\ell+1}) 
    \begin{pmatrix}
 -z_n \\
 1
    \end{pmatrix}
\bra{\delta_n}\frac{1}{1+F_n\tilde{F}_n}F_{n+1}\ket{\delta}-
    \begin{pmatrix}
 1 \\
 0
    \end{pmatrix}
\bra{\delta}\tilde{F}_n\frac{1}{1+F_n\tilde{F}_n}F_{n+1}\ket{\delta}
    \end{split}
\end{equation}
Isolating the second coefficient, we obtain

\begin{equation}
\label{eq:exact-expression-ztilden}
    \begin{split}
        \tilde{z}_n=-\frac{\bra{\delta}\frac{1}{1+F_n\tilde{F}_n}F_{n+1}\ket{\delta}}{1+\bra{\delta}\frac{F_n\tilde{F}_n}{1+F_n\tilde{F}_n}\ket{\delta}}
    \end{split}
\end{equation}
which we have not been able to further simplify. Isolating the first coefficient, we obtain

\begin{equation}
    \begin{split}
   -\sum_{\ell=n+1}^{\infty}z_\ell \tilde{z}_{\ell}     &= \bra{\delta}\tilde{F}_n\frac{1}{1+F_n\tilde{F}_n}F_{n+1}\ket{\delta} \\ 
    \end{split}
\end{equation}
which is nontrivial from the exact expressions of $z_n, \tilde{z}_n$ obtained in Eqs.~\eqref{eq:exact-expression-zn}--\eqref{eq:exact-expression-ztilden}.  As a consequence we have that
\begin{equation}
\label{eq:optimal-density-fredholm}
\begin{split}
   - z_n \tilde{z}_{n}  &=  \bra{\delta}\tilde{F}_n\frac{1}{1+F_n\tilde{F}_n}F_{n+1}\ket{\delta}- \bra{\delta}\tilde{F}_{n-1}\frac{1}{1+F_{n-1}\tilde{F}_{n-1}}F_{n}\ket{\delta}\\
   &=  \Delta^+ \bra{\delta}\tilde{F}_{n-1}\frac{1}{1+F_{n-1}\tilde{F}_{n-1}}F_{n}\ket{\delta}\\
\end{split}
\end{equation}
where $\Delta^+ f_n=f_{n+1}-f_n$ is the finite difference operator.

\subsubsection{Additional identities}
If we combine Eqs.~\eqref{eq:exact-expression-ztilden} and \eqref{eq:exact-expression-zn-3e}, we further obtain that 
    \begin{equation}
        z_n\tilde{z}_n = \bra{\delta}\frac{1}{1+F_n\tilde{F}_n}F_{n+1}\ket{\delta}\bra{\delta}\tilde{F}_n\frac{1}{1+F_{n-1}\tilde{F}_{n-1}}\ket{\delta} 
    \end{equation}
which is highly non-trivial from Eq.~\eqref{eq:optimal-density-fredholm}. Furthermore, Eqs.~\eqref{eq:exact-expression-zn}--\eqref{eq:exact-expression-ztilden} bring the expressions for $z_n$ and $\tilde{z}_n$ on a more symmetric setting, indeed

        \begin{equation}
    \begin{split}
        \tilde{z}_n&=-\frac{\bra{\delta}\frac{1}{1+F_n\tilde{F}_n}F_{n+1}\ket{\delta}}{1+\bra{\delta}\frac{F_n\tilde{F}_n}{1+F_n\tilde{F}_n}\ket{\delta}}\\
         z_n&=- \bra{\delta}\tilde{F}_n\frac{1}{1+F_{n-1}\tilde{F}_{n-1}}\ket{\delta} (1+\bra{\delta}\frac{F_n\tilde{F}_n}{1+F_n\tilde{F}_n}\ket{\delta})
    \end{split}
\end{equation}

\subsection{Check of the time derivative of $z_n$}
\label{subsec:sanity-check-yo-system}

In this Section, we verify by an algebraic method that $z_n$ indeed verifies the system of Eqs.~\eqref{eq:YO-polymer-saddle}. To this aim, we require three prelimiary results. From Equations~\eqref{eq:time-deriv-fourier-reflec-coeff}, we recall the time derivatives of the Fourier operators
        \begin{equation}
    \begin{split}
        \p_t F_n &= F_n-F_{n+1}\\
        \p_t \tilde{F}_n &= \tilde{F}_{n-1}-\tilde{F}_n\\
    \end{split}
    \end{equation}
    
   In particular, the combination of the two for any $n,m \in \Z$ leads to
    \begin{equation}
        \begin{split}
       \p_t (F_n \tilde{F}_m) &= (F_n-F_{n+1})\tilde{F}_m +F_n (\tilde{F}_{m-1}-\tilde{F}_m)\\
       & =F_n \tilde{F}_{m-1} - F_{n+1}\tilde{F}_m\\
       &=F_{n+1}\ket{\delta}\bra{\delta}\tilde{F}_m
       \end{split}
    \end{equation}
which is a rank-one operator. Furthermore, from the definition of the product of the operators $F_n$ and $\tilde{F}_n$ from Eq.~\eqref{eq:def-product-discrete-operator}, we obtain the following relation for the finite difference 

\begin{equation}
\label{eq:finite-difference-discrete-kernel}
    F_n \tilde{F}_n = F_{n+1} \tilde{F}_{n+1}+ F_{n+1}\ket{\delta}\bra{\delta} \tilde{F}_{n+1} \Longrightarrow \Delta^+ F_n \tilde{F}_n = - F_{n+1}\ket{\delta}\bra{\delta} \tilde{F}_{n+1} 
\end{equation}
where $\Delta^+ f_n=f_{n+1}-f_n$ so that the finite difference is a rank-one operator. More generally this relation remains valid for different indices

\begin{equation}
    F_{n} \tilde{F}_m = F_{n+1} \tilde{F}_{m+1}+ F_{n+1}\ket{\delta}\bra{\delta} \tilde{F}_{m+1} \Longrightarrow \Delta^+ F_n \tilde{F}_m = - F_{n+1}\ket{\delta}\bra{\delta} \tilde{F}_{m+1} 
\end{equation}
 The following derivative of the resolvant appearing in the expression for $z_n$ is obtained as follows
\begin{equation}
\begin{split}
    \p_t \frac{1}{1+F_n \tilde{F}_n} &= -\frac{1}{1+F_n \tilde{F}_n}\p_t (F_n \tilde{F}_n) \frac{1}{1+F_n \tilde{F}_n}\\
    &= -\frac{1}{1+F_n \tilde{F}_n}F_{n+1}\ket{\delta}\bra{\delta}\tilde{F}_n \frac{1}{1+F_n \tilde{F}_n}\\
\end{split}
\end{equation}

Equipped with these results, we calculate the time derivative of $z_n$

\begin{equation}
    \begin{split}   
    \p_t z_n &=  -  \p_t\bra{\delta}\tilde{F}_n\frac{1}{1+F_n\tilde{F}_n}\ket{\delta}\\
    &=  -  \bra{\delta}\p_t\tilde{F}_n\frac{1}{1+F_n\tilde{F}_n}\ket{\delta}-\bra{\delta}\tilde{F}_n\p_t\frac{1}{1+F_n\tilde{F}_n}\ket{\delta}\\
     &= -z_n -   \bra{\delta}\tilde{F}_{n-1}\frac{1}{1+F_n\tilde{F}_n}\ket{\delta}+\bra{\delta}\tilde{F}_n\frac{1}{1+F_n\tilde{F}_n}F_{n+1}\ket{\delta}\bra{\delta}\tilde{F}_n\frac{1}{1+F_n\tilde{F}_n}\ket{\delta}\\
     &= -z_n -   \bra{\delta}\tilde{F}_{n-1}\frac{1}{1+F_n\tilde{F}_n}\ket{\delta}-\bra{\delta}\tilde{F}_n\frac{1}{1+F_n\tilde{F}_n}F_{n+1}\ket{\delta}z_n\\
    \end{split}
\end{equation}
Using that $F_n\tilde{F}_n=F_{n-1}\tilde{F}_{n-1}-F_n\ket{\delta}\bra{\delta}\tilde{F}_n$ alongside the Sherman-Morrison formula on the second term of the right hand side, we obtain
\begin{equation}
    \begin{split}   
    \p_t z_n        &= -z_n -   \bra{\delta}\tilde{F}_{n-1}\frac{1}{1+F_{n-1}\tilde{F}_{n-1}-F_n\ket{\delta}\bra{\delta}\tilde{F}_n}\ket{\delta}-\bra{\delta}\tilde{F}_n\frac{1}{1+F_n\tilde{F}_n}F_{n+1}\ket{\delta}z_n\\
              &= -z_n +z_{n-1}-\frac{\bra{\delta}\tilde{F}_{n-1}\frac{1}{1+F_{n-1}\tilde{F}_{n-1}}F_n\ket{\delta}\bra{\delta}\tilde{F}_n\frac{1}{1+F_{n-1}\tilde{F}_{n-1}}\ket{\delta}}{1-\bra{\delta}\tilde{F}_n\frac{1}{1+F_{n-1}\tilde{F}_{n-1}}F_n\ket{\delta}}   -\bra{\delta}\tilde{F}_n\frac{1}{1+F_n\tilde{F}_n}F_{n+1}\ket{\delta}z_n\\
    \end{split}
\end{equation}

Using Eq.~\eqref{eq:exact-expression-zn-2e}, we recognise in the right hand side the expression of $z_n$ and we use Eq.~\eqref{eq:optimal-density-fredholm} to express the remaining difference as the product $z_n \tilde{z}_n$ and obtain

\begin{equation}
    \begin{split}   
    \p_t z_n          &= -z_n +z_{n-1}+\left(\bra{\delta}\tilde{F}_{n-1}\frac{1}{1+F_{n-1}\tilde{F}_{n-1}}F_n\ket{\delta}-\bra{\delta}\tilde{F}_n\frac{1}{1+F_n\tilde{F}_n}F_{n+1}\ket{\delta}\right)z_n\\
      &= -z_n +z_{n-1}+z_n^2\tilde{z}_n\\
    \end{split}
\end{equation}

\subsection{Discrete Fredholm framework for the solution of the non-linear system}

The operator manipulations done so far can be related to a more fundamental object which is the Fredholm determinant of the $F_n\tilde{F}_n$ defined as
\begin{equation}
    D_n=\Det(I+F_n\tilde{F}_n)_{\ell^2(\mathbb{N}^*)}
\end{equation}
and we expect the operators to live to act on $\ell^2(0,\infty)$. Let us develop some algebraic framework for such object similarly to what was done in Refs.~\cite{krajenbrink2020painleve,bothner2022riemann} for the Fredholm determinant of products of continuous Hankel operators.\\

The finite difference relation \eqref{eq:finite-difference-discrete-kernel} implies a multiplicative recursion for the Fredholm determinant $D_n$. Using the matrix determinant lemma, we have that

    \begin{equation}
    \begin{split}
    D_{n+1}=\Det(I+F_{n+1}\tilde{F}_{n+1})&=\Det(I+ F_n \tilde{F}_n-F_{n+1}\ket{\delta}\bra{\delta} \tilde{F}_{n+1})\\
    &=\Det(I+ F_n \tilde{F}_n)(1-\bra{\delta} \tilde{F}_{n+1}\frac{I}{I+ F_n \tilde{F}_n}F_{n+1}\ket{\delta})\\
    &=D_n\left(1-\bra{\delta} \tilde{F}_{n+1}\frac{I}{I+ F_n \tilde{F}_n}F_{n+1}\ket{\delta}\right)
    \end{split}
\end{equation}

This recursion can also be written backwards
\begin{equation}
    \begin{split}
    D_n=\Det(I+ F_n \tilde{F}_n)&=\Det(I+F_{n+1}\tilde{F}_{n+1}+F_{n+1}\ket{\delta}\bra{\delta} \tilde{F}_{n+1})\\
    &=\Det(I+F_{n+1}\tilde{F}_{n+1})(1+\bra{\delta} \tilde{F}_{n+1}\frac{I}{I+ F_{n+1} \tilde{F}_{n+1}}F_{n+1}\ket{\delta})\\
    &=D_{n+1}\left(1+\bra{\delta} \tilde{F}_{n+1}\frac{I}{I+ F_{n+1} \tilde{F}_{n+1}}F_{n+1}\ket{\delta}\right)
    \end{split}
\end{equation}
which implies the identity

\begin{equation}
   \left(1+\bra{\delta} \tilde{F}_{n+1}\frac{I}{I+ F_{n+1} \tilde{F}_{n+1}}F_{n+1}\ket{\delta}\right)\left(1-\bra{\delta} \tilde{F}_{n+1}\frac{I}{I+ F_n \tilde{F}_n}F_{n+1}\ket{\delta}\right) =1
\end{equation}

The multiplicative recursion between implies that a suitable object to study is actually the logarithm of the Fredholm determinant and thus we obtain that
\begin{equation}
\begin{split}
    \Delta^+ \log D_n  &= \log \left(1-\bra{\delta} \tilde{F}_{n+1}\frac{I}{I+ F_n \tilde{F}_n}F_{n+1}\ket{\delta}\right) \\
    &= -\log \left(1+\bra{\delta} \tilde{F}_{n+1}\frac{I}{I+ F_{n+1} \tilde{F}_{n+1}}F_{n+1}\ket{\delta}\right)
\end{split}
\end{equation}

We also have from Eq.~\eqref{eq:Fredholm-normalisation-1} the relation between the logarithmic finite-difference of the Fredholm determinant and the variables $\{ z_n, \tilde{z}_n \}$ of our problem.
\begin{equation}
    \log\frac{ D_{n+1}}{ D_n }=\Delta^+ \log D_n =\sum_{\ell=n+1}^{\infty} \log (1-\tilde{z}_\ell z_{\ell+1})
\end{equation}
which implies that the logarithmic second-order finite-difference of the Fredholm determinant reads
\begin{equation}
    \begin{split}
    (\Delta^+)^2 \log D_n  &=\Delta^+\sum_{\ell=n+1}^{\infty} \log (1-\tilde{z}_\ell z_{\ell+1})\\
    &=-\log (1-\tilde{z}_{n+1} z_{n+2})\\
    \end{split}
\end{equation}
Exponentiating this identity, we obtain
\begin{equation}
    \frac{D_{n+1}^2}{D_{n+2}D_{n}} = 1-\tilde{z}_{n+1} z_{n+2}
\end{equation}

This identity is akin to one obtained in the continuous setting of the weak noise theory of KPZ \cite[Eq.~(13)]{UsWNT2021}. We can expect as in Refs.~\cite{krajenbrink2020painleve,bothner2022riemann} that a hierarchy of functions can be constructed which will verify a differential recursion. This precise recursion is left for future work. The first step towards this construction would be to study the finite difference of the resolvant of $F_n \tilde{F}_n$, the starting formula were presented in Section~\ref{subsubsec:exact-sol-zn} and we recall them here for completeness.

\begin{equation}
\begin{split}
    \frac{1}{1+F_{n}\tilde{F}_{n}}&=\frac{1}{1+F_{n+1} \tilde{F}_{n+1}+ F_{n+1}\ket{\delta}\bra{\delta} \tilde{F}_{n+1}} \\ 
    &=\frac{1}{1+F_{n+1} \tilde{F}_{n+1}}-\frac{1}{1+\bra{\delta}\tilde{F}_{n+1}\frac{1}{1+F_{n+1}\tilde{F}_{n+1}}{F}_{n+1}\ket{\delta}}\frac{1}{1+F_{n+1} \tilde{F}_{n+1}}F_{n+1}\ket{\delta}\bra{\delta} \tilde{F}_{n+1}\frac{1}{1+F_{n+1} \tilde{F}_{n+1}}\\ 
\end{split}
\end{equation}
and

\begin{equation}
\begin{split}
    \frac{1}{1+F_{n+1}\tilde{F}_{n+1}}&=\frac{1}{1+F_{n} \tilde{F}_{n}- F_{n+1}\ket{\delta}\bra{\delta} \tilde{F}_{n+1}} \\ 
    &=\frac{1}{1+F_{n} \tilde{F}_{n}}+\frac{1}{1-\bra{\delta}\tilde{F}_{n+1}\frac{1}{1+F_{n}\tilde{F}_{n}}{F}_{n+1}\ket{\delta}}\frac{1}{1+F_{n} \tilde{F}_{n}}F_{n+1}\ket{\delta}\bra{\delta} \tilde{F}_{n+1}\frac{1}{1+F_{n} \tilde{F}_{n}}\\ 
\end{split}
\end{equation}

\section{Limit to the classical Toda system}

The weak noise theory of the O'Connell-Yor polymer can be seen as a generalisation of the classical Toda system \cite{toda1967vibration}. This can be seen from the Lax pair of the model as in Ref.~\cite{kuznetsov2000quantum} but also directly from the dynamical system \eqref{eq:YO-polymer-saddle}. To see this, consider the following Cole-Hopf type change of variable
\begin{equation}
\begin{split}
    z_n(t)&=\alpha e^{h_n(t)+ \alpha^2 t}\\
    \tilde{z}_n(t) &= e^{-h_n(t)- \alpha^2 t}(\alpha +p_n(t))
\end{split}    
\end{equation}
Note that the Jacobian of this change of variable is a constant. Then, the dynamics verified by the two fields $\{ h_n, p_n \}$ is
\begin{equation}
\begin{split}
    \p_t h_n &= -1+ e^{h_{n-1}-h_n}+ \alpha p_n + a_n \\
    \p_t p_n & = e^{h_{n-1}-h_n}(\alpha+p_n)-e^{h_{n}-h_{n+1}}(\alpha+p_{n+1})
\end{split}
\end{equation}
Rescaling $t=\frac{\tau}{\alpha}$, and $a_n= \alpha \hat a_n$, the action $S_0$ of the O'Connell-Yor polymer 
\eqref{dynOY} 
reads in terms of the new variables 

\bea  
  &&  S_0=S[h,p] = \int \rmd \tau \sum_{n =1}^{N}\bigg[ \alpha  \left(1-e^{h_{n-1}(\tau )-h_n(\tau
   )}-\frac{1}{2}  p_n(\tau )^2+ p_n(\tau ) \p_\tau h_n(\tau )- \hat a_n p_n(\tau) \right) \\
   && +\alpha ^2 \p_\tau h_n(\tau )+\left(1-e^{h_{n-1}(\tau )-h_n(\tau )}\right) p_n(\tau
   )+\frac{\alpha ^3 }{2} - \alpha^2 \hat a_n \bigg]
\eea  
Note that the terms proportional to $\alpha^2$ and $\alpha^3$ are respectively is a total derivative
and a constant, hence they do not contribute to the equations of motion. 
Taking $\alpha\to \infty$ we find the classical Toda action, and the associated equations of motion (i.e. saddle point equations) 
\begin{equation}
\begin{split}
    \p_\tau h_n &= p_n + \hat a_n \\
    \p_\tau p_n & = e^{h_{n-1}-h_n}-e^{h_{n}-h_{n+1}}
\end{split}
\end{equation}

This result brings a few comments:
\begin{enumerate}
    \item Since we have solved the weak noise theory of the O'Connell-Yor polymer using the discrete Fredholm determinant framework in Section~\ref{sec:GLM-eq-solution}, our result is also extended for the Toda lattice.
    \item Note that our Lax pair formulation has allowed us to include drifts $\hat a_n$ in the Toda system. It would be interesting
    to understand their role in the dynamics.
    \item Finally, the Toda system has been related historically to the QR decomposition in linear algebra, see e.g. Refs.~\cite{symes1982qr,deift1983ordinary}. It would be extremely interesting to relate the O'Connell-Yor dynamics  \eqref{eq:YO-polymer-saddle} as an extension of the flow of the QR decomposition. Furthermore, as we expect the existence of an integrable time discretisation of the O'Connell-Yor dynamics and thus of the Toda model, understanding the discrete flow could also shed some new light on linear algebra algorithms. We leave these questions open for a future work. Furthermore, the interpretation of the drifts $a_n$ in these algorithms have to be understood. 
\end{enumerate}

\section{Lax pair duality}
The Lax pair we have studied so far involved a two-dimensional vector with two complementary fields. In this section for notation reasons, we will denote these components as
\begin{equation}
    \vec{v}_n=(\Phi_n,\Psi_n)^\intercal
\end{equation}
We will show that the original $2\times 2$ Lax pair is dual to an infinite Lax pair, using the method of Ref.~\cite{sklyanin2000backlund,kuznetsov2000quantum}. To this aim, consider the following infinite vectors
\begin{equation}
    \vec{\Phi}= 
    \begin{pmatrix}
        \Phi_{-\infty}\\ 
        \vdots \\
        \Phi_\ell \\
        \vdots \\
        \Phi_{+\infty}
    \end{pmatrix}, \quad \vec{\Psi}= 
    \begin{pmatrix}
        \Psi_{-\infty}\\ 
        \vdots \\
        \Psi_\ell \\
        \vdots \\
        \Psi_{+\infty}
    \end{pmatrix}
\end{equation}
Our aim will be to show the existence of two matrices $\mathcal{L}_+$ and $\mathcal{L}_-$ such that
\begin{equation}
\label{eq:dual-lax-spectral}
    \mathcal{L}_+\vec{\Psi}=\lambda^2 \vec{\Psi}, \quad    \mathcal{L}_-\vec{\Psi}=\lambda^2 \vec{\Psi}
\end{equation}
and the existence of two matrices $\mathcal{M}_+$ and $\mathcal{M}_-$ ensuring the Lax evolutions
\begin{equation}
\label{eq:infinite-lax-formulation}
    \p_t\mathcal{L}_++[\mathcal{L}_+,\mathcal{M}_+ ]=0, \quad \p_t\mathcal{L}_-+[\mathcal{L}_-,\mathcal{M}_- ]=0 \, .
\end{equation}
This Lax formulation has been used recently in the context of the Toda lattice, the Ablowitz-Ladik model and other models to study the relation between Generalised Gibbs Ensembles and Random Matrix Theory \cite{grava2023discrete,mazzuca2023equilibrium,spohn2020generalized,spohn2021hydrodynamic,spohn2022hydrodynamic}. The derivation of the Lax formulation \eqref{eq:infinite-lax-formulation} from a $2\times 2$ Lax matrix problem will follow the lines of Ref.~\cite{kuznetsov2000quantum}. \\

To start, we re-write the time equation of the Lax pair, i.e. $\p_t v_n=U_n v_n$ as
\begin{equation}
\label{eq:vec-lax-time-eq}
\begin{split}
    \p_t \vec{\Phi} &= \frac{\lambda^2-1}{2}\vec{\Phi}+\mathrm{Diag}(-z_{n-1}) \vec{\Psi}\\
     \p_t \vec{\Psi} &= \mathrm{Diag}(\tilde{z}_{n}) \vec{\Phi}-\frac{\lambda^2-1}{2}\vec{\Psi}\\
\end{split}
\end{equation}
where $\mathrm{Diag}(\tilde{z}_{n}) $ is an infinite diagonal matrix with $(n,n)$ element equal to $\tilde{z}_{n}$ and $\mathrm{Diag}(-z_{n-1})$ has its $(n,n)$ element equal to $-z_{n-1}$.\\

Concerning the space equation of the Lax pair, i.e. $v_{n+1}=L_n v_n$, we first rescale the components 
\begin{equation}
     \vec{\Phi} \to  \mathrm{Diag}(\lambda^{-n})\vec{\Phi}, \quad \vec{\Psi} \to  \mathrm{Diag}(\lambda^{-n})\vec{\Psi}
\end{equation}
the space Lax equation component by component is then
\begin{equation}
\label{eq:Lax-Duality1}
    \begin{split}
        \Phi_{n+1}&=\Phi_n+z_n \Psi_n\\
        \Psi_{n+1}&=-\tilde{z}_n\Phi_n+(\lambda^2-z_n\tilde{z}_n) \Psi_n\\
    \end{split}
\end{equation}
So to define the spectral problem \eqref{eq:dual-lax-spectral} for $\vec{\Psi}$ we will eliminate $\Phi_n$ in the second equation of \eqref{eq:Lax-Duality1}. We have two possibilities
\begin{enumerate}
    \item we can either invert the first equation of Eq.~\eqref{eq:Lax-Duality1} as
\begin{equation}
\label{eq:dual-inversion-1}
    \Phi_n=\Phi_{-\infty}+\sum_{\ell=-\infty}^{n-1} z_\ell \Psi_{\ell}
\end{equation}
in which case we obtain that
\begin{equation}
\label{eq:dual-inversion-11}
   \Psi_{n+1}=-\tilde{z}_n\sum_{\ell=-\infty}^{n} z_\ell \Psi_{\ell}-\tilde{z}_n\Phi_{-\infty}+\lambda^2 \Psi_n 
\end{equation}
Given the asymptotic data of the scattering problem, see Eqs.~\eqref{bc01}--\eqref{bc02}, we can choose $\Phi_{\pm\infty}=0$ and rewrite \eqref{eq:dual-inversion-11} in matrix form as $\mathcal{L}_+\vec{\Psi}=\lambda^2 \vec{\Psi}$, where the matrix $\mathcal{L}_+$ reads

\begin{equation}
\label{eq:def-L+}
\mathcal{L}_+= \mathrm{Diag}(\tilde{z}_n)
    \begin{pmatrix}
1 & 0 &  &  &  &  \\
 & \ddots & \ddots &  & \boxed{\mathbf{0}} &  \\
 &  & 1 & 0 &  &  \\
 &  &  & 1 & \ddots &  \\
 & \boxed{\mathbf{1}} &  &  & \ddots & \\
 &  &  &  &  & 1
\end{pmatrix}\mathrm{Diag}(z_n)
+
\begin{pmatrix}
0 & 1 &  &  &  &  \\
 & \ddots & \ddots &  & \boxed{\mathbf{0}} &  \\
 &  & 0 & 1 &  &  \\
 &  &  & 0 & \ddots &  \\
 & \boxed{\mathbf{0}} &  &  & \ddots & 1\\
 &  &  &  &  & 0
\end{pmatrix}
\end{equation}
The last matrix in Eq.~\eqref{eq:def-L+} has the first upper band filled with $1$ and zeroes everywhere else. The intermediate matrix between the two diagonal matrices is lower-triangular filled with $1$ on the diagonal and everywhere below.\\

Let us look at the compatibility equation now by taking the time derivative of Eq.~\eqref{eq:dual-lax-spectral}
\begin{equation}
\begin{split}
        \p_t\mathcal{L}_+\vec{\Psi}+(\mathcal{L}_+-\lambda^2)\p_t \vec{\Psi}&=    \p_t\mathcal{L}_+\vec{\Psi}+(\mathcal{L}_+-\lambda^2)\mathrm{Diag}(\tilde{z}_{n}) \vec{\Phi}\\
        &=0
    \end{split}
\end{equation}
where we have used the second line of Eq.~\eqref{eq:vec-lax-time-eq} together with Eq.~\eqref{eq:dual-lax-spectral}.
We now need to use the inversion between $\vec{\Phi}$ and $\vec{\Psi}$ from Eq.~\eqref{eq:dual-inversion-1} to obtain

\begin{equation}
    \vec{\Phi}=
  \begin{pmatrix}
0 &  &  &  &  &  \\
 1& \ddots &  &  & \boxed{\mathbf{0}} &  \\
 & \ddots & 0 &  &  &  \\
 &  & 1 & 0 &  &  \\
 & \boxed{\mathbf{1}} &  & \ddots & \ddots & \\
 &  &  &  & 1 & 0 
\end{pmatrix}
  \mathrm{Diag}(z_{n})  \vec{\Psi}:= B \, \mathrm{Diag}(z_{n})  \vec{\Psi}
\end{equation}
where the scalar matrix $B$ is a lower-triangular matrix filled with $0$ on the diagonal and $1$ everywhere below. 
\begin{equation}
\begin{split}
        \p_t\mathcal{L}_+\vec{\Psi}+(\mathcal{L}_+-\lambda^2)\p_t \vec{\Psi}&=    \p_t\mathcal{L}_+\vec{\Psi}+(\mathcal{L}_+-\lambda^2)\mathrm{Diag}(\tilde{z}_{n})  B \, \mathrm{Diag}(z_{n})  \vec{\Psi}\\
        &=\left[\p_t\mathcal{L}_++[\mathcal{L}_+,\mathrm{Diag}(\tilde{z}_{n})  B \, \mathrm{Diag}(z_{n}) ] \right]\vec{\Psi}\\
        &=0
    \end{split}
\end{equation}
where we obtained the commutator was obtained using again Eq.~\eqref{eq:dual-lax-spectral}. This leads to the definition of the matrix $\mathcal{M}_+$ as
\begin{equation}
    \mathcal{M}_+ = \mathrm{Diag}(\tilde{z}_{n})  B \, \mathrm{Diag}(z_{n})
\end{equation}
\item or we invert the first equation of Eq.~\eqref{eq:Lax-Duality1} as
\begin{equation}
\label{eq:dual-inversion-2}
    \Phi_n=\Phi_{+\infty}-\sum_{\ell=n}^{+\infty} z_\ell \Psi_{\ell}
\end{equation}
in which case we obtain
\begin{equation}
     \Psi_{n+1}=\tilde{z}_n\sum_{\ell=n+1}^{+\infty} z_\ell \Psi_{\ell}-\tilde{z}_n\Phi_{+\infty}+\lambda^2 \Psi_n
\end{equation}

Suppose $\Phi_{+\infty}=0$, we then have the matrix representation $\mathcal{L}_-\vec{\Psi}=\lambda^2 \vec{\Psi}$ where the $\mathcal{L}_-$ reads

\begin{equation}
\label{eq:def-L-}
\mathcal{L}_-= -\mathrm{Diag}(\tilde{z}_n)
    \begin{pmatrix}
0 & 1 &  &  &  &  \\
 & \ddots & \ddots &  & \boxed{\mathbf{1}} &  \\
 &  & 0 & 1 &  &  \\
 &  &  & 0 & \ddots &  \\
 & \boxed{\mathbf{0}} &  &  & \ddots & 1\\
 &  &  &  &  & 0 
\end{pmatrix}
\mathrm{Diag}(z_n)
+
\begin{pmatrix}
0 & 1 &  &  &  &  \\
 & \ddots & \ddots &  & \boxed{\mathbf{0}} &  \\
 &  & 0 & 1 &  &  \\
 &  &  & 0 & \ddots &  \\
 & \boxed{\mathbf{0}} &  &  & \ddots & 1\\
 &  &  &  &  & 0
\end{pmatrix}
\end{equation}
The last matrix in \eqref{eq:def-L-} is upper-triangular with the first band filled with $1$ and the rest is $0$. The intermediate matrix between the two diagonal matrices is upper-triangular filled with $0$ on the diagonal and $1$ everywhere above. Let us look at the compatibility equation now by taking the time derivative of Eq.~\eqref{eq:dual-lax-spectral}
\begin{equation}
\begin{split}
        \p_t\mathcal{L}_-\vec{\Psi}+(\mathcal{L}_--\lambda^2)\p_t \vec{\Psi}&=    \p_t\mathcal{L}_-\vec{\Psi}+(\mathcal{L}_--\lambda^2)\mathrm{Diag}(\tilde{z}_{n}) \vec{\Phi}\\
        &=0
    \end{split}
\end{equation}
We now need to use the inversion between $\vec{\Phi}$ and $\vec{\Psi}$ from Eq.~\eqref{eq:dual-inversion-2} to obtain

\begin{equation}
    \vec{\Phi}=
 -     \begin{pmatrix}
1 &  &  &  &  &  \\
 0& \ddots &  &  & \boxed{\mathbf{1}} &  \\
 & \ddots & 1 &  &  &  \\
 &  & 0 & 1 &  &  \\
 & \boxed{\mathbf{0}} &  &\ddots  & \ddots & \\
 &  &  &  & 0 & 1 
\end{pmatrix}
  \mathrm{Diag}(z_{n})  \vec{\Psi}:= C \, \mathrm{Diag}(z_{n})  \vec{\Psi}
\end{equation}
where the scalar matrix $C$ is upper-triangular filled with $1$ on the diagonal and everywhere above. It leads to
\begin{equation}
\begin{split}
        \p_t\mathcal{L}_-\vec{\Psi}+(\mathcal{L}_--\lambda^2)\p_t \vec{\Psi}&=    \p_t\mathcal{L}_-\vec{\Psi}+(\mathcal{L}_--\lambda^2)\mathrm{Diag}(\tilde{z}_{n})  C \, \mathrm{Diag}(z_{n})  \vec{\Psi}\\
        &=\left[\p_t\mathcal{L}_-+[\mathcal{L}_-,\mathrm{Diag}(\tilde{z}_{n})  C \, \mathrm{Diag}(z_{n}) ] \right]\vec{\Psi}\\
        &=0
    \end{split}
\end{equation}
from which we deduce the definition of the matrix $\mathcal{M}_-$
\begin{equation}
    \mathcal{M}_- = \mathrm{Diag}(\tilde{z}_{n})  C \, \mathrm{Diag}(z_{n})
\end{equation}
\end{enumerate}

\subsubsection{Comment on the conserved quantities from the Lax duality}

Since $\mathcal{L}_+$ and $\mathcal{L}_-$ verify the Lax equations

\begin{equation}
    \p_t\mathcal{L}_++[\mathcal{L}_+,\mathcal{M}_+ ]=0, \quad \p_t\mathcal{L}_-+[\mathcal{L}_-,\mathcal{M}_- ]=0 \, .
\end{equation}

It follows that their traces are conserved quantities, i.e. for any $n \in \N$
\begin{equation}
    \p_t \Tr (\mathcal{L}_+^n)=0, \quad \p_t \Tr (\mathcal{L}_-^n)=0 \, .
\end{equation}
References \cite{grava2023discrete,mazzuca2023equilibrium,spohn2020generalized,spohn2021hydrodynamic,spohn2022hydrodynamic} have used these conserved quantities to construct equilibrium ensembles for integrable models and random matrix ensembles. A notable remark is that the first conserved quantity $C_0$ also termed log-intensity in \cite{spohn2020generalized} seems missing from this construction. A combination of the kind 
\begin{equation}
    \frac{\mathcal{L}_+^n-\mathcal{L}_-^n}{n}
\end{equation}
in $n\to 0$ limit might allow to retrieve it.\\

As a final comment, let us mention that the local densities induced by the traces of the Lax matrices have recently been interpreted using random walks \cite{spohn2023hydrodynamic}. Indeed, defining the local densities as 
\begin{equation}
    Q_j^{[n,+]}=(\mathcal{L}_+^n)_{j,j}, \quad Q_j^{[n,-]}=(\mathcal{L}_-^n)_{j,j}, \quad j\in \N
\end{equation}
they be expanded as follows
\begin{equation}
  Q_j^{[n,+]}=\sum_{j_1=0}^{\infty}\dots \sum_{j_{n-1}=0}^{\infty}(\mathcal{L}_+)_{j,j_1}(\mathcal{L}_+)_{j_1,j_2}\dots (\mathcal{L}_+)_{j_{n-1},j}
\end{equation}
and similarly for $Q_j^{[n,-]}$. Each contribution $(\mathcal{L}_+)_{j_1,j_2}$ can be seen as the weight to go from site $j_1$ to site $j_2$ and the total weight $Q_j^{[n,+]}$ is the weight of starting from site $j$ and returning to site $j$ in $n$ steps. It would be interesting to extend this interpretation in the current context and characterise the geometry and properties of the random walk.

\section{Gauge equivalence of the Lax problem with a classical integrable implicit spin chain}
\subsection{Towards a Lax pair for a spin system}

Following the construction of Ishimori \cite{ishimori1982integrable} which relates the Ablowitz-Ladik system to the Haldane-Ishimori-Skylanin spin chain (an integrable version discretisation of the classical Landau-Lifschitz chain) we relate our system \eqref{eq:YO-polymer-saddle} to a semi-discrete spin model. To this aim we define a gauge transform $g_n$ so that

\begin{equation}
    \vec{v}_n = g_n \vec{w}_n
\end{equation}
which modifies the Lax pair in \eqref{eq:Lax-compatibility-supp-mat} as $(L_n,U_n)\to (\hat{L}_n,\hat{U}_n)$ as
\begin{equation}
\begin{split}
    \hat{L}_n&=g_{n+1}^{-1}L_n g_n\\
    \hat{U}_n &= g_n^{-1}U_n g_n -g_n^{-1}\p_t g_n 
    \end{split}
\end{equation}

This is a priori valid for an arbitrary $2\times 2$ gauge $g_n$ which we now specify. Here we define $g_n$ as the solution of \eqref{eq:Lax-compatibility-supp-mat} for $\lambda=1$ and zero drift $\{a_n=0 \}$ and for convenience we rewrite \eqref{eq:supp-mat-factorisation-lax-Ln} as $L_n = \mathsf{A}_n \mathsf{B}_n \mathsf{C}_n$ 
with
\begin{equation}
   \mathsf{A}_n= 
    \begin{pmatrix}
1  &   0 \\ 
-\frac{g}{2}\tilde{z}_n &  1
\end{pmatrix}
   \quad  \mathsf{B}_n =\begin{pmatrix}
\frac{1}{\lambda} &   0 \\ 
0 &  \lambda 
\end{pmatrix} \quad \mathsf{C}_n =\begin{pmatrix}
1  &   z_{n} \\ 
0 &  1
\end{pmatrix}
\end{equation}
We have that $g_n$ verifies the compatible system
\begin{equation}
\label{eq:supp-mat-evolution-gauge}
    \begin{split}
        g_{n+1}&= \begin{pmatrix}
1  &   0 \\ 
-\frac{g}{2}\tilde{z}_n &  1
\end{pmatrix}
\begin{pmatrix}
1  &   z_{n} \\ 
0 &  1
\end{pmatrix} g_n = \mathsf{A}_n \mathsf{C}_n g_n \\ 
\p_t g_n &= \begin{pmatrix}
0  &   -z_{n-1} \\ 
&\\
\frac{g}{2}\tilde{z}_n &  0
\end{pmatrix} g_n 
    \end{split}
\end{equation}
From the evolution of the gauge $g_n$ we can obtain the modified Lax pair in the gauge frame. The space Lax matrix $\hat{L}_n$ reads
\begin{equation}
\begin{split}
    g_{n+1}^{-1}L_n g_n &= g_n^{-1}\mathsf{C}_n^{-1}\mathsf{A}_n ^{-1} \mathsf{A}_n \mathsf{B}_n \mathsf{C}_ng_n  \\
    &= g_n^{-1}\mathsf{C}_n^{-1} \mathsf{B}_n \mathsf{C}_ng_n  \\
\end{split}
\end{equation}
We then decompose the $\mathsf{B}_n$ as
\begin{equation}
    \begin{split}
        \mathsf{B}_n 
&=\frac{\lambda+1/\lambda}{2}I_2+ \frac{-\lambda+1/\lambda}{2} \sigma_3
    \end{split}
\end{equation}
where $I_2$ is the $2\times 2$ identity matrix and $\sigma_3$ is the Pauli matrix $
   \sigma_3 = \begin{pmatrix}
1  &   0 \\ 
0 &  -1
\end{pmatrix}$. This yields

\begin{equation}
\begin{split}
 \hat{L}_n=  g_{n+1}^{-1}L_n g_n &=\frac{\lambda+1/\lambda}{2}I_2+ \frac{-\lambda+1/\lambda}{2}g_n^{-1}\mathsf{C}_n^{-1} \sigma_3\mathsf{C}_ng_n   \\
\end{split}
\end{equation}
As in Ref.~\cite{ishimori1982integrable}, we interpret the last term as a spin operator and to this aim we define the spin matrix $S_n$ as
\begin{equation}
\label{eq:def-spin-1}
    S_n = g_n^{-1}\mathsf{C}_n^{-1} \sigma_3\mathsf{C}_ng_n 
\end{equation}
to rewrite the Lax matrix in the following form
\begin{equation}
    \hat{L}_n=\frac{\lambda+1/\lambda}{2}I_2+ \frac{-\lambda+1/\lambda}{2}S_n   \\ 
\end{equation}
Before investigating the second Lax matrix $\hat{U}_n$, let us describe a few properties of the spin matrix $S_n$.
It is direct to see that 
\begin{equation}
\label{eq:properties-spin-matrix}
    S_n^2=I_2, \quad \Tr S_n=0, \quad \Det S_n=-1 \, .
\end{equation}
From \eqref{eq:supp-mat-evolution-gauge}, we see that the spin allows for different rewritings.
\begin{equation}
    S_n = g_{n+1}^{-1}\mathsf{A}_n \sigma_3 \mathsf{A}_n^{-1}g_{n+1} =   g_{n+1}^{-1}\mathsf{A}_n \sigma_3\mathsf{C}_ng_n  = g_n^{-1}\mathsf{C}_n^{-1} \sigma_3 \mathsf{A}_n^{-1}g_{n+1}
\end{equation}

The second time Lax matrix is given explicitly as
\begin{equation}
\label{eq:supp-mat-dual-spin}
\begin{split}
  \hat{U}_n=   g_n^{-1}U_n g_n -g_n^{-1}\p_t g_n &= \frac{\lambda^2-1}{2} g_n^{-1} \sigma_3 g_n\\
    \end{split}
\end{equation}
From the right hand side of \eqref{eq:supp-mat-dual-spin}, we define a second spin matrix
\begin{equation}
 \tilde{S}_n = g_n^{-1} \sigma_3 g_n   
\end{equation}
which has the same properties as $S_n$ of \eqref{eq:properties-spin-matrix}. Note that the gauge transformation preserved the following properties of the Lax matrices
\begin{equation}
    \Det \hat{L}_n=1 \, ,\quad \Tr \hat{U}_n=0
\end{equation}

\subsection{Compatibility equation and spin dynamics}
The compatibility equation of the gauged Lax pair reads
\begin{equation}
     \p_t \hat{L}_n=\hat{U}_{n+1}\hat{L}_n -\hat{L}_n \hat{U}_n
\end{equation}
Computing explicitly this compatibility equation, we observe that in order not to have a dependency on the spectral parameter $\lambda$ in the final equation, we are required to have the identity

\begin{equation}
\label{eq:supp-mat-compatibility-spin-cancellation}
  \tilde{S}_{n+1} S_n- S_n \tilde{S}_{n}=\tilde{S}_{n+1}-\tilde{S}_{n}  
\end{equation}
This identity is indeed true and can be computed by brute-force by expanding the spin matrices with their representations, e.g. \eqref{eq:def-spin-1}, by factorising to the left $g_{n+1}^{-1}$ and to the right $g_n$. From this identity, we deduce the evolution equation for the spin $S_n$
\begin{equation}
\label{eq:supp-mat-evolution-eq-spin}
\begin{split}
    \p_t S_n     &= -(\tilde{S}_{n+1}-\tilde{S}_{n}) \\
    \end{split}
\end{equation}
The spins $\tilde{S}_{n}$ thus carry the interpretation of a local spin current.

\subsection{A few properties to describe the spin system}
\subsubsection{Discrete parallel frame}

Following Ref.~\cite{hoffmann2008discrete}, it is possible to interpret the spin dynamics using a discrete regular curve $\gamma_n$ so that 
    \begin{equation}
    S_n=\gamma_{n+1}-\gamma_n \, .
\end{equation}
The spin dynamics is equivalent to the compatible system
\begin{equation}
    \begin{split}
        \p_t \gamma_n&=-\tilde{S}_n\\
        \Delta^+ \gamma_n &= S_n
    \end{split}
\end{equation}
Since we have from the normalisation of the spin $|\gamma_{n+1}-\gamma_n|^2=1$, $\gamma_n$ is said to be an arclength parametrized curve and $g_n$ can be interpreted as a discrete parallel frame \cite{hoffmann2008discrete}. \\
\subsubsection{Stereographic representation and mapping from spin matrix to spin vector}
Moreover, if we define the spin matrix $S_n$ in its stereographic representation \cite{papanicolaou1987complete} as
\begin{equation}
    S_n = \frac{1}{1+w_n \tilde{w}_n}
    \begin{pmatrix}
        1-w_n \tilde{w}_n& 2w_n\\
        2\tilde{w}_n & -1+w_n \tilde{w}_n
    \end{pmatrix}
\end{equation}
then we can represent it in a spin vector form as
\begin{equation}
    \vec{S}_n = \frac{1}{1+w_n \tilde{w}_n}
    \begin{pmatrix}
        w_n+\tilde{w}_n\\
      \I(  w_n-\tilde{w}_n)\\
        1-w_n \tilde{w}_n
    \end{pmatrix}
\end{equation}
where we endow the vectors with the canonical real inner product to preserve the norm $|\vec{S}_n|^2=1$. Note that since the coefficients $ \{w_n, \tilde{w_n} \}$ are real, this inner product is not compact, this can be seen by removing the complex $\I$ in the definition of the vector and define a signed inner product with signature $(+,-,+)$. A consequence of that is that two spins can have an inner product equal to 1 without equal, e.g. if $\tilde{w}_n=0$ for two different vectors, then their inner product is automatically equal to 1 even if the respective coefficient $w_n$ is arbitrary. This problem is solved in the case of complex fields when $\tilde{w}_n = w_n^*$, i.e. the coefficients are complex conjugate.\\

In the current context, we introduce the spin vectors to relate some conserved densities and currents of the model \eqref{eq:YO-polymer-saddle} to spin quantities as in the work of Ref.~\cite{zakharov1979equivalence}. For a generic traceless matrix endowed with the vector representation

\begin{equation}
    M=
    \begin{pmatrix}
        a & b\\
        c &-a
    \end{pmatrix}, \quad
    \vec{M}=
    \begin{pmatrix}
        (b+c)/2\\
        \I(b-c)/2\\
        a
    \end{pmatrix}
\end{equation}
we have the identity between the signed norm and the determinant $   |\vec{M}|^2=-\det(M)=\Tr(M^2)/2$. From a direct calculation, using \eqref{eq:YO-polymer-saddle} and \eqref{eq:supp-mat-evolution-gauge}, we have that

\begin{equation}
\begin{split}
    \p_t S_n &= -2 g_{n+1}^{-1}
    \begin{pmatrix}
        0 & z_n \\
        \frac{g}{2}\tilde{z}_n & 0
    \end{pmatrix}g_n\\
    \p_t \tilde{S}_n &= -2 g_{n}^{-1}
    \begin{pmatrix}
        0 & z_{n-1} \\
        \frac{g}{2}\tilde{z}_n & 0
    \end{pmatrix}g_n
\end{split}
\end{equation}

Taking the determinants of these equations and using the matrix determinant - spin identity, we obtain
\begin{equation}
\label{eq:supp-mat-duality-spin-conservedquantities}
\begin{split}
    |\p_t \vec{S}_n|^2 &=- \det (\p_t S_n)=2g z_n \tilde{z}_n\\
    |\p_t \vec{\tilde{S}}_n|^2 &=- \det (\p_t \tilde{S}_n)=2g z_{n-1} \tilde{z}_n
    \end{split}
\end{equation}
These expressions are exactly the densities and currents derived in Eq.~\eqref{eq:conserved-quantities-laurent-suppmat}.\\

On top on the vector norms obtained, we complement this discussion by providing a few vector inner products. Using the evolution equation for $S_n$  \eqref{eq:supp-mat-evolution-eq-spin} as well as \eqref{eq:supp-mat-duality-spin-conservedquantities}, this yields the following inner products
\begin{equation}
\begin{split}
    (\vec{\tilde{S}}_{n+1}-\vec{\tilde{S}}_{n})\cdot \vec{\tilde{S}}_{n+1}&=gz_n \tilde{z}_n\\
    (\vec{\tilde{S}}_{n+1}-\vec{\tilde{S}}_{n})\cdot \vec{\tilde{S}}_{n}&=-gz_n \tilde{z}_n\\
\end{split}
\end{equation}
Finally, we also have for any two spins $    \vec{S}_n \cdot\vec{\tilde{S}}_m=\frac{1}{2}\Tr(S_n \tilde{S}_m)$ 
and therefore from Eq.~\eqref{eq:supp-mat-compatibility-spin-cancellation}, we obtain
\begin{equation}
     \vec{S}_n \cdot(\vec{\tilde{S}}_{n+1}-\vec{\tilde{S}}_n)=0
\end{equation}
which is also interpreted as $\vec{S}_n \cdot \p_t \vec{S}_n=0$, hence the infinitesimal variation of the spin $S_n$ is orthogonal to the direction of the spin, which is coherent with the conservation of the norm of the spin.

\subsubsection{In search of a Hamiltonian}

At this stage, it is natural to ask whether there exists an explicit Hamiltonian $\mathcal{H}$ such that the spin evolution can be written as

\begin{equation}
\label{eq:hamiltonian-spin-evolution}
    \p_t \vec{S}_n = - \vec{S}_n \wedge \frac{\delta \mathcal{H}}{\delta \vec{S_n}} \quad  \Longleftrightarrow  \quad 
    \vec{S}_n \wedge \frac{\delta \mathcal{H}}{\delta \vec{S_n}} =\vec{\tilde{S}}_{n+1}-\vec{\tilde{S}}_{n}
\end{equation}

We do expect that the Hamiltonian $\mathcal{H}$ matches a combination of the conserved quantities of the O'Connell-Yor polymer weak noise theory after a suitable change of variable from the fields $\{z_n , \tilde{z}_n\}$ to the spins $\{\vec{S}_n \}$, see notably the conserved quantities $C_0, C_1, \tilde{C}_1, \tilde{C}_2 $ in Section~\ref{subsec:supp-mat-conserved-quantities}.\\

Since both sides of the equations in  \eqref{eq:hamiltonian-spin-evolution} are orthogonal to $\vec{S}_n$, we introduce the Frenet frame which can be used to decompose the vectors in a convenient basis. This frame is described by three vectors $\{t_n, b_n, m_n \}$ defined as
\begin{equation}
    \{t_n= \vec{S}_n, \quad b_n=\frac{\vec{S}_{n-1} \wedge \vec{S}_n}{|\vec{S}_{n-1} \wedge \vec{S}_n|}, \quad m_n = b_n \wedge t_n \}
\end{equation}

 The relation between $S_n$ and $\tilde{S}_n$ is still implicit at this stage but there are instances where this is easily explicit. Indeed, in the non-interacting cases, we have that 
\begin{enumerate}
    \item If $\tilde{z}_n=0$, $S_n=\tilde{S}_{n+1}$ due to the fact that $\mathsf{A}_n$ is the identity. In which case we obtain
\begin{equation}
\begin{split}
    \p_t \vec{S}_n     &= \vec{S}_{n-1}-\vec{S}_n \\
    \end{split}
\end{equation}
    \item If $z_n=0$, $S_n=\tilde{S}_{n}$ due to the fact that $\mathsf{C}_n$ is the identity. In which case we obtain
    \begin{equation}
\begin{split}
    -\p_t \vec{S}_n     &= \vec{S}_{n+1}-\vec{S}_{n} \\
    \end{split}
\end{equation}
\end{enumerate}
These equations are reminiscent of the system \eqref{eq:YO-polymer-saddle} in the non-interacting case. More generally, we leave it as an open question to decompose the vector $\tilde{S}_n$ in the Frenet frame and find the general Hamiltonian $\mathcal{H}$.\\

Finally, the spin model related to the Ablowitz-Ladik model was recently used  in Ref.~\cite{angelopoulos2018invariant} to investigate its invariant measures. It would be interesting to see whether some techniques of Ref.~\cite{angelopoulos2018invariant} extend in our context of the weak noise theory of the O'Connell-Yor polymer.

\section{Small-time limit of the Fredholm determinant result for the O'Connell-Yor polymer}

In this section we start from the formula for the generating function of $Z^\beta_N(t)$, the OY partition sum studied in \cite{imamura2016determinantal} which is  is identical to ours if we identify $\beta = \sqrt{\varepsilon}$. We study its weak noise limit $\varepsilon \ll 1$, leading to a conjectural form for $\Psi_N(\Lambda)$
which agrees with the one derived in the text using inverse scattering. The manipulations in this
Appendix are quite heuristic but they have the merit to show that the algebraic structure which emerges from 
the Fredholm determinant is similar to the one derived in the text from first principles by the inverse scattering method. The method applied in this Appendix is similar to the one used in Ref.~\cite[Section X of the Supp.Mat]{krajenbrink2023crossover} in the context of the crossover from the large deviations of the macroscopic fluctuation theory to the weak noise theory of the KPZ equation.\\ 

The starting identity, given in Refs.~\cite{BorodinMacdo} and \cite[Proposition 14]{imamura2016determinantal} reads
\begin{equation}
    \E \left[e^{-\frac{e^{-\beta u}Z^\beta_N(t)}{\beta^{2(N-1)}}} \right] = \Det(I+L)_{L^2(C_0)}
\end{equation}
where the kernel is written as
\begin{equation}
\begin{split}
    L(v,v')&=\int_{\I \R+\delta}\frac{\rmd w}{2\I \pi} \frac{\pi}{\sin(\pi(v'-w))} \frac{e^{\varepsilon w^2 t/2-w \tilde u}}{e^{\varepsilon v'^2 t/2-v'\tilde u}}\frac{1}{w-v}\frac{\Gamma(v')^N}{\Gamma(w)^N}\\
    \end{split}
\end{equation}
with $C_0$ a contour enclosing the origin with radius $r<1 /2$ and $r<\delta<1 -r$. The measure over $C_0$ is  $\rmd v/2\I \pi$ and we further have the constraint that
\begin{equation}
    0<\Re(w-v')<1
\end{equation}
The identification with our variables reads
\be 
z_N(t) = e^{- (1 + \frac{\varepsilon}{2}) t} Z_N(t) \quad, \quad Z_N(t) = Z^{\beta=\sqrt{\varepsilon}}_N(t)
\ee 
as well as
\begin{equation}
    \beta = \sqrt{\varepsilon}
\end{equation}
so that the Fredholm determinant identity yields

\begin{equation}
\label{eq:supp-mat-fred-det-sasamoto}
    \E \left[e^{- \varepsilon^{N-1} e^{- \tilde u + (1 + \frac{\varepsilon}{2}) t} z_N(t)}  \right] = \Det(I+L)_{L^2(C_0)}
\end{equation}
To complete our identification, we further introduce the relation
\begin{equation}
\label{eq:identification-variables-supp-mat}
    \varepsilon^{N-1} e^{- \tilde u + (1 + \frac{\varepsilon}{2})} = \frac{\Lambda}{\varepsilon}
\end{equation}
Endowed with the identification, we now transform the Fredholm identity \eqref{eq:supp-mat-fred-det-sasamoto} to derive the large deviation function $\Psi_N(\Lambda)$. The first step is the introduction of a factorization of the kernel $L(v,v')$ as

\begin{equation}
\begin{split}
    L(v,v')    &= \int_{\I \R+\delta}\frac{\rmd w}{2\I \pi}A(v,w)\tilde{A}(w,v')
    \end{split}
\end{equation}
where have defined the kernels
\begin{equation}
    A(v,w)=\frac{1}{v-w}, \quad \tilde{A}(w,v')=\frac{\pi}{\sin(\pi(w-v'))}  \frac{e^{\varepsilon w^2 t/2-w\tilde{u}}}{e^{\varepsilon v'^2 t/2-v'\tilde{u}}}\frac{\Gamma(v')^N}{\Gamma(w)^N}
\end{equation}

Introducing the identity
\begin{equation}
    \frac{\pi}{\sin(\pi s)}z^s = \int_\R \rmd r \, \frac{z}{z+e^{-r}}e^{-sr}, \quad 0<\Re(s)<1
\end{equation}
which we apply using
\begin{equation}
    s=w-v', \quad z=e^{-\tilde u}
\end{equation}
This allows to re-write the kernel $\tilde{A}(w,v')$ and to factorize it as
\begin{equation}
\begin{split}
    \tilde{A}(w,v')&=\int_\R \rmd r \, \frac{1}{1+e^{ \tilde u-r}}  \frac{e^{\varepsilon w^2 t/2-wr}}{e^{\varepsilon v'^2 t/2-v'r}}\frac{\Gamma(v')^N}{\Gamma(w)^N}\\
    &= \int_\R \rmd r \,  \sigma(r)A_1(w,r)A_2 (r,v')
    \end{split}
\end{equation}
where we have defined 
\begin{equation}
    \sigma(r)= \frac{1}{1+e^{\tilde{u}-r}}, \quad A_1(w,r) =  \frac{1}{\Gamma(w)^N } e^{\varepsilon w^2 t/2-wr}, \quad A_2(r,v')= e^{-\varepsilon v'^2 t/2+v'r}\Gamma(v')^N
\end{equation}

To summarise, the Fredholm manupulations we have done lead to the factorisation
\begin{equation}
\begin{split}
    \Det(1+L)_{L^2(C_0)}&=\Det(1+AA_1\sigma A_2)_{L^2(C_0)}\\
    &=\Det(1+\sigma A_2 AA_1)_{L^2(\R)}\\
    \end{split}
\end{equation}
where from the first to the second line we have used Sylvester's identity $\Det(I+AB)=\Det(I+BA)$. This last Fredholm determinant has the typical structure for which the first cumulant method, developed in \cite{KrajLedou2018,ProlhacKrajenbrink,krajenbrink2019beyond} to study the relevant asymptotics (here small $\varepsilon$), applies.
This allows to formally interpret this Fredholm as an expectation value of a determinantal point process $\{a_\ell\}_{\ell \geq 1}$ which correlations are controlled by the kernel $A_2 AA_1$. Hence

\begin{equation}
\label{eq:supp-mat-fredholm-det-manip-1}
\begin{split}
    \Det(1+L)_{L^2(C_0)}    &= \mathbb{E}\left[\prod_{\ell=1}^{+\infty}(1-\sigma(a_\ell))\right]\\
    &= \mathbb{E}\left[\prod_{\ell=1}^{+\infty}e^{-\varphi(a_\ell)}\right]\\
    \end{split}
\end{equation}
where we have introduced $e^{-\varphi}=1-\sigma$ to interpret \eqref{eq:supp-mat-fredholm-det-manip-1} as a linear statistics of the point process over the observable
\begin{equation}
\varphi(r)=\log(1+e^{r-\tilde{u}})=-\textrm{Li}_1(-e^{r-\tilde{u}})
\end{equation}

The first cumulant approximation applied to the expectation value \eqref{eq:supp-mat-fredholm-det-manip-1} asserts \cite[Section 6]{KrajLedou2018} that as some parameter goes to infinity (here it will be $1/\varepsilon$, see below),
we expect the point process to self-average, i.e.
\begin{equation}
  \Det(1+L)_{L^2(C_0)}=\Det(1+\sigma A_2 AA_1)_{L^2(\R)}= \mathbb{E}\left[\prod_{\ell=1}^{+\infty}e^{-\varphi(a_\ell)}\right]\sim e^{\Tr(\varphi A_2 A A_1)} 
\end{equation}
The quantity of interest only involves the diagonal part of the kernel $A_2 A A_1$ and we have to evaluate 

\begin{equation}
    \begin{split}
        \Tr(\varphi A_2 A A_1)&= \int_\R \rmd r \int_{\I \R+\delta}\frac{\rmd w}{2\I \pi} \int_{C_0}\frac{\rmd v'}{2\I \pi}\, \textrm{Li}_1(-e^{r-\tilde{u}}) e^{\varepsilon w^2 t/2- \varepsilon v'^2 t/2}\frac{\Gamma(v')^N}{\Gamma(w)^N }\frac{e^{-(w-v')r}}{w-v'}
    \end{split}
\end{equation}
taking into account that the measure on the variables $v'$ is $\frac{\rmd v'}{2 \I \pi}$. Since $0<\Re(w-v')<1$, we can proceed to an integration by part with respect to the integration variable $r$ to obtain 
\begin{equation}
    \begin{split}
        \Tr(\varphi A_2 A A_1)&= \int_\R \rmd r \int_{\I \R+\delta}\frac{\rmd w}{2\I \pi} \int_{C_0}\frac{\rmd v'}{2\I \pi}\, \textrm{Li}_2(-e^{r-\tilde{u}}) e^{\varepsilon w^2 t/2-\varepsilon v'^2 t/2}\frac{\Gamma(v')^N}{\Gamma(w)^N }  e^{-(w-v')r}
    \end{split}
\end{equation}
The function $\textrm{Li}_2$ denotes the dilogarithm.\\

As a summary, after replacing the variable $\tilde{u}$ by the identification \eqref{eq:identification-variables-supp-mat}, we obtain that

\begin{equation}
    \mathbb{E}\left[e^{-\frac{\Lambda}{\varepsilon}z_N(1)} \right] \sim e^{-\Tr(-\varphi A_2 A A_1)} 
\end{equation}
and upon taking $t=1$ in the kernel, we have
\begin{equation}
    \begin{split}
      -  \Tr(\varphi A_2 A A_1)&= -\int_\R \rmd r \int_{\I \R+\delta}\frac{\rmd w}{2\I \pi} \int_{C_0}\frac{\rmd v'}{2\I \pi}\, \textrm{Li}_2(-e^{-(1 +\frac{\varepsilon }{2})}   \varepsilon^{-N} \Lambda e^{r}) \frac{e^{\varepsilon w^2 /2-wr}}{e^{\varepsilon v'^2  /2-v'r}}\frac{\Gamma(v')^N}{\Gamma(w)^N } \\
      &=-\int_\R \rmd r \, \textrm{Li}_2(-\Lambda e^{r}) I(r+1+\varepsilon /2+N\log \varepsilon)
    \end{split}
\end{equation}
from the first to the second line, we have shifted the variable $r$ to absorb the additional factors in the dilogarithm.  We further perform the change of variable
\begin{equation}
    w=\frac{\tilde{w}}{\varepsilon}, \quad v'=\frac{\tilde{v}'}{\varepsilon}
\end{equation}
and subsequently drop the tilde. Upon this change, the integral $I$ reads
\begin{equation}
\begin{split}
     I(R=r+1 +\varepsilon /2+N\log \varepsilon) 
     &= \frac{1}{\varepsilon^2}\int_{\I \R+\delta'}\frac{\rmd w}{2\I \pi} \int_{C'_0}\frac{\rmd v'}{2\I \pi}\,  \frac{e^{\frac{1}{\varepsilon}(w^2 /2-wR)}}{e^{\frac{1}{\varepsilon}(v'^2  /2-v'R)}}\frac{\Gamma(v'/\varepsilon)^N}{\Gamma(w/\varepsilon)^N }\\
     \end{split}
\end{equation}
which we formally write in the following form
\begin{equation}
\label{eq:sup-mat-fred-det-rate-func}
     I(R=r+1 +\varepsilon /2+N\log \varepsilon) = \frac{1}{\varepsilon^2}\int_{\I \R+\delta'}\frac{\rmd w}{2\I \pi} \int_{C'_0}\frac{\rmd v'}{2\I \pi}\,  \frac{e^{\frac{1}{\varepsilon}\Phi(w)}}{e^{\frac{1}{\varepsilon}\Phi(v')}}
\end{equation}
Since the parameter $\varepsilon$ is taken to  be small, it is natural to introduce the function $\Phi(w)$ to obtain a rate function. Its expression reads
\begin{equation}
    \Phi(w)=\frac{w^2}{2}-wR -\varepsilon N \log \Gamma(\frac{w}{\varepsilon}) 
\end{equation}
We now evaluate the two integrals in $I$ using the saddle point method controlled by the rate $1/\varepsilon$. 
From the asymptotics of the $\Gamma$ function
\begin{equation}
    \log \Gamma(z) = z \log z -z -\frac{1}{2}\log z + \frac{1}{2}\log 2\pi + \frac{1}{12z} -\frac{1}{360 z^3} + \mathcal{O}(1/z^5) 
\end{equation}
alongside the expression $R=r+1 +\varepsilon /2+N\log \varepsilon$ one obtains in the limit $\varepsilon \to 0$ 
\begin{equation}
    \Phi(w)= \frac{w^2}{2}+(N-1-r) w- N w \log (w)+\frac{1}{2} \varepsilon  (N \log (w)-N \log (\varepsilon )-N \log (2 \pi )-w)+\mathcal{O}(\varepsilon^2)
\end{equation}
At leading order, we re-write $\Phi(w)$

\be 
\frac{1}{\varepsilon}\Phi(w) =\frac{1}{\varepsilon}(\phi(w)-rw) - \frac{1}{2} w +\frac{N}{2} \log(w)  -\frac{N}{2} \log (2 \pi \varepsilon)+\mathcal{O}(\varepsilon)
\ee 
and obtain the rate function $\phi(w)$ as
\be 
\phi(w) = \frac{w^2}{2}+ (N-1)w  - N w \log w
\ee 
Its derivative reads 
\be 
\phi'(w) = w- 1 - N \log w
\ee 
The saddle point equation is therefore
\be 
\label{eq:saddle-point0sup-mat-for-contour}
\phi'(w)=r \Longleftrightarrow  e^r = w^{-N} e^{w-1}
\ee 

At the saddle point for $w$ and $v'$ the subdominant terms as well as the 
constants compensates between the two integrals in \eqref{eq:sup-mat-fred-det-rate-func} and we have the following estimate
\begin{equation}
    I(R)\simeq \frac{1}{2\I \pi \varepsilon} \frac{1}{\phi''(w(r))}
\end{equation}
so that
\begin{equation}
\label{eq:sup-mat-fred-det-first-cum-saddle}
    \begin{split}
      -  \Tr(\varphi A_2 A A_1)   &=-\frac{1}{ 2\I \pi \varepsilon}\int_\gamma \rmd r \, \textrm{Li}_2(-\Lambda e^{r}) \frac{1}{\phi''(w(r))}
    \end{split}
\end{equation}
To make this saddle point easily attainable, one way is to deform the integration contour of $r$ which is not $\R$ anymore but the image of \eqref{eq:saddle-point0sup-mat-for-contour} as $v'$ varies along $C_0'$, which we call $\gamma$. We have also assumed that the integration contour of $w$ could be deformed to be folded around $C_0'$.\\

It is possible to further simplify the estimate of the first cumulant in \eqref{eq:sup-mat-fred-det-first-cum-saddle} by proceeding to a change of variable $r \to w$ using the saddle point equation $\phi'(w(r))=r$. The Jacobian of this change of variable reads
\begin{equation}
    \phi''(w(r))\frac{\rmd w}{\rmd r}=1, \quad \frac{\rmd r}{\phi''(w(r))}=\rmd w
\end{equation}
To summarize, the first cumulant of the Fredholm determinant reads in the small $\varepsilon$-limit
\begin{equation}
    \begin{split}
      -  \Tr(\varphi A_2 A A_1)   &=-\frac{1}{\varepsilon }\int_{C_0'} \frac{\rmd w}{2\I \pi} \, \textrm{Li}_2(-\Lambda w^{-N} e^{w-1}) 
    \end{split}
\end{equation}
and our calculation have given us the following large deviation principle
\begin{equation}
    \mathbb{E}\left[e^{-\frac{\Lambda}{\varepsilon}z_N} \right] \sim e^{-\frac{1}{\varepsilon}\Psi(\Lambda)}
\end{equation}
where the final large deviation function reads
\begin{equation}
   \Psi(\Lambda) = - \int_{C_0'} \frac{\rmd w}{2\I \pi} \, \mathrm{Li}_2(-\Lambda w^{-N} e^{w-1}) \, .
\end{equation}
This agrees with the result of the main text \eqref{eq:resultPsi} using the inverse scattering method. 

\section{The Lambert W function} \label{app:lambert}
We introduce the Lambert $W$ function \cite{corless1996lambertw} which we use extensively throughout this work. Consider the function defined on $\mathbb{C}$ by $f(z)=ze^z$, the $W$ function is composed of all inverse branches of $f$ so that $W(z e^z)=z$. It does have two real branches, $W_0$ and $W_{-1}$ defined respectively on $[-e^{-1},+\infty[$ and $[-e^{-1},0[$. On their respective domains, $W_0$ is strictly increasing and $W_{-1}$ is strictly decreasing. By differentiation
 of $W(z) e^{W(z)}=z$, one obtains a differential equation valid for all branches of $W(z)$
\begin{equation} \label{derW} 
\frac{\rmd W}{\rmd z}(z)=\frac{W(z)}{z(1+W(z))}
\end{equation}
Concerning their asymptotics, $W_0$ behaves logarithmically for large argument $W_0(z)\simeq_{z\to +\infty} \log(z)-\log \log (z)$ and is linear for small argument  $W_0(z) =_{z \to 0} z-z^2+\mathcal{O}(z^3)$. $W_{-1}$ behaves logarithmically for small argument $W_{-1}(z)\simeq_{z \to 0^-} \log(-z)-\log(-\log(-z))$. Both branches join smoothly at the point $z=-e^{-1}$ and have the value $W(-e^{-1})=-1$. These remarks are summarized on Fig.~\ref{fig:Lambert}. More details on the
other branches, $W_k$ for integer $k$, can be found in \cite{corless1996lambertw}.

\begin{figure}[h!] 
\begin{center}
\includegraphics[width = 0.6\linewidth]{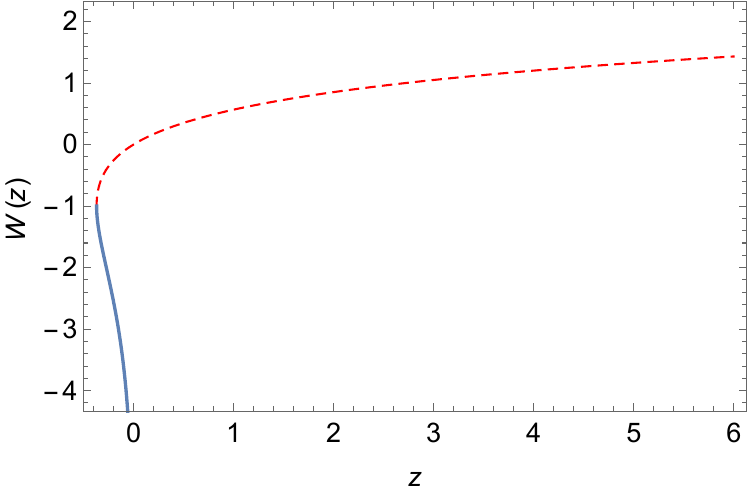}
\caption{The Lambert function $W$. The dashed red line corresponds to the branch $W_0$ whereas the blue line corresponds to the branch $W_{-1}$. }
\label{fig:Lambert}
\end{center}
\end{figure}

\end{widetext}

\end{document}

%% file: bibliography_orderedV1.tex
\newpage{\pagestyle{empty}\cleardoublepage}

%% file: PRL_v2_wnt-YOpolymer.bbl
\begin{thebibliography}{100}
\bibitem{footnotecontour} Recall that for $\Lambda>\Lambda^*_N$ the integration contour must be 
elongated along the imaginary direction to avoid the branch cuts of
the logarithm, see \cite{SM}.
\end{thebibliography}
